\documentclass[10pt,twocolumn,twoside,journal]{ieeetran}
\usepackage{amsmath,amssymb,amsfonts}
\usepackage{graphicx}
\usepackage{textcomp}

\usepackage[backend=bibtex,style=numeric,giveninits=true,maxnames=4,sorting=none,doi=false,isbn=false,url=true,eprint=false]{biblatex}

\usepackage{graphicx}
\usepackage{amsmath}
\usepackage[caption=false,font=footnotesize]{subfig}
\usepackage{fixltx2e}
\usepackage{stfloats}
\usepackage{url}
\usepackage{varwidth}
\usepackage{lipsum}
\usepackage{xcolor}
\usepackage{algorithm}
\usepackage{rotating}
\usepackage{algorithmicx}
\usepackage{algpseudocode}
\usepackage{csquotes}
\usepackage{amsfonts}
\usepackage{bm}
\usepackage{float}
\usepackage{gensymb}
\usepackage{placeins}
\usepackage[export]{adjustbox}
\usepackage{soul}
\usepackage{hyperref}
\usepackage{fancyhdr}
\renewcommand\st[1]{\unskip}
\bibliography{IEEEabrv,bibliography.bib}

\begin{document}
\title{WNet: A data-driven dual-domain denoising model for sparse-view computed tomography with a trainable reconstruction layer}

\author{Theodor Cheslerean-Boghiu, Felix C. Hofmann, Manuel Schulthei\ss, Franz Pfeiffer, Daniela Pfeiffer, \\and Tobias Lasser \IEEEmembership{Member,~IEEE,}
\thanks{Manuscript received 10 May 2022; revised 31 October 2022 and 2 January 2023; accepted 10 January 2023. Date of publication 26 January 2023; date of current version 20 February 2023. This work was partially supported by the German Federal Ministry of Health (2520DAT920) and funded by the Federal Ministry of Education and Research (BMBF) and the Free State of Bavaria under the Excellence Strategy of the Federal Government and the L\"ander, by the German Research Foundation (DFG) under GRK2274, as well as by the Technical University of Munich \- Institute for Advanced Study.}
\thanks{This work involved human subjects or animals in its research. Approval of all ethical and experimental procedures and protocols was granted by the Ethics Commission of the Medical Faculty, Technical University of Munich, Germany, Application No. 87/18S, and performed in line with the Declaration of Helsinki.}
\thanks{T. Cheslerean-Boghiu and T. Lasser are with the Computational Imaging and Inverse Problems, Department of Informatics, School of Computation, Information, and Technology and Munich Institute of Biomedical Engineering of the Technical University of Munich, Germany (email: theo.cheslerean@tum.de, lasser@in.tum.de).}
\thanks{F. Hofmann and D. Pfeiffer are with the Department of Diagnostic and Interventional Radiology, School of Medicine Klinikum rechts der Isar, Technical University of Munich, Germany (email: felix.hofmann@tum.de, daniela.pfeiffer@tum.de).}
\thanks{M. Schulthei\ss \, and F. Pfeiffer are with the Chair of Biomedical Physics, Department of Physics, School of Natural Sciences, Technical University of Munich, Germany (email: manuel.schultheiss@tum.de, franz.pfeiffer@tum.de).}
\thanks{D. Pfeiffer and F. Pfeiffer are also with the Institute for Advanced Study, Technical University of Munich.}
\thanks{This article has supplementary downloadable material available at https://doi.org/10.1109/TCI.2023.3240078, provided by the authors.}
}

\maketitle

\newcommand{\associateed}[1]{\textcolor{black}{#1}}
\newcommand{\reviewerone}[1]{\textcolor{black}{#1}}
\newcommand{\reviewertwo}[1]{\textcolor{black}{#1}}
\newcommand{\reviewerthree}[1]{\textcolor{black}{#1}}
\newcommand{\reviewerfour}[1]{\textcolor{black}{#1}}
\newcommand{\correction}[1]{\textcolor{black}{#1}}

\newcommand{\correctionsecond}[1]{\textcolor{black}{#1}}
\newcommand{\reviewerfoursecond}[1]{\textcolor{black}{#1}}


\fboxsep=0pt
\fboxrule=1pt

\begin{abstract}
Deep learning based solutions are being succesfully implemented for a wide variety of applications. Most notably, clinical use-cases have gained an increased interest and have been the main driver behind some of the cutting-edge data-driven algorithms proposed in the last years. For applications like sparse-view tomographic reconstructions, where the amount of measurement data is small in order to keep acquisition \reviewerfour{\st{times}time} short and radiation dose low, reduction of the streaking artifacts has prompted the development of data-driven denoising algorithms with the main goal of obtaining diagnostically viable images with only a subset of a full-scan data. We propose WNet, a data-driven dual-domain denoising model which contains a trainable reconstruction layer for sparse-view artifact denoising. Two encoder-decoder networks perform denoising in both sinogram- and reconstruction-domain simultaneously, while a third layer implementing the Filtered Backprojection algorithm is sandwiched between the first two and takes care of the reconstruction operation. We investigate the performance of the network on sparse-view chest CT scans, and we highlight the added benefit of having a trainable reconstruction layer over the more conventional fixed ones. We train and test our network on two clinically relevant datasets and we compare the obtained results with three different types of sparse-view CT denoising and reconstruction algorithms.

\end{abstract}

\begin{IEEEkeywords}
sparse-view computed tomgography, deep learning, precision learning, dual-domain, trainable kernel
\end{IEEEkeywords}
\IEEEpeerreviewmaketitle

\thispagestyle{fancy}
\cfoot{\small \copyright \copyright 2023 IEEE. Personal use of this material is permitted. Permission from IEEE must be obtained for all other uses, in any current or future media, including reprinting/republishing this material for advertising or promotional purposes, creating new collective works, for resale or redistribution to servers or lists, or reuse of any copyrighted component of this work in other works.}

\section{Introduction} \label{section:introduction}

\IEEEPARstart{X}{-ray} Computed Tomography (CT) is a type of non-destructive testing widely employed in medical imaging applications for diagnostic purposes. The use of X-radiation as the imaging medium, however, poses a challenging risk to patients exposing them to ionizing radiation. 
\reviewerfour{\st{Thus, one of the main goals of today's research for X-ray CT is the reduction of the radiation exposure the patient is subjected to. }Thus, one of the main goals of today's research for X-ray CT is to reduce radiation exposure to patients.} 
One approach geared towards lowering the dose is sparse-view CT (SVCT), which \reviewerfour{\st{aims at decreasing}aims to decrease} the number of \correction{\st{exposures to be measured} measured exposures}. \correction{\st{However} Unfortunately}, while this approach lowers the total radiation dose a patient is exposed to, an insufficiently sized set of projections has been shown to generate streaking artifacts in images reconstructed with conventional analytical methods \cite{hdnet}. 

\subsection{Classical Tomographic Reconstruction Techniques}
Analytical methods like the Filtered Backprojection (FBP) \cite{tv-why-sidky} for \reviewerfour{\st{two-dimensional}parallel-beam} tomographic reconstruction and the FDK algorithm \cite{projdomweights} for the \reviewerfour{\st{three-dimensional}cone-beam} tomographic reconstruction case have been shown to perform well when full-view measurement data is available. \correction{\st{As the sparsity of the projections becomes higher}With a higher projection sparsity}, such methods generate unwanted streaks in the final image \correction{\st{which lower}diminishing} their diagnostic value \cite{dataconsistlimited}. \correction{\st{Recently}Current} research has been done towards training the FDK filtering weights with some success in de-streaking limited angle CT reconstructions using a data-driven fine-tuned sinogram filtering step \cite{projdomweights}.

Iterative reconstruction techniques have been proposed to replace analytical ones for SVCT. Such methods allow for the incorporation of a-priori information about the underlying problem, in the form of penalization terms, with the development of various regularization techniques for compressed sensing like total variation (TV) \cite{tv-chambolle,tv-sidky,tv-why-sidky} and its versions \cite{tv-adaptive-liu,tv-nonlocal-zhang,l0-li}, or even algorithms which enforce sparsity in other domains like wavelets \cite{tv-wavelets} or curvelets \cite{curvelets}. \correction{\st{The}A} major disadvantage of the iterative methods, however, is the high computational cost of the repeated forward and backward projection operations during the iterative update steps \cite{framelet}\reviewerfour{\st{and the necessity of tuning several hyperparameters (e.g. regularization strength and number of iterations until convergence is achieved) which are required for good performance of the algorithm.}. Additionally, \correction{empirically} tuning several hyperparameters (e.g. regularization strength, loss balancing, and number of iterations until convergence is achieved) is required for the algorithm to perform well.}

\subsection{Deep Learning based Denoising of Sparse-view CT}

Deep learning (DL) methods have shown excellent performance in a multitude of different areas like classification \cite{dl-classification}, denoising \cite{dl-denoising}, or superresolution \cite{dl-superresolution} for clinical applications. 
Studies have shown that DL methods naturally learn representative low, middle and high level features in the given training dataset \cite{dl-vis-features} to \reviewerfour{\st{generate labels or correctly perform denoising for new inputs which}correctly perform denoising for new inputs. This} makes them the perfect candidate to be used towards reducing or (ideally) eliminating regular streaking artifacts \correction{\st{for} present in} SVCT \correction{reconstructions}.

A nice overview of state-of-the-art DL-based methods for medical image processing is provided by Maier et al. \cite{gentleintrodl}, where the authors mention several competing solutions for SVCT, many of whom are based on the UNet or GAN architectures. Some of the first instances of such de-streaking algorithms can be traced back to FBPConvNet \cite{fbpconvnet}, which, as a standard encoder-decoder model, takes a sparse-view FBP reconstruction and is trained towards producing a de-streaked version of it. Zhang et al. extended FBPConvNet with residual connections, DD-Net \cite{ddnet}, which has shown greater capability of suppressing noise and artifacts while recovering contrast information. Han et al. \cite{framelet} argued that conventional CNNs assume local invariance in the input image but the global nature of the streaking artifacts requires CNN architectures with large receptive fields. They proposed an encoder-decoder framelet-domain denoising method which overcomes the limitation of the small receptive fields imposed by the convolutional operation. Lee et al. \cite{wavelet} improved on the framelet-domain architecture with sparsity enforcing wavelet-based data-driven denoising to enlarge the receptive field of conventional convolutions. More recently, generator-discriminator architectures, like the Generative Adversarial Networks, have been shown to produce superior results to UNet-based model both for sparse-view and for low-dose CT, in part thanks to their usage of feature-extracting adversarial loss \correction{\st{compared to a more conventional}instead of conventional} pixel-based loss \cite{wgan,lowdosewasserstein}.

The major limitation to these methods, however, remains the fact that artifact denoising is performed directly on the reconstructed image. This single-domain approach disregards any prior knowledge encoded in the original measurements. Dong et al. \cite{sino-unet} proposed to perform sinogram-domain instead of image-domain denoising and then reconstruct an artifact-free image. However, this approach is again limited to performing in a single domain. New studies have shown that cascaded CNNs performing denoising in both projection and reconstruction greatly improve the quality of the reconstruction over single-domain approaches \cite{comparison,dronenet}.

\subsection{Combined Denoising-Reconstruction Algorithms}

Precision Learning is a concept where known tomographic operators are included into the training process, thus enabling both measurement-domain and reconstruction-domain feature learning. Embedding the operator in the network usually leads to a reduction in the total number of trainable parameters and to a lower maximal training error \cite{precisionlearning_v2}. 

Dual-domain denoising architectures for SVCT that take advantage of this concept have been proposed and successfully employed towards reducing streaking artifacts \cite{hdnet}. The insertion of a reconstruction module into the network enables the full use of information in both projection and image domain \cite{dudonet}. The majority of the proposed networks with such a feature use a fixed FBP/FDK layer as the domain-transform operation inserted between two encoder-decoder networks used to extract features in the sinogram- and reconstruction-domain: \cite{hdnet} for sparse-view CT, \cite{dualdomainhelical} for helical CT, \cite{dataconsistlimited} for limited angle CT, and \cite{dudonet} for metal artifact reduction. \reviewerfoursecond{More recently, transformer-based architectures have been shown to outperform CNNs in image classification and denoising applications, one of such architectures being proposed by Shi et al. \cite{svcttrans}, who use a transformer encoder to extract features in image domain and a transformer decoder to link those features to the ones extracted from the sinogram. Another transformer-based architecture proposed by Wang et al. \cite{dudotrans} uses a Swin Transformer to merge sinogram and image features.}

The filters used by \reviewerfoursecond{all the aforementioned} networks in the FBP/FDK layers are fixed, discretized analytical filters originally developed for full-view CT. Such filters are, however, sensitive to noise and discretization errors \cite{precisionlearning}. As none of these filters are optimal for SVCT, learning specialized filters for individual CT use-cases like SVCT is required. \reviewerfoursecond{Initial development of custom FBP filters was proposed by Pelt et al.~\cite{datadepfiltering} who proved the quality improvement of data-dependent FBP filters optimized for individual CT cases.} The first steps towards enabling \correction{DL-based} data-driven fine-tuning of the reconstruction step can be found in the works of Syben et al.~\cite{precisionlearning}, who proposed doing data-driven optimization to lower the discretization artifacts for conventional CT, and W\"urfl et al.~\cite{projdomweights}, who proposed learning the filter weights of the FDK operation for limited-angle CT. \reviewerfoursecond{Jiao et al.~\cite{dualcnn} used a CNN network to learn the FBP filtering operation for the SVCT scenario.}

Going even further, recent developments have re-introduced the regularized iterative reconstruction techniques with data-driven generated priors like the DRONE model or the DL-PICCS algorithm \cite{dronenet,dl-piccs}. While such architectures show great performance for SVCT, their data-driven aspect is only one part of the whole pipeline, while the iterative reconstruction algorithm governs the total computational cost. \correction{This impacts its applicability to 3D CT where thousands of slices must be reconstructed in a short period of time.} Additionally, it also requires optimization of several hyperparameters on a case by case basis to ensure convergence.

In this study we propose an improved dual-domain SVCT denoising hybrid network, in which the projection- and image-domain denoising operations are performed by two encoder-decoder networks while the reconstruction is performed in a FBP layer with a trainable filtering operation. We expand on the methods we previously proposed in Boghiu et al. \cite{dl-svct-fully3d}. We show that allowing fine-tuning of the FBP filter minimizes discretization and interpolation errors in the sinogram. 

\reviewerfoursecond{Current methods of sinogram upsampling rely on conventional interpolation methods (nearest neighbor interpolation \cite{dronenet}, custom bilinear interpolation \cite{hdnet}) or on a single-step DL-based upsampling-denoising method like the one proposed by Chao et al.~\cite{cbctdual}}. In contrast, we employ a \enquote{geometry-aware} interpolation method to great effect in producing sinograms at 4x the angular resolution of a sparse sinogram improving on the 3x upsampling limitation of the sparse sinogram mentioned in \cite{dronenet}. Two encoder-decoder networks are used to successfully suppress streaking artifacts in both domains, generating competitive results to the state-of-the-art.

Compared to current data-driven algorithms for the same purpose, the use of the fine-tuned filter significantly reduces upsampling artifacts and allows for the optimal amount of information to be passed to the image-domain denoising module. \reviewerfoursecond{Jiao et al. \cite{dualcnn} showed that a CNN-network is capable of learning the FBP filtering operation of the sinogram. Their sinograms do not require, however, any denoising, as they do not perform any measurement upsampling.}

\reviewerone{Furthermore, we show that the our WNet architecture diminishes the need for complex training and inference operations used by the state-of-the-art. On one hand, we don't need WGAN based denoising modules with discriminative loss which reduces the total number of trainable parameters. On the other hand, doing iterative reconstruction based on deep learning priors requires empirical hyperparameter optimization on a case-by-case basis \cite{dronenet}, which can potentially limit the generalizability of the model. WNet does not rely on any iterative denoising after the inference step as the dependency between real measurements and the learned output is embedded inside the network.}

This paper is structured as follows: Section \ref{section:methods} contains a brief description of the sparse-view tomographic problem and introduces the proposed dual-domain architecture. Then, in Section \ref{section:results} we present the results obtained with the proposed method and we compare against several state-of-the-art methods in the literature. \reviewerfour{We discuss the performance of the proposed method and its competitors in Section \ref{section:comparison} while} in Section \ref{section:interpretability} we dive into an interpretability analysis of our network, where we take apart each of its modules and investigate their performance and role in the whole reconstruction pipeline. \reviewerfour{We also take a brief look into the generalization performance of our network in Section \ref{section:discussiongeneralization} and then offer an outlook to future studies on this topic in Section \ref{section:futurework}. }\correction{\st{Lastly, in Section we sum up the findings of this paper.}Finally, we sum up the findings of this paper in Section \ref{section:conclusion}.}

\section{Methods} \label{section:methods}
\begin{figure*}[!t]
\begin{minipage}{\linewidth}
  		\centering
  		\centerline{\includegraphics[width=\linewidth]{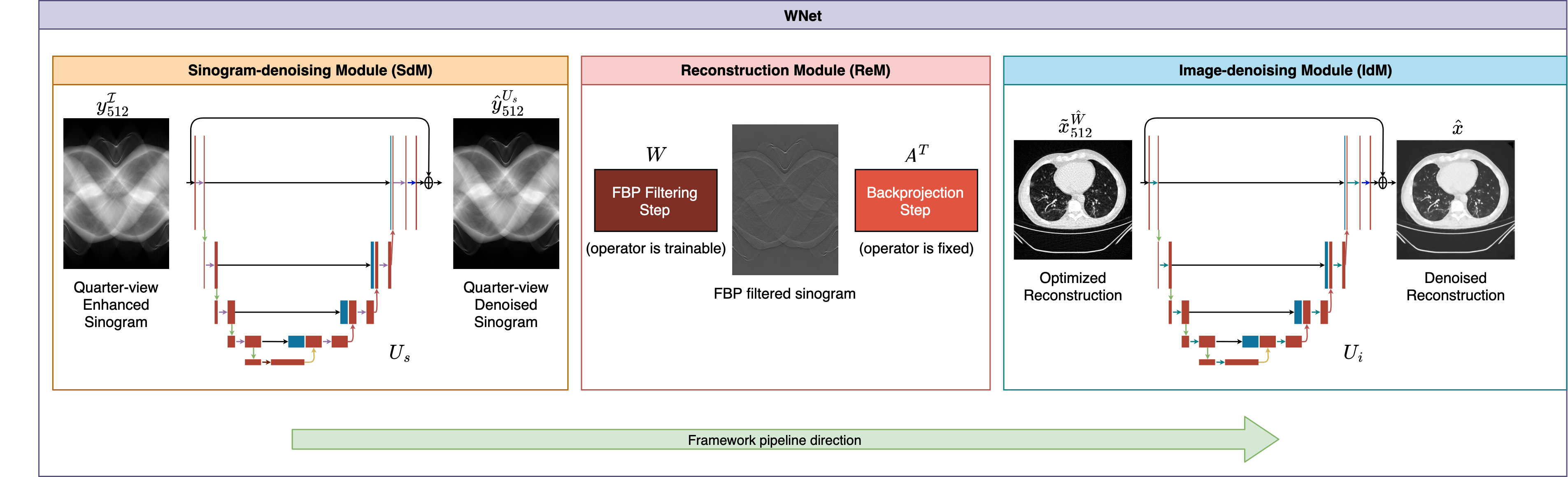}} 
  		\centerline{(a) WNet Architecture}\medskip
\end{minipage}
\\
\vspace{1em}
\begin{minipage}{0.49\linewidth}
  		\centering
  		\centerline{\includegraphics[width=\linewidth]{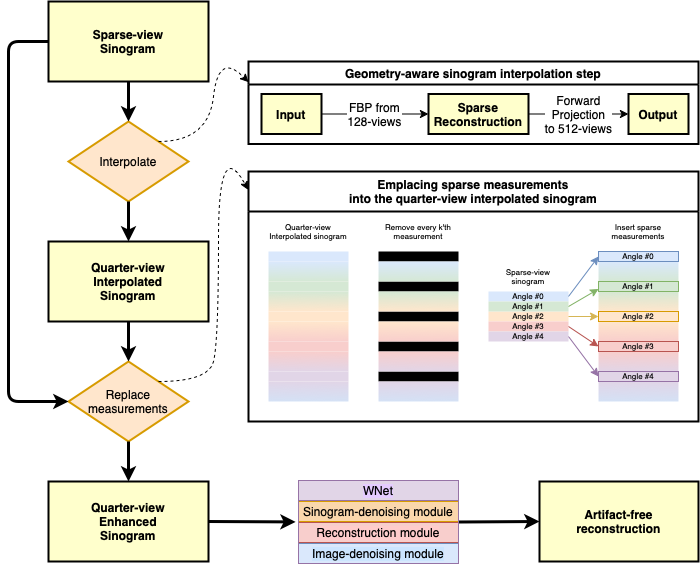}} 
  		\centerline{(b) Reconstruction Pipeline}\medskip
\end{minipage}
\begin{minipage}{0.49\linewidth}
  		\centering
  		\centerline{\includegraphics[width=\linewidth]{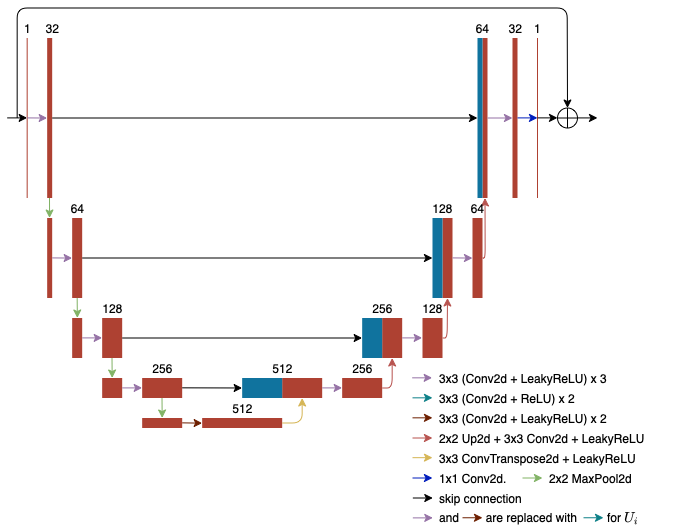}} 
  		\centerline{(c) UNet Architecture}\medskip
\end{minipage}
\caption{\textbf{(a) WNet Architecture} Architecture of the joint dual-domain denoising and reconstruction network WNet which is composed of three modules: Sinogram-denoising module (SdM), Reconstruction Module (ReM), and Image-denoising Module (IdM) \textbf{(b) Reconstruction Pipeline} The WNet model in the context of the whole reconstruction pipeline; the preprocessing stage of the initial sparse-view sinogram consists of a geometry-aware interpolation followed by a measurement consistency step \textbf{(c) UNet Architecture} UNet-based architecture of the SdM and IdM in the WNet model with individual building blocks highlighted as arrows of different colors. }
\label{fig:wnet}
\end{figure*}

{\color{black}
SVCT is a special case of regular X-ray CT where a significantly smaller set of measurements is used to reconstruct the imaged specimen. The low number of measurements increases the ill-posedness of the tomographic inverse problem, which leads to a decrease in quality in the reconstructed image. This is reflected in the presence of streaking and discretization artifacts in the result \cite{pyronn}. In this section we present the mathematical details of the tomographic inverse problem and we \correction{\st{follow-up}continue} with a motivation for the dual-domain denoising approach.

\subsection{Sparse-view Computed Tomography}

Given a set of measurements $y_{k} \in \mathbb{R}^{k \cdot M}$, where $k \in \mathbb{N} $ denotes the number of angles employed to obtain the measurements \correction{with} $M \in \mathbb{N} $ number of detector elements, we define the following discrete forward model for a tomographic reconstruction setup:
\begin{equation} \label{eq:forward-model}
    A_{k} \cdot x = y_{k}
\end{equation}
where $x \in \mathbb{R}^{N}$ is the measured volume (with unknown inner structure) and $ A \in \mathbb{R}^{N \times (k \cdot M)}$ an operator encoding information about the X-ray image formation process and the acquisition geometry. \correction{\st{We have a set of measurements $y_k$, obtained with a CT machine or a C-arm setup, which is used to reconstruct the unknown $x$. }}

There are various methods to solve this inverse problem, \correction{\st{however}but} we will focus on the Filtered Backprojection (FBP) reconstruction \associateed{with the Ram-Lak filter}. We solve eq. (\ref{eq:forward-model}) for $x$ using the FBP algorithm \cite{history-ct}:
\begin{equation}
    \tilde{x}_{k} = A_{k}^{T} W y_{k}
\end{equation}
where $A_{k}^{T}$ is the backprojection operation (defined as the transpose of the forward operator) and $W \in \mathbb{R}^{(k\cdot M) \times {k\cdot M}}$ a convolution operator describing the filtering operation performed on the \reviewerthree{\st{raw}post-log} sinogram values. 

\subsection{Artifact Suppression and Reconstruction} \label{section:artifact_suppression}
As eq. (\ref{eq:forward-model}) is underdetermined for small $k$ (e.g. $32 \le k \le 128$) the reconstruction $\tilde{x}_{k}$ will contain unwanted artifacts (streaks) \cite{sr-svct}. A denoising operation $U_i : \mathbb{R}^{N} \mapsto \mathbb{R}^{N} $ takes an input plagued by streaking artifacts (like $\tilde{x}_{k}$) and generates an output with less or (ideally) no artifacts, while recovering original information hidden behind the streaks. 

A data-driven approach to finding an optimal $U_i$ translates to minimizing the following user-defined loss function:
\begin{equation} \label{eq:best-ui-tmp}
    \underset{U_i}{\text{arg min}} \, \mathcal{L}(U_{i}(\tilde{x}_{k}), \bm{x})
\end{equation}
where $\bm{x}$ is usually an approximation of the original solution.

All the loss functions $\mathcal{L}_{*}$ mentioned throughout this manuscript are Huber losses given by the following formula:
\reviewertwo{
\begin{equation}
    \mathcal{L}_{s/r/i}(x,y) = 
    \begin{cases} 
        \frac{1}{2} (x-y)^2 \text{, if } |(x-y)|<1 \\
        |(x-y)| - \frac{1}{2} \text{, otherwise}
    \end{cases}
\end{equation}}
\reviewerfoursecond{The Huber loss was chosen as it combines the fast initial convergence of the L1 norm in early epochs \cite{whyl1}, while for later epochs, where the loss is in the vicinity of 0, the L2 norm ensures \enquote{nice} and continuous gradients \cite{whyhuber}.}

As this approach performs the denoising process in the reconstruction domain, \reviewerthree{\st{prior}original} information encoded in $y_{k}$ is possibly filtered out through the FBP reconstruction step. Finding a way to restore image information in the measurement domain set will ensure that an operator like $U_i$ has more information to work with in the image domain. This means that we ought to first find a mapping from $y_{k}$ to $y_{K}$, with $K = C \cdot k$, ($C \in \mathbb{N}, C>1$), and then apply the $U_i$ denoising operation to the corresponding FBP reconstruction $\tilde{x}_{K} = A_{K}^{T} W y_{K}$ such that we obtain a reconstruction with less artifacts thanks to the increased sampling rate of the sinogram \cite{dronenet}. 

Hereby, we introduce a \enquote{geometry-aware} interpolation operation implemented as an FBP operation on the original $y_k$ followed by a forward projection \reviewerthree{to a K-view sinogram}. The interpolated sinogram is subjected to a measurement consensus step $\mathcal{C}(\cdot)$ where the original measurements from $y_{k}$ are inserted into the obtained $K$-view sinogram $y_{K}^{\text{int}} \in \mathbb{R}^{K \cdot M}$ at the corresponding angular locations \reviewertwo{(See Fig. \ref{fig:wnet}-b for a visualization of the consensus step)}:
\begin{equation} \label{eq:enhanced_sinogram}
    y_{K}^{\mathcal{I}} = \mathcal{I}(y_{k}) = \mathcal{C}(\underbrace{A_{K} A_{k}^{T} W y_{k}}_{=y_{K}^{\text{int}}})
\end{equation}

The \correction{consensus} step $\mathcal{C}(\cdot)$ only allows for the preservation of original information in the interpolated sinogram. A denoising operator $U_s : \mathbb{R}^{K \cdot M} \mapsto \mathbb{R}^{K \cdot M}$ is then required to properly handle the interpolation and discretization artifacts in the generated $y_{K}^{\mathcal{I}}$ sinogram ($y_{K}^{U_s} = U_s(\mathcal{I}(y_{k}))$), which can be obtained by minimizing the following loss function:
\begin{equation} \label{best-us}
    \bm{\hat{U}_s} = \underset{U_s}{\text{arg min}} \, \mathcal{L}_s(U_s(\mathcal{I}(y_{k})),\bm{y_{K}})
\end{equation}
subject to  $(y_{K}^{U_s})_{i \cdot k,*} = (y_{k})_{i,*}$ for every $i \in \mathbb{N}$ enforced by $\mathcal{C}$. $\bm{y_K} \in \mathbb{R}^{K \cdot M}$ is a real measured $K$-view sinogram used as the ground truth to compute the loss function. In other words, we train a denoising operator $U_s$ towards removing the interpolation artifacts produced by $\mathcal{I}$ while preserving the originally measured angles ($y_{k}$) in the denoised sinogram ($y_{K}^{U_s}$).

Given the discrete nature of the tomographic inverse problem in eq. (\ref{eq:forward-model}), a reconstruction generated with the FBP method is prone to certain corruptions commonly called cupping and \reviewerfoursecond{DC} shift artifacts \cite{precisionlearning}. Since widely used filters like the Ram-Lak or Cosine filters have been proven to perform well when the set of measurements is sufficiently large \cite{fbp-filters}, FBP for SVCT with these types of filters is not ideal. \reviewerfoursecond{Shi et al. \cite{schemefilterct} showed that specific CT use-cases (e.g. SVCT) require optimized filters, while Syben et al.~\cite{precisionlearning} proposed to learn such filters in a data-driven process}. Since we use the FBP algorithm to compute a reconstruction from $y_K^{U_s}$ we can optimize the FBP filtering operator $W$ towards reducing the aforementioned artifacts by minimizing the following loss function:
\begin{equation} \label{best-w}
    \bm{\hat{W}} = \underset{W}{\text{arg min}} \, \mathcal{L}_r(A_{K}^{T} W \bm{\hat{U}_s}(\mathcal{I}(y_k)),x)
\end{equation}

Having an optimal sinogram denoising operator $\bm{\hat{U}_s}$ and an optimal convolution operator $\bm{\hat{W}}$ the last step is to apply the operator $U_i$, introduced earlier, on the improved reconstruction $\tilde{x}_{K}^{\bm{\hat{W}}}$. An optimal $U_i$ minimizes the following loss:
\begin{equation} \label{best-ui}
    \bm{\hat{U}_i} = \underset{U_i}{\text{arg min}} \, \mathcal{L}_i(U_{i} (A_{K}^{T} \bm{\hat{W}}  \bm{\hat{U}_s}(\mathcal{I}(y_k)),\bm{x})
\end{equation}
Putting everything together, we finally obtain a dual-domain optimally denoised reconstruction:
\begin{equation}
    \bm{\hat{x}} = \bm{\hat{U}_i} (A_{K}^{T} \bm{\hat{W}} \bm{\hat{U}_s}(\mathcal{I}(y_k))) \approx x
\end{equation}

\subsection{Model Architecture} \label{section:model}

Fig. \ref{fig:wnet}a shows the proposed WNet model architecture consisting of its three main modules while Fig. \ref{fig:wnet}b highlights every step involved in the WNet reconstruction pipeline.

The first module of the WNet, the Sinogram-denoising Module (SdM), uses an encoder-decoder type of network $U_s$ to perform sinogram-domain denoising. The goal is to filter out the sparse-view and upsampling artifacts present in the interpolated sinogram $y_{K}^{\mathcal{I}}$ which was obtained with the geometry-aware interpolation operation (Fig. \ref{fig:wnet}b) from eq. \ref{eq:enhanced_sinogram}. The output of a trained $\bm{\hat{U}_s}$ is the denoised sinogram $y_{K}^{\bm{\hat{U}_s}}$ \reviewerone(also subject to the consensus step)}.

\correction{\st{The interpolation step, a fixed operation, is performed in the preprocessing stage of the input. It consists of a forward projection to a $K$-view sinogram performed after a Ram-Lak filtered backprojection reconstruction from a $k$-view sinogram. Additionally, the measurement consensus step $\mathcal{C}$ guarantees that the original measurements $y_{k}$ are contained in the interpolated $K$-view sinogram $y_{K}$.}}

The output of $\bm{\hat{U}_s}$ is then fed to the next module, the Reconstruction Module (ReM), which handles the FBP reconstruction of $y_{K}^{\bm{\hat{U}_s}}$. The ReM has two steps:

\reviewerfour{
\begin{enumerate}
\item a trainable convolution operation which uses the trained kernel $\bm{\hat{W}}$ to filter the input sinogram $y_{K}^{\bm{\hat{U}_s}}$ and generate $y_K^{\bm{\hat{W}}}$. 
\item succeded by a fixed tomographic backprojection operation, modeled as the transpose of the forward operator introduced in eq. (\ref{eq:forward-model}), $A^T$, producing $\tilde{x}_K^{\bm{\hat{W}}}$
\end{enumerate}
}

The third and final module, the Image-denoising Module (IdM), contains a second encoder-decoder network $U_i$ which is employed towards suppressing the sparse artifacts in $\hat{x}_K^{\bm{\hat{W}}}$, generating a final denoised image $\bm{\hat{x}}$. The UNet-based architecture of both encoder-decoder networks $U_s$ and $U_i$ is visualized in Fig. \ref{fig:wnet}c.

\newcommand\x{0.32}

\begin{figure}[!t]
\centering
\begin{varwidth}{0.02\linewidth}
\hfill \vfill
\end{varwidth}
\begin{varwidth}{0.97\linewidth}
	\centering
	\begin{varwidth}{0.32\linewidth}
	  		\centering
	  		\centerline{\footnotesize Ground Truth}\smallskip
	  		\centerline{\includegraphics[width=\linewidth]{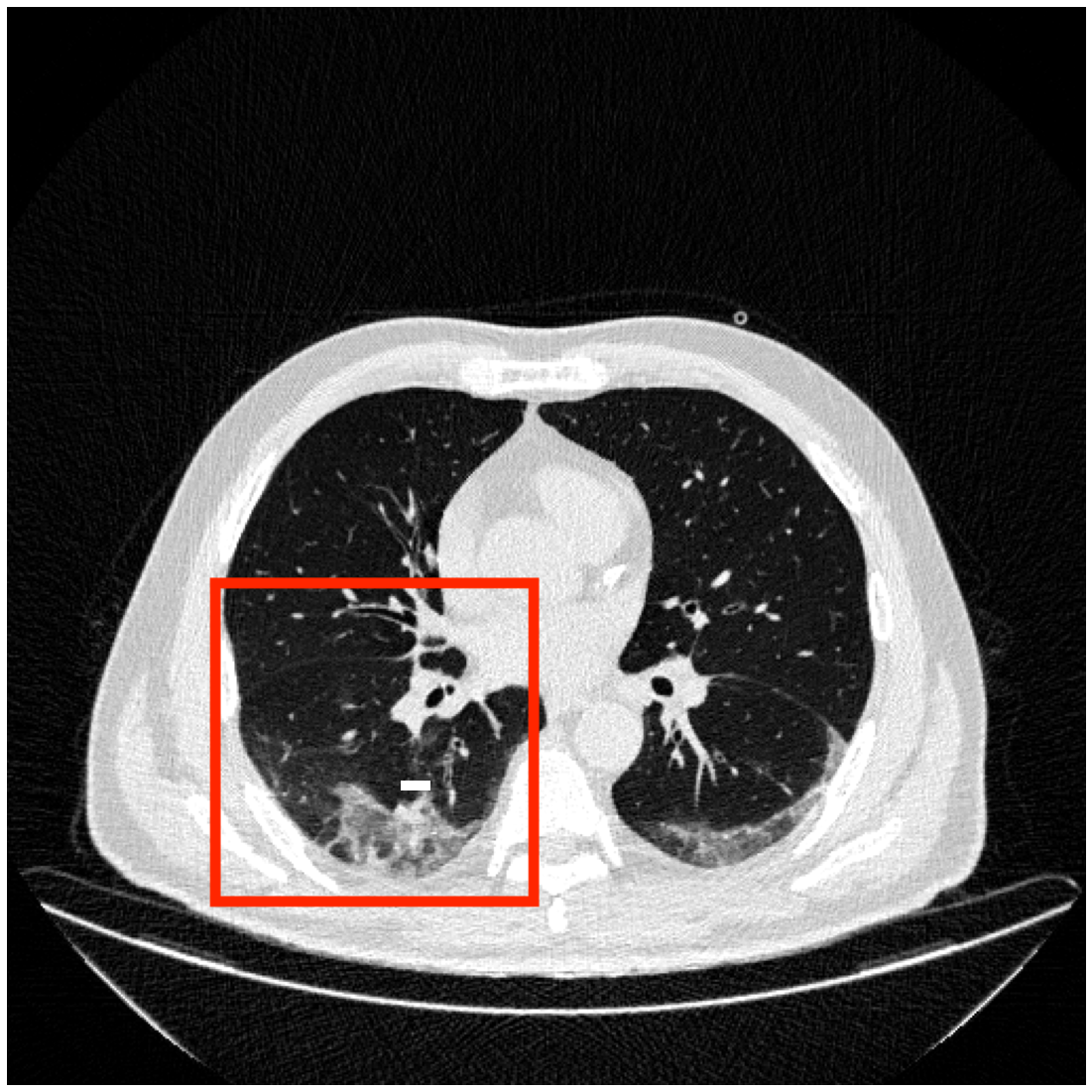}} 
	  		\smallskip
	\end{varwidth}
\end{varwidth}
\\
\vspace{1em}
\begin{varwidth}{0.02\linewidth}
  		\begin{turn}{90}
  		\footnotesize Reconstruction slice Covid-19 patient
  		\end{turn}
\end{varwidth}
\begin{varwidth}{0.97\linewidth}
    \begin{varwidth}{\x\linewidth}
      		\centering
      		\centerline{\footnotesize FBP}\smallskip
      		\centerline{\includegraphics[width=\linewidth]{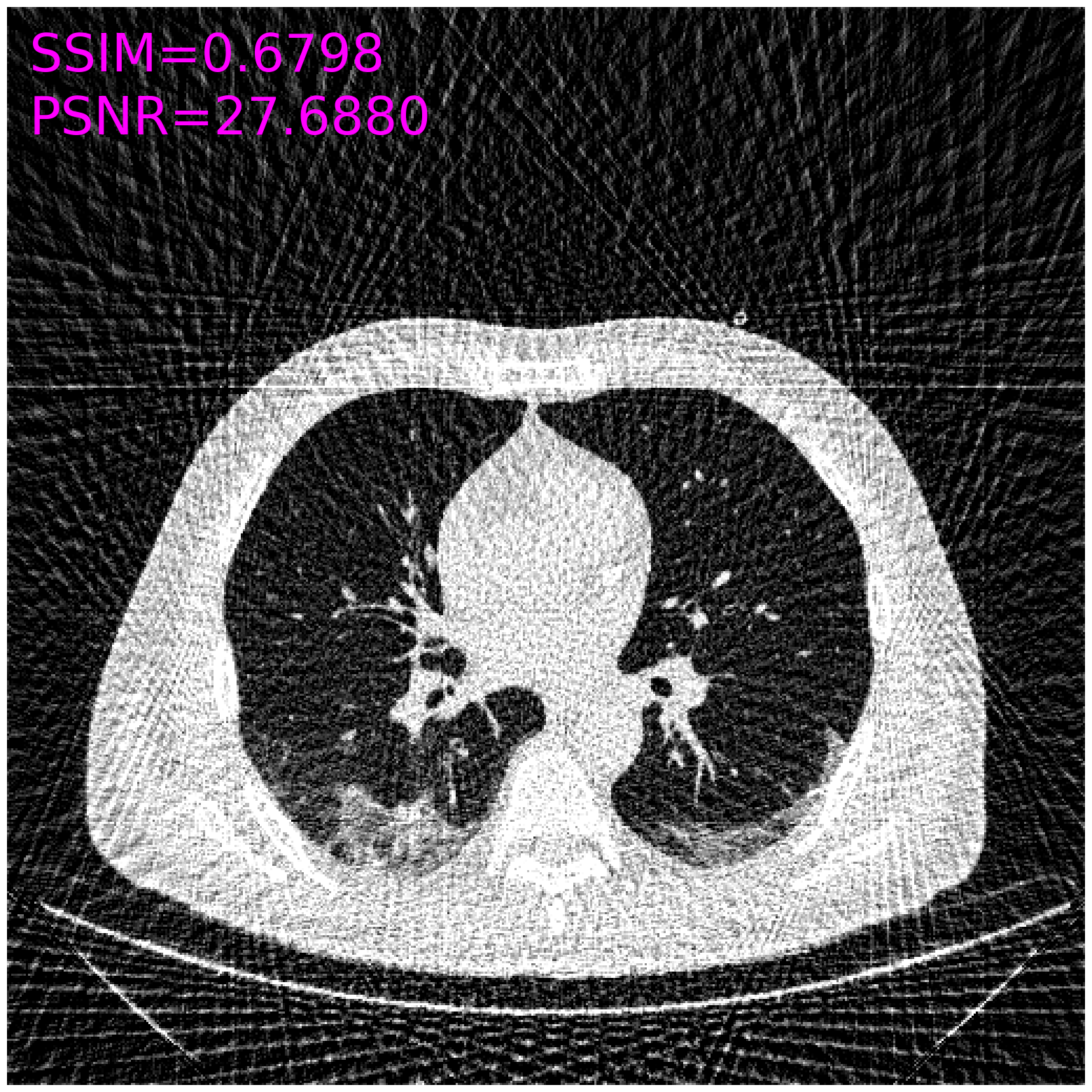}} 
      		\vfill
    \end{varwidth}
    \begin{varwidth}{\x\linewidth}
      		\centering
      		\centerline{\footnotesize WLS+TV}\smallskip
      		\centerline{\includegraphics[width=\linewidth]{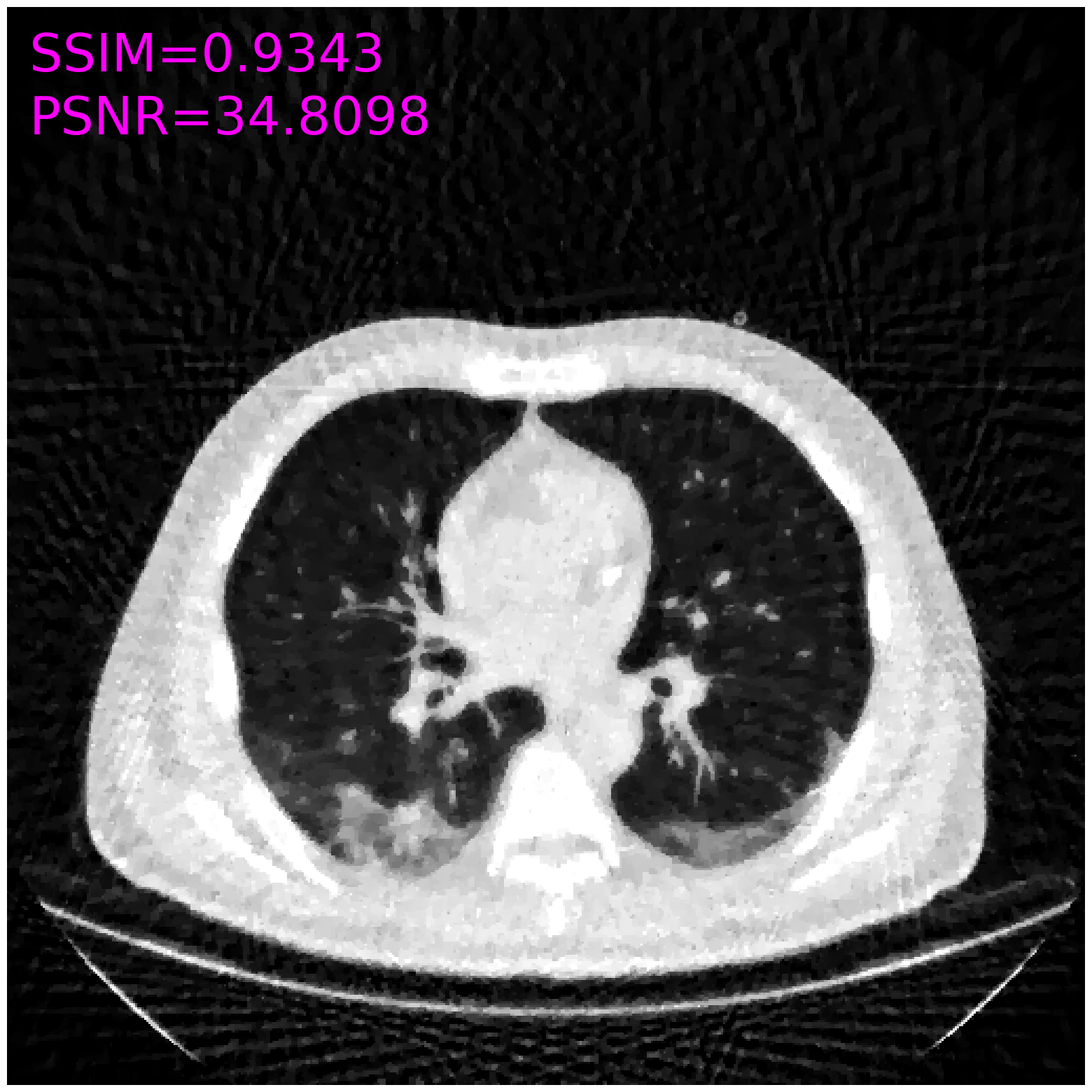}} 
    \end{varwidth}
    \begin{varwidth}{\x\linewidth}
      		\centering
      		\centerline{\footnotesize FBPConvNet}\smallskip
      		\centerline{\includegraphics[width=\linewidth]{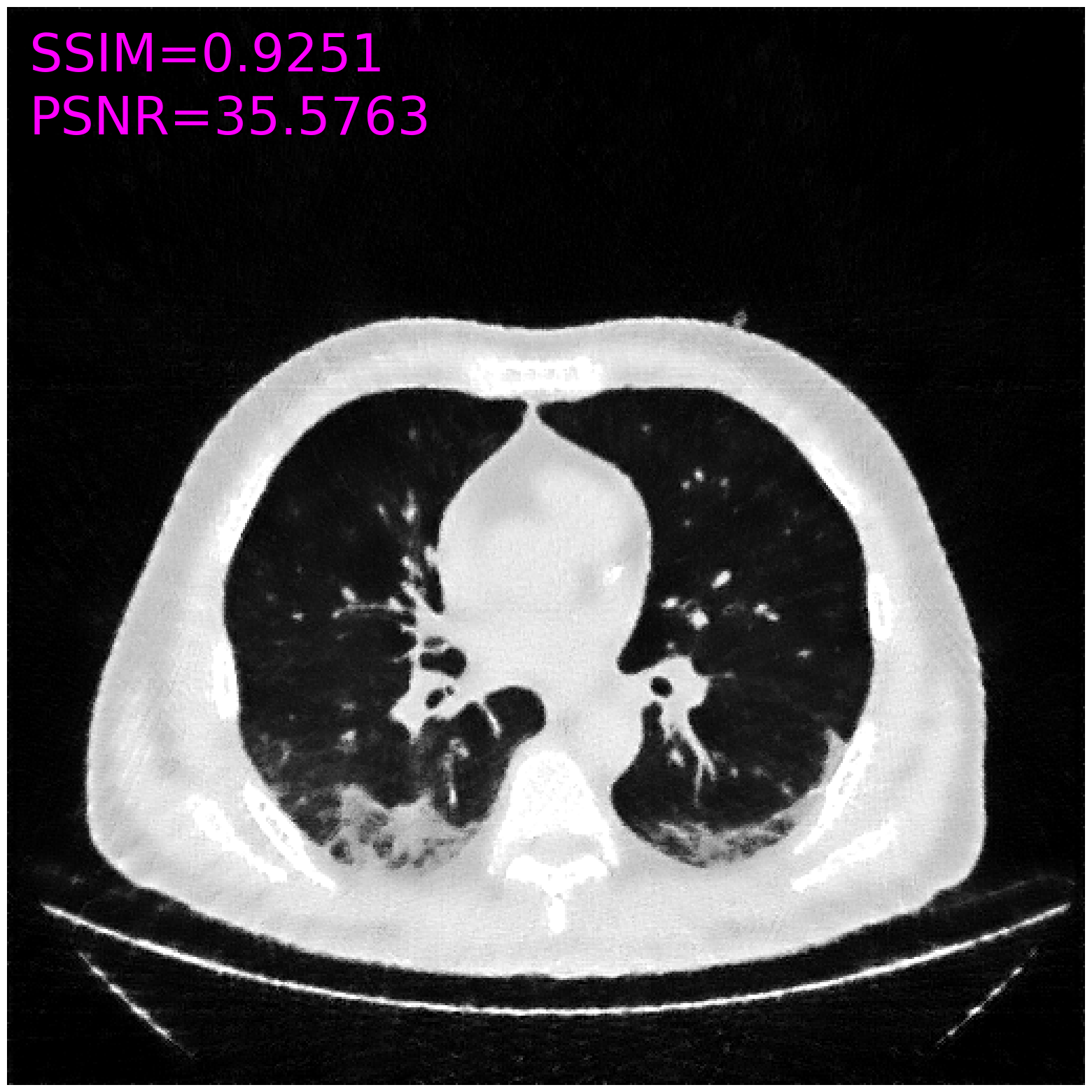}} 
    \end{varwidth}
    \\
    \begin{varwidth}{\x\linewidth}
      		\centering
      		\centerline{\includegraphics[width=\linewidth]{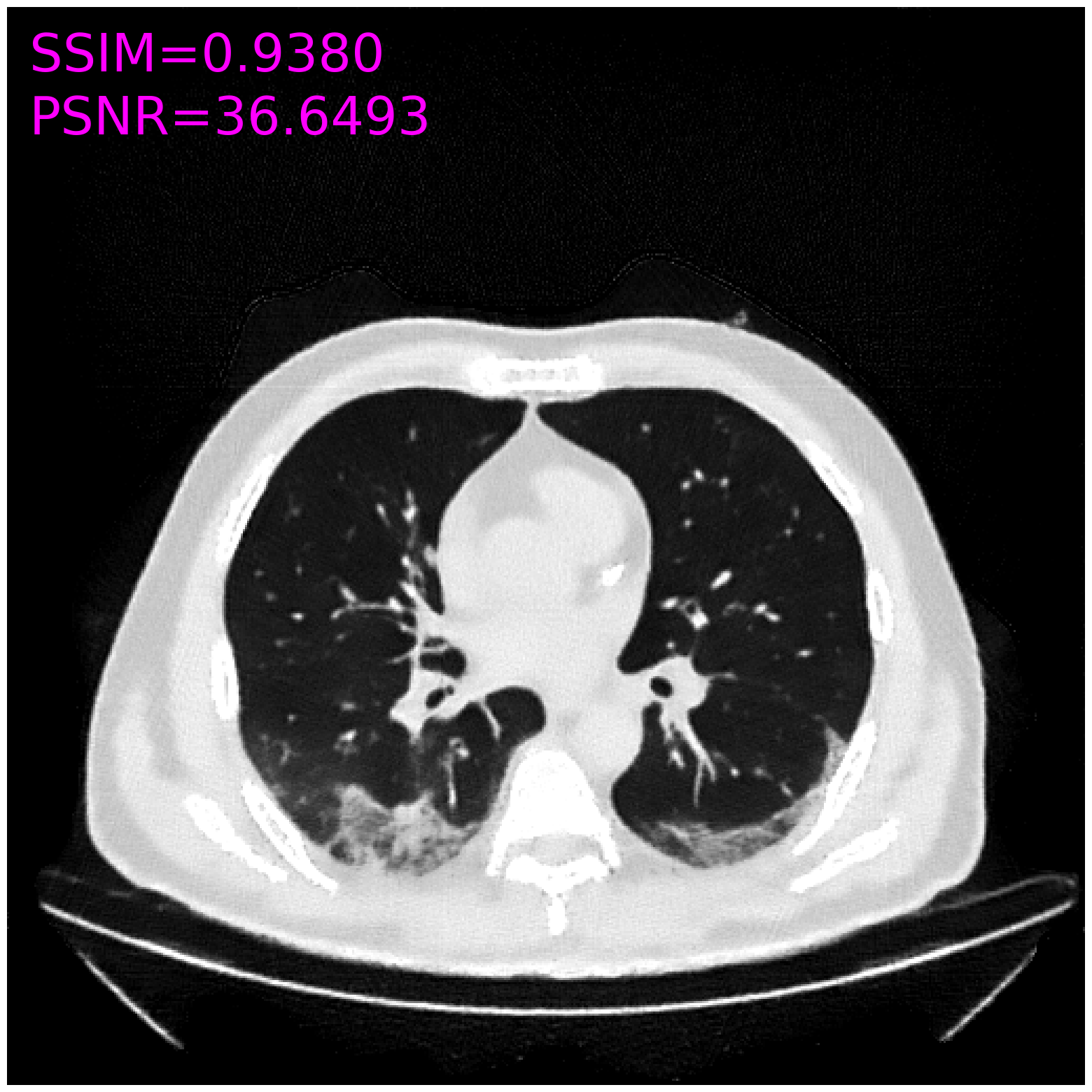}} 
      		\centerline{\footnotesize DD-Net}\smallskip
    \end{varwidth}
    \begin{varwidth}{\x\linewidth}
      		\centering
      		\centerline{\includegraphics[width=\linewidth]{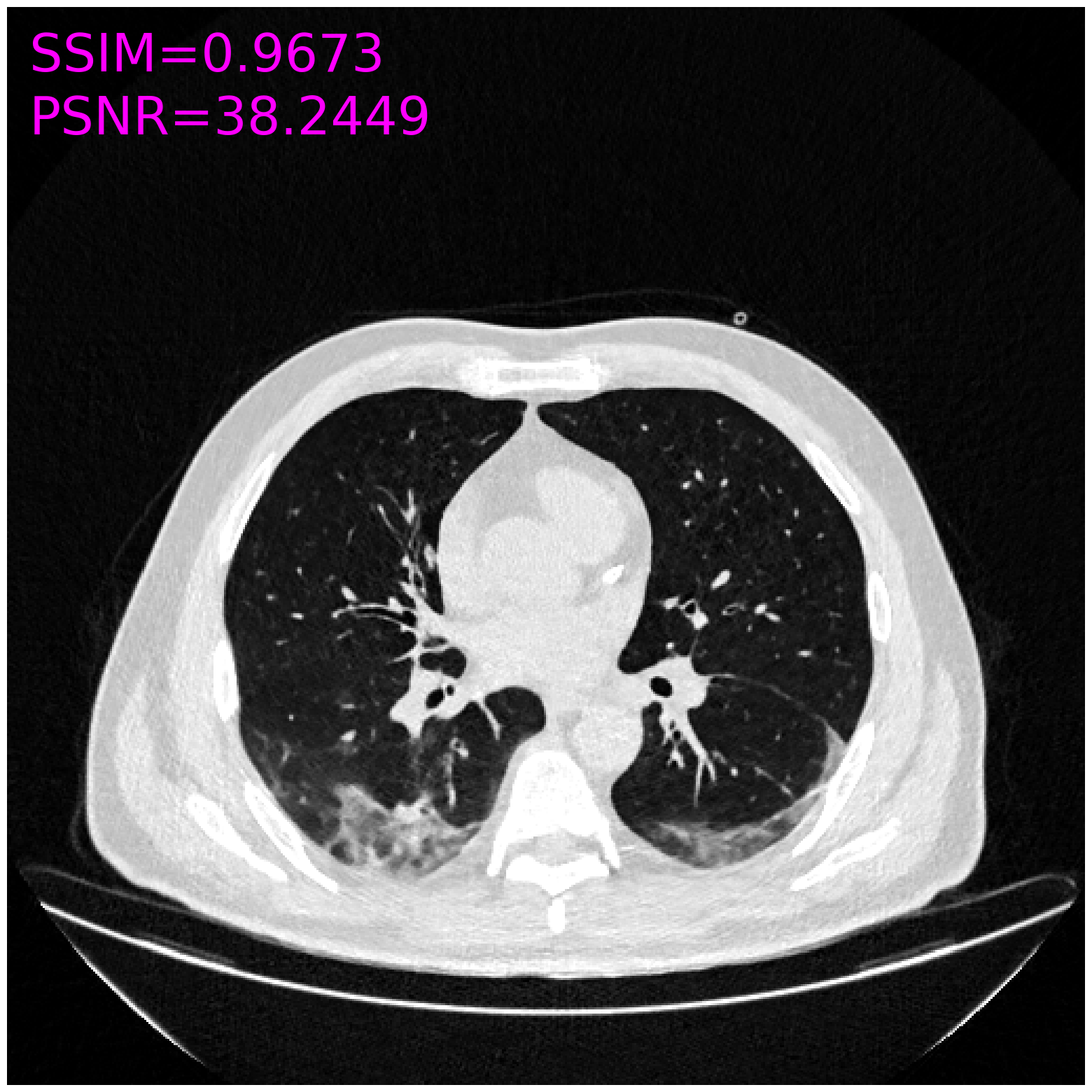}} 
      		\centerline{\footnotesize DRONE}\smallskip
    \end{varwidth}
    \begin{varwidth}{\x\linewidth}
      		\centering
      		\centerline{\includegraphics[width=\linewidth]{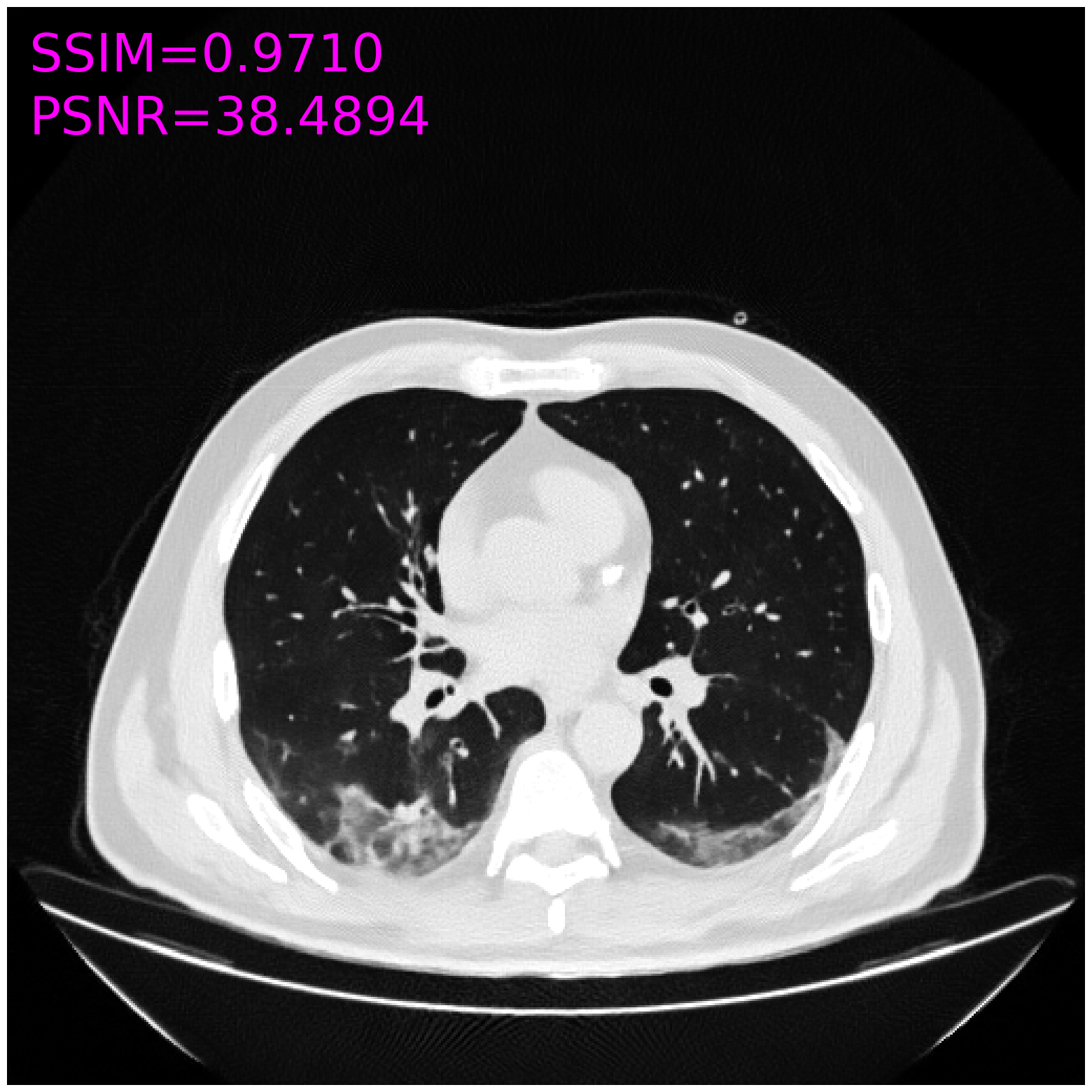}} 
      		\centerline{\footnotesize WNet (ours)}\smallskip
    \end{varwidth}
\end{varwidth}
\\
\vspace{1em}
\begin{varwidth}{0.02\linewidth}
  		\begin{turn}{90}
  		\footnotesize Diff. Recon
  		\end{turn}
\end{varwidth}
\begin{varwidth}{0.97\linewidth}
    \begin{varwidth}{\x\linewidth}
      		\centering
      		\centerline{\footnotesize FBP}\smallskip
      		\centerline{\includegraphics[width=\linewidth]{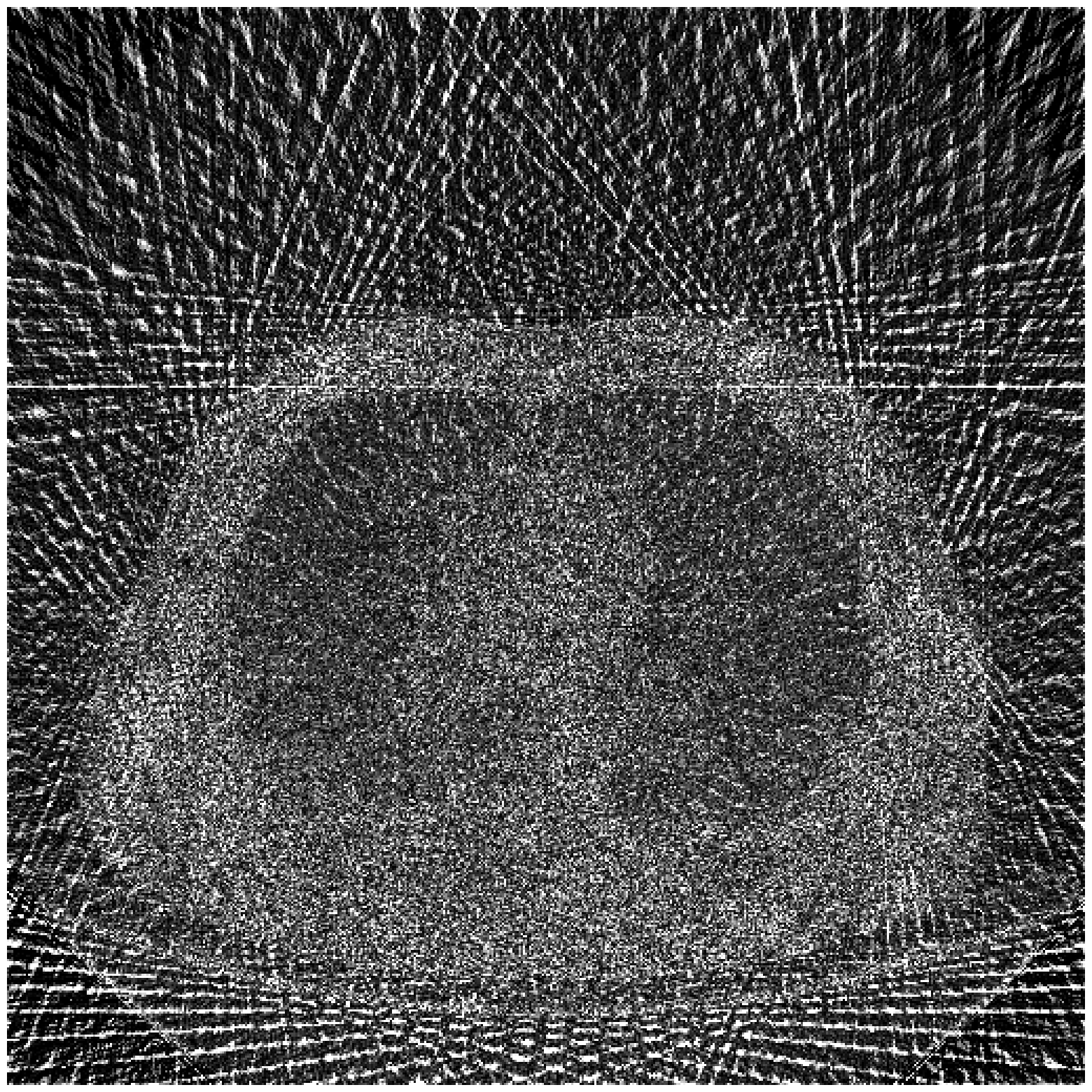}} 
    \end{varwidth}
    \begin{varwidth}{\x\linewidth}
      		\centering
      		\centerline{\footnotesize WLS+L1}\smallskip
      		\centerline{\includegraphics[width=\linewidth]{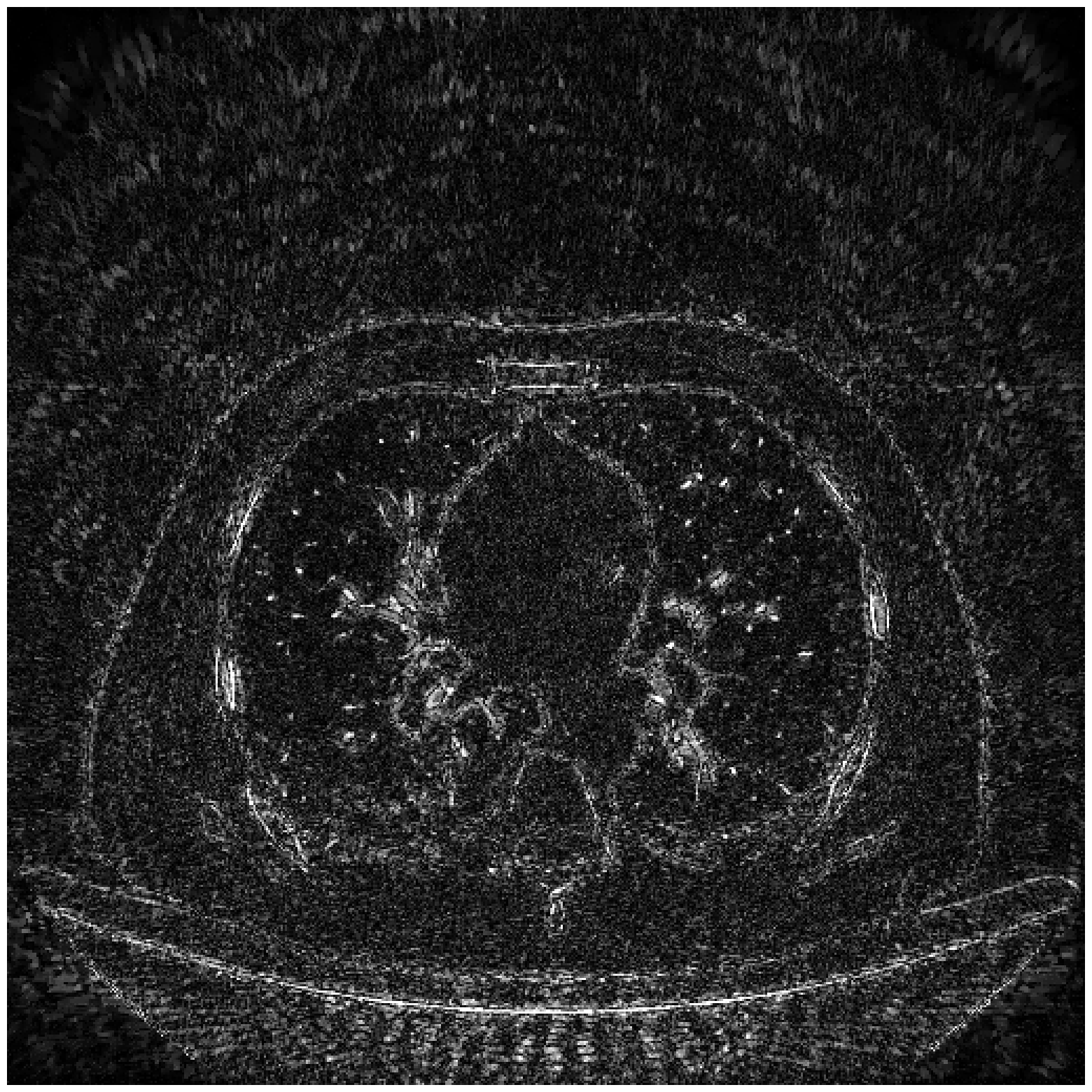}} 
    \end{varwidth}
    \begin{varwidth}{\x\linewidth}
      		\centering
      		\centerline{\footnotesize FBPConvNet}\smallskip
      		\centerline{\includegraphics[width=\linewidth]{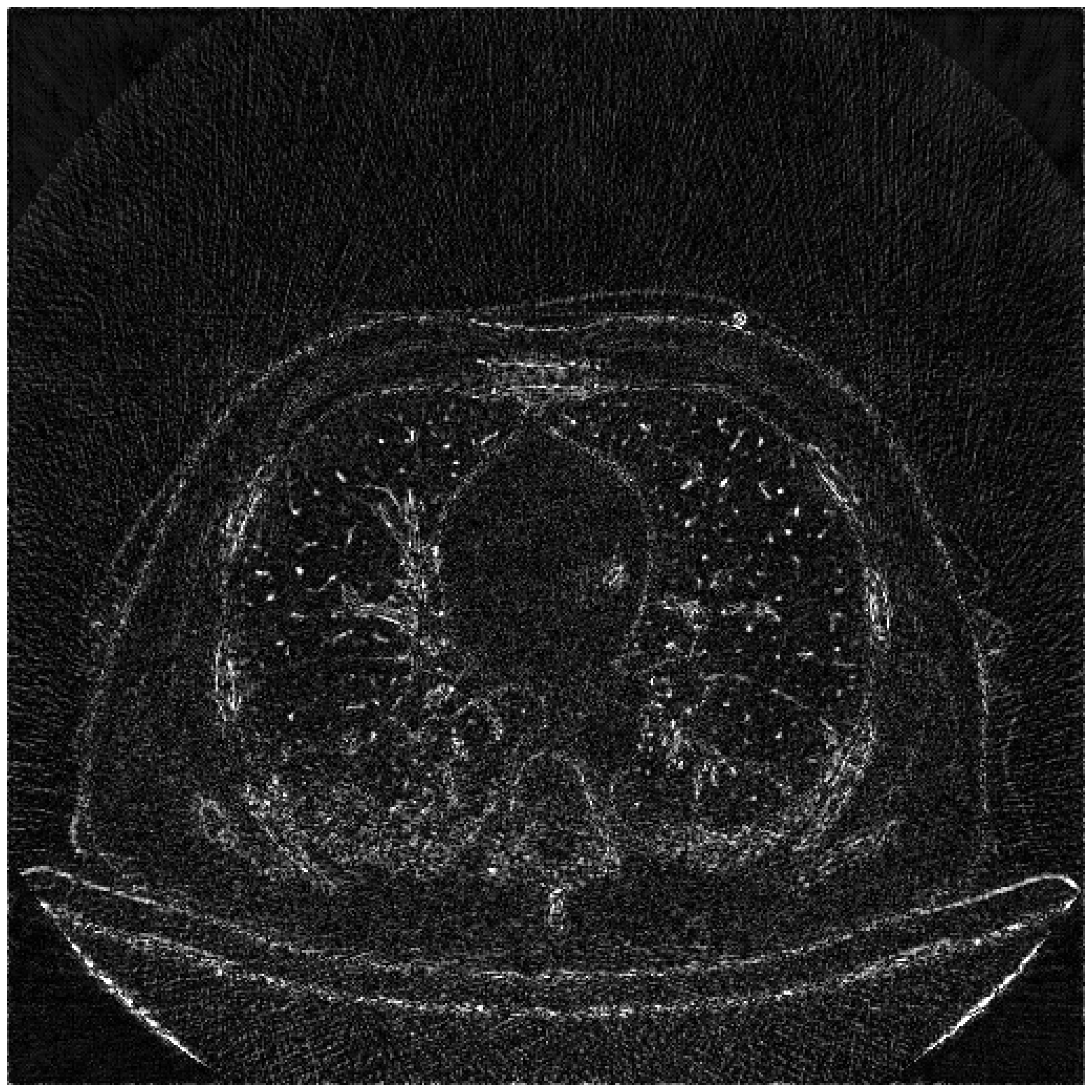}} 
    \end{varwidth}
    \\
    \begin{varwidth}{\x\linewidth}
      		\centering
      		\centerline{\includegraphics[width=\linewidth]{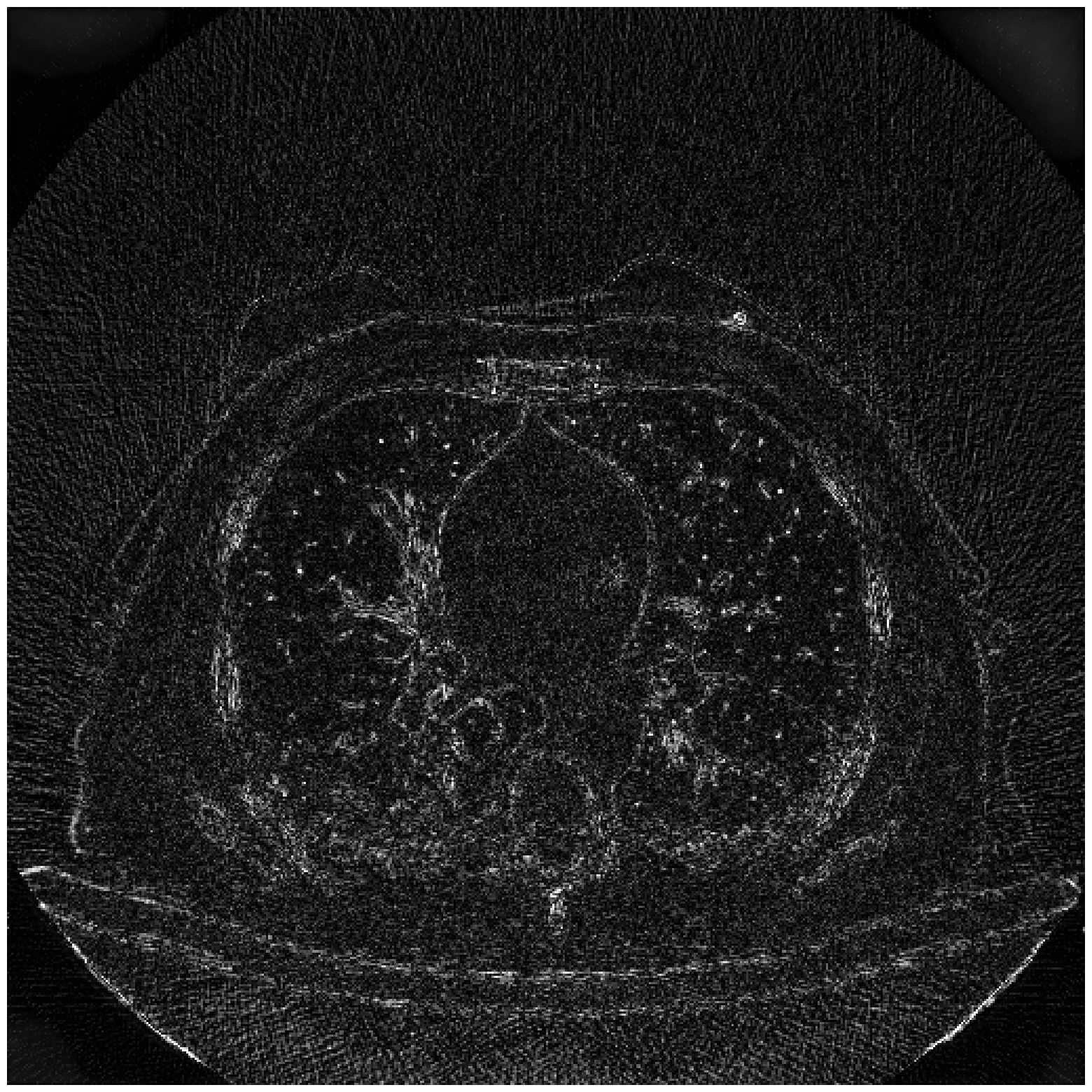}} 
      		\centerline{\footnotesize DD-Net}\smallskip
    \end{varwidth}
    \begin{varwidth}{\x\linewidth}
      		\centering
      		\centerline{\includegraphics[width=\linewidth]{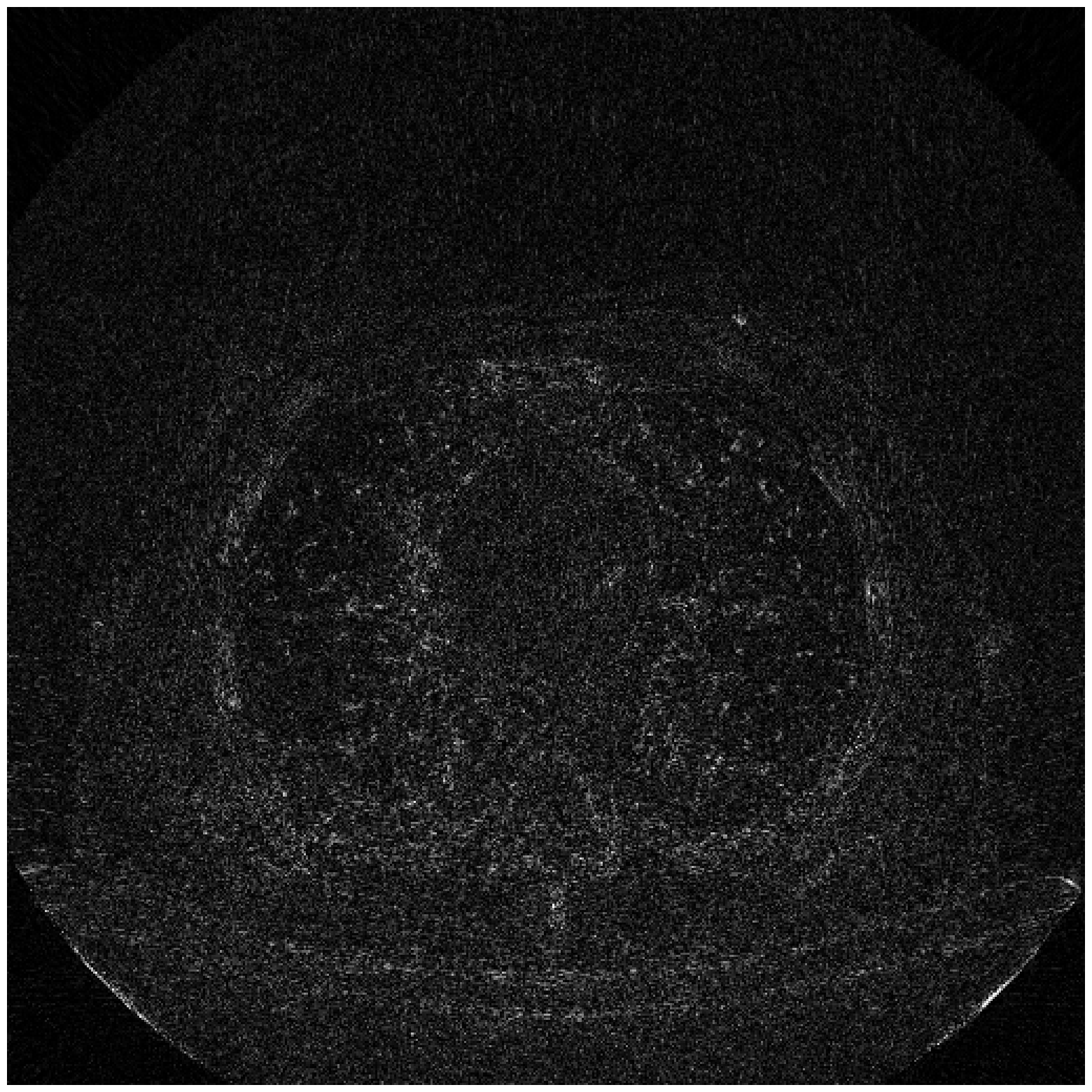}} 
      		\centerline{\footnotesize DRONE}\smallskip
    \end{varwidth}
    \begin{varwidth}{\x\linewidth}
      		\centering
      		\centerline{\includegraphics[width=\linewidth]{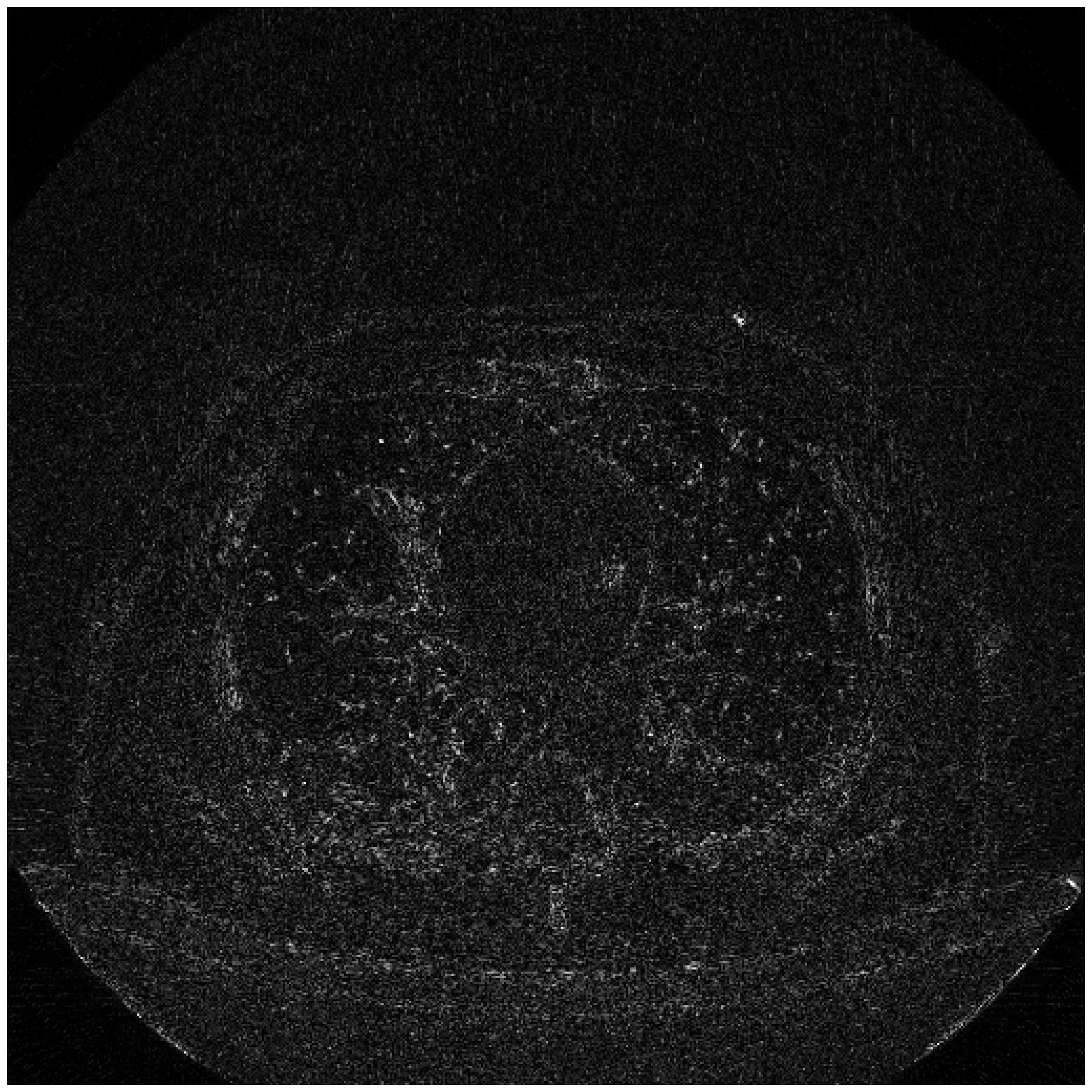}} 
      		\centerline{\footnotesize WNet (ours)}\smallskip
    \end{varwidth}
\end{varwidth}

\caption{Reconstruction results of a clinically relevant slice for determining COVID-19 pneumonia in a patient selected out of the test dataset. The methods used were Filtered Backprojection (FBP, see Section \ref{section:comparison}), Weighted Least Squares with Total Variation Regularization (WLS+TV, see Section \ref{section:comparison}), FBPConvNet (see \cite{fbpconvnet}), DD-Net (see \cite{ddnet}), DRONE (see \cite{dronenet}) and WNet (ours, see Figure \ref{fig:wnet} and Section \ref{section:model}). FBP, WLS+L1, DRONE and WNet used the same forward and backward tomographic operators described in Section \ref{section:initialization}. \textbf{(top set)} Reconstructed slices showing the chest area of the patient. The SSIM and PSNR metrics computed for the individual reconstruction with respect to the \enquote{Ground Truth} were added in the top left corner (magenta). The display window for all reconstructions is [-1024,150] HU. \textbf{(bottom set)} Difference images to the ground truth corresponding to the output of each method. The display window for all generated difference images is [0,400] HU.}
\label{fig:main-results-1}
\end{figure}

\begin{figure}[!t]
\centering
\begin{varwidth}{\x\linewidth}
  		\centering
  		\centerline{\footnotesize Ground Truth}\smallskip
  		\centerline{\includegraphics[width=\linewidth]{figures/groundtruth_patch_10}} 
\end{varwidth}
\\
\vspace{1em}
\begin{varwidth}{0.02\linewidth}
  		\begin{turn}{90}
  		\footnotesize ROI
  		\end{turn}
\end{varwidth}
\begin{varwidth}{0.97\linewidth}
    \begin{varwidth}{\x\linewidth}
      		\centering
      		\centerline{\footnotesize FBP}\smallskip
      		\centerline{\includegraphics[width=\linewidth]{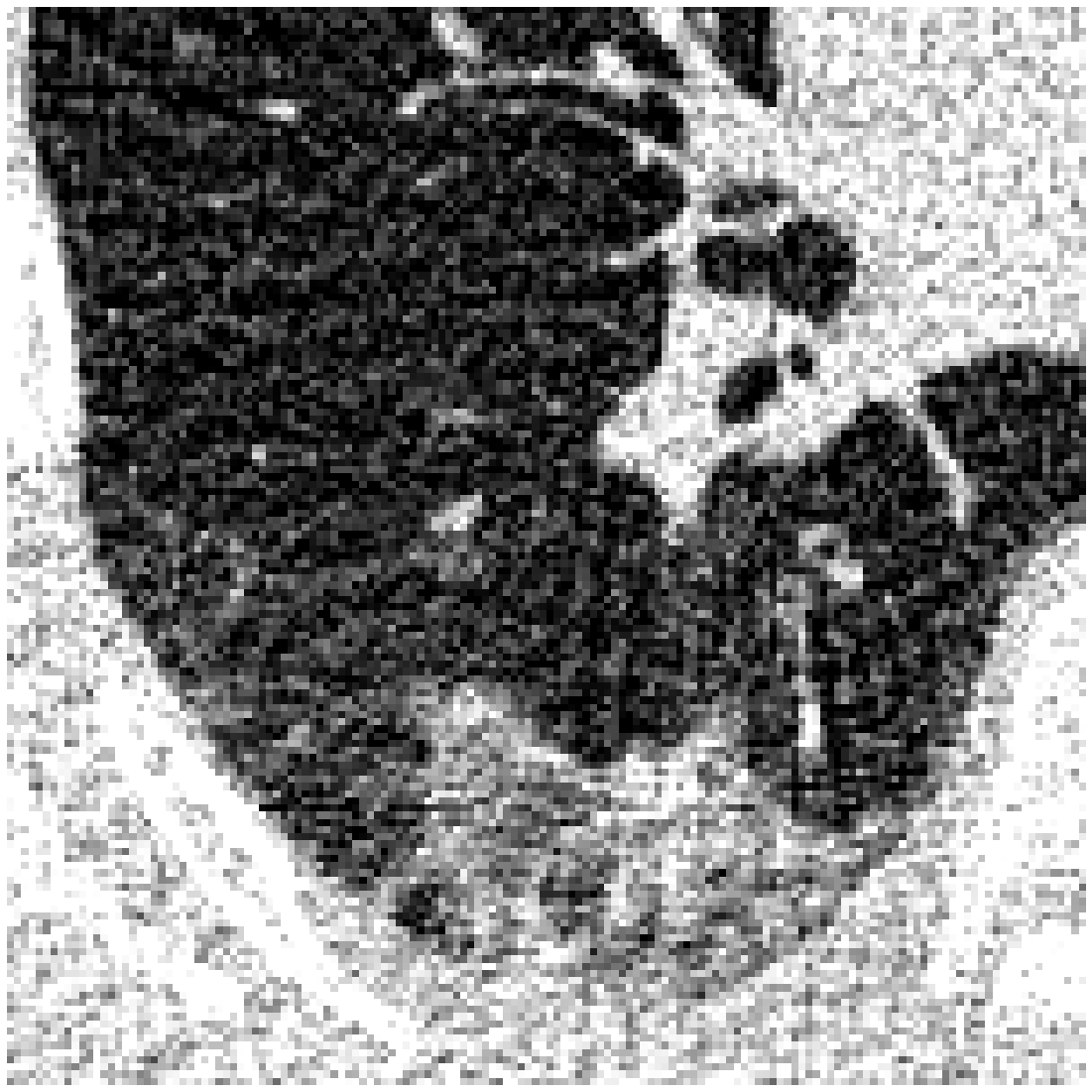}} 
    \end{varwidth}
    \begin{varwidth}{\x\linewidth}
      		\centering
      		\centerline{\footnotesize WLS+TV}\smallskip
      		\centerline{\includegraphics[width=\linewidth]{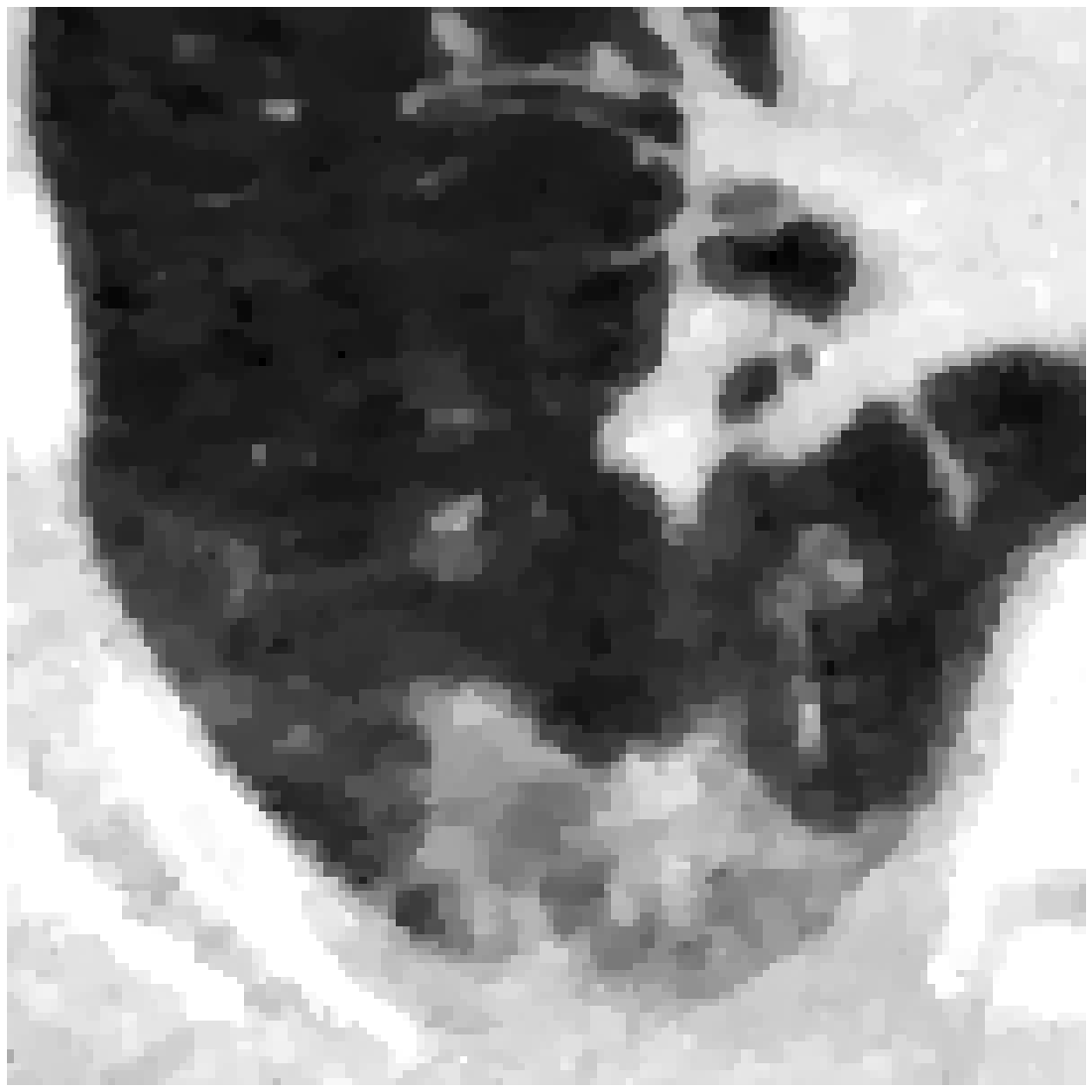}} 
    \end{varwidth}
    \begin{varwidth}{\x\linewidth}
      		\centering
      		\centerline{\footnotesize FBPConvNet}\smallskip
      		\centerline{\includegraphics[width=\linewidth]{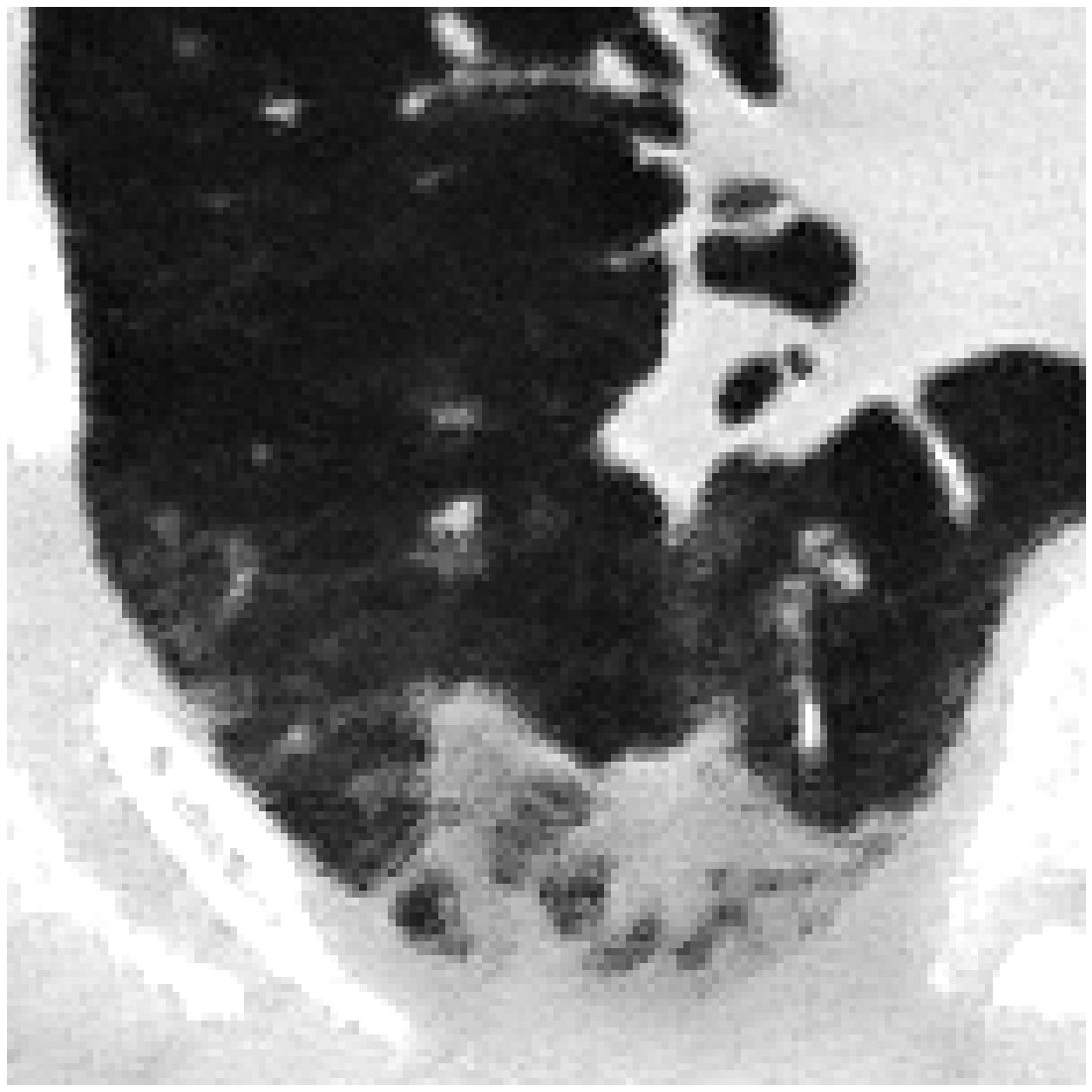}} 
    \end{varwidth}
    \\
    \begin{varwidth}{\x\linewidth}
      		\centering
      		\centerline{\includegraphics[width=\linewidth]{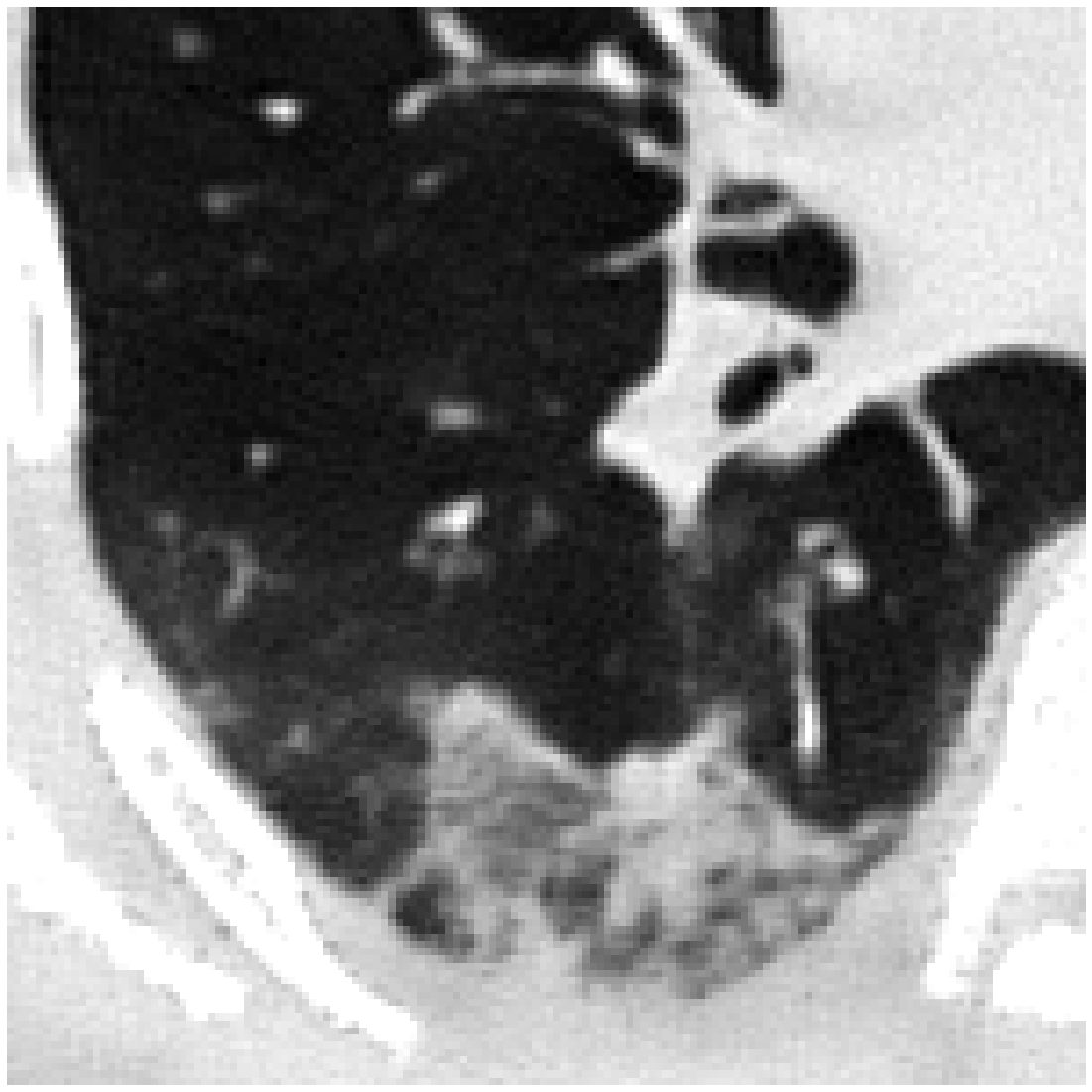}} 
      		\centerline{\footnotesize DD-Net}\smallskip
    \end{varwidth}
    \begin{varwidth}{\x\linewidth}
      		\centering
      		\centerline{\includegraphics[width=\linewidth]{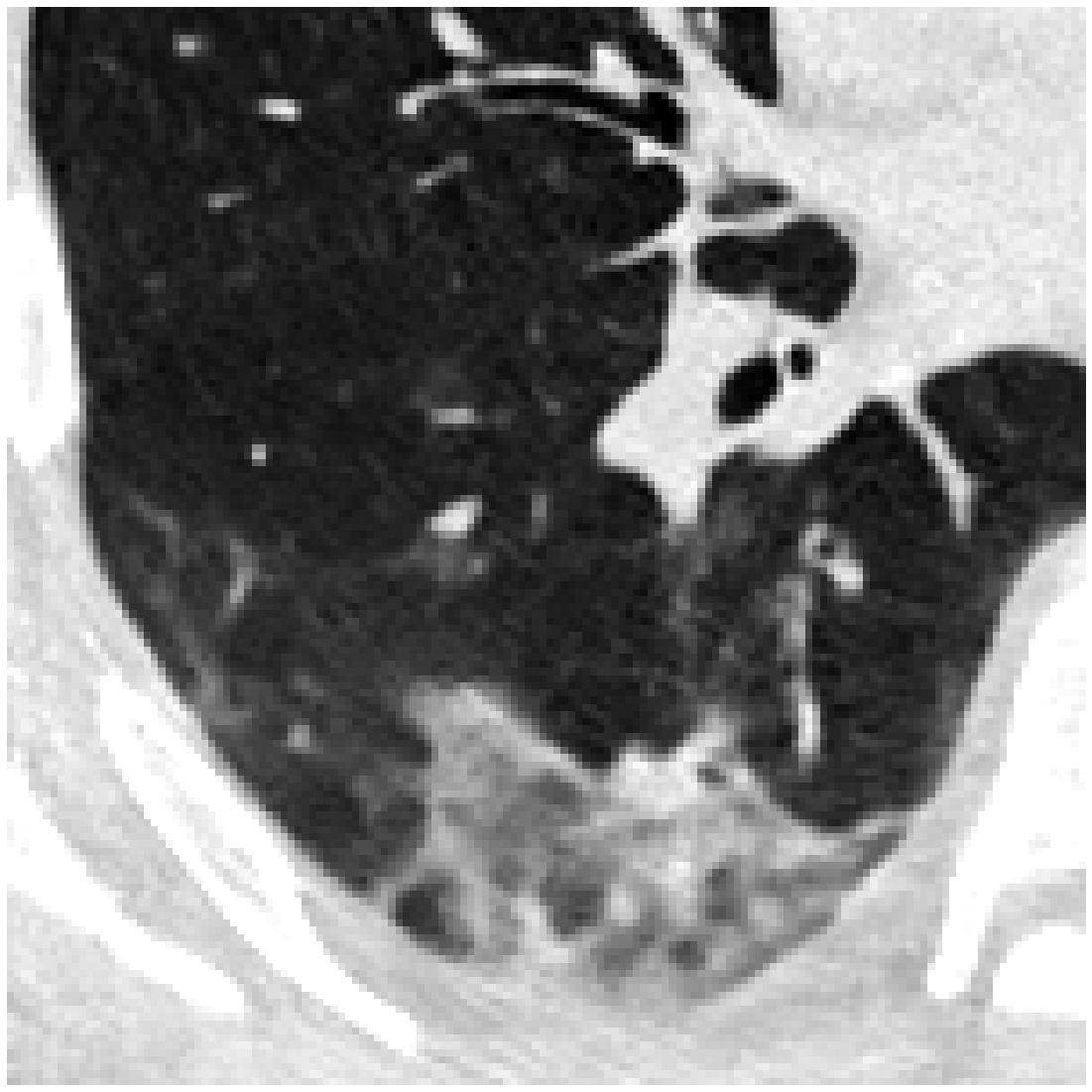}} 
      		\centerline{\footnotesize DRONE}\smallskip
    \end{varwidth}
    \begin{varwidth}{\x\linewidth}
      		\centering
      		\centerline{\includegraphics[width=\linewidth]{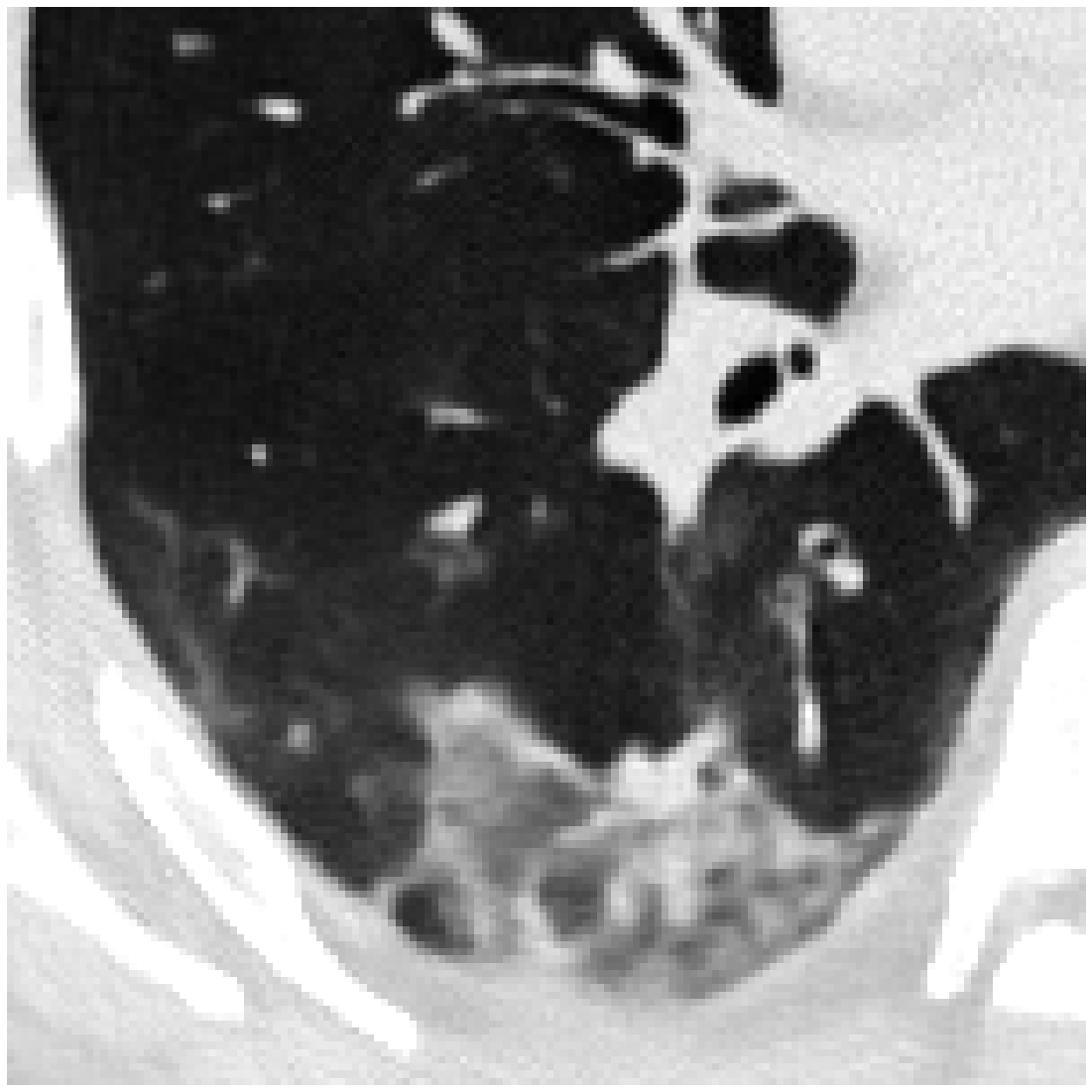}} 
      		\centerline{\footnotesize WNet (ours)}\smallskip
    \end{varwidth}
\end{varwidth}
\vspace{1em}
\\
\begin{varwidth}{0.02\linewidth}
  		\begin{turn}{90}
  		\footnotesize Diff. ROI
  		\end{turn}
\end{varwidth}
\begin{varwidth}{0.97\linewidth}
    \begin{varwidth}{\x\linewidth}
      		\centering
      		\centerline{\footnotesize FBP}\smallskip
      		\centerline{\includegraphics[width=\linewidth]{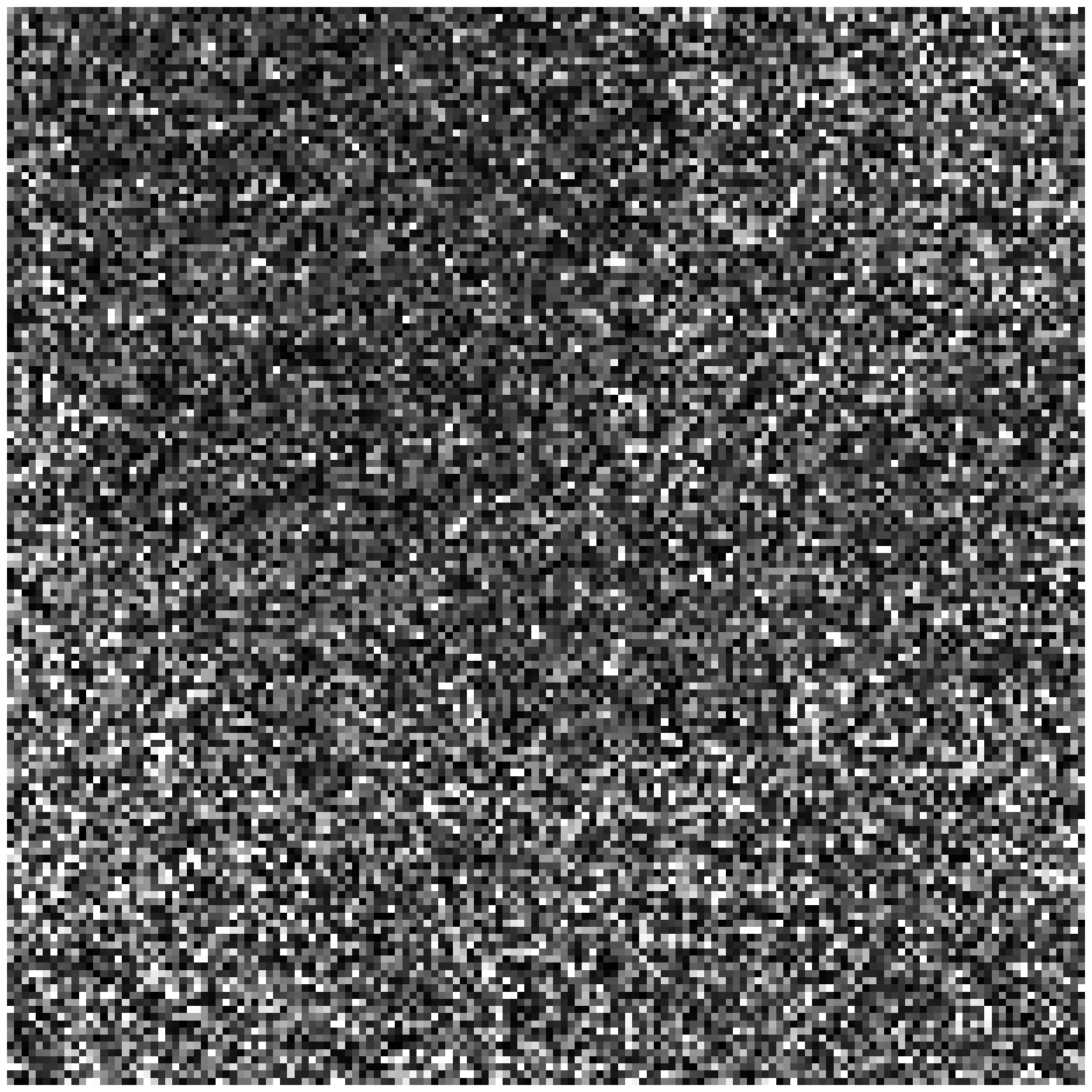}} 
      		\vfill
    \end{varwidth}
    \begin{varwidth}{\x\linewidth}
      		\centering
      		\centerline{\footnotesize WLS+L1}\smallskip
      		\centerline{\includegraphics[width=\linewidth]{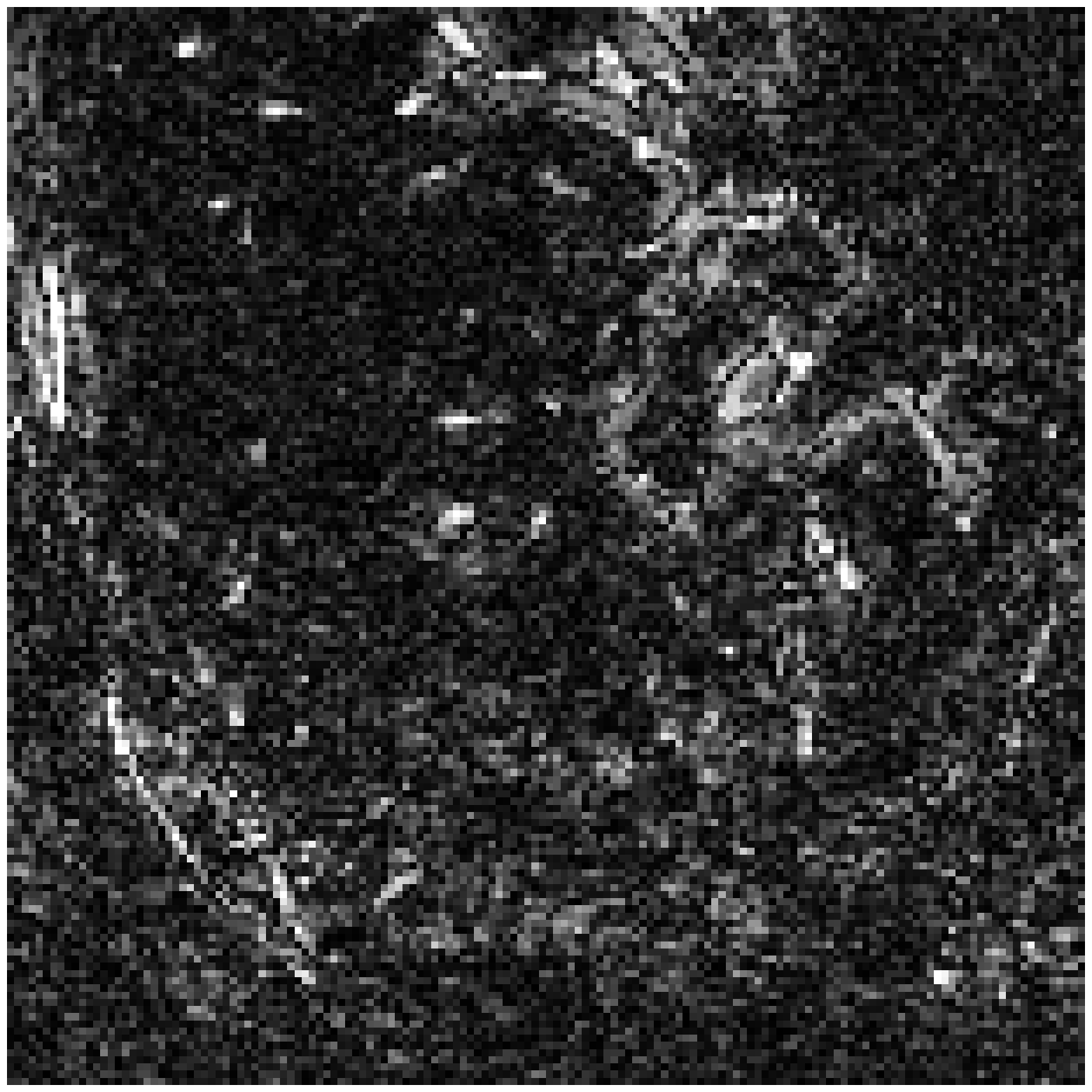}} 
    \end{varwidth}
    \begin{varwidth}{\x\linewidth}
      		\centering
      		\centerline{\footnotesize FBPConvNet}\smallskip
      		\centerline{\includegraphics[width=\linewidth]{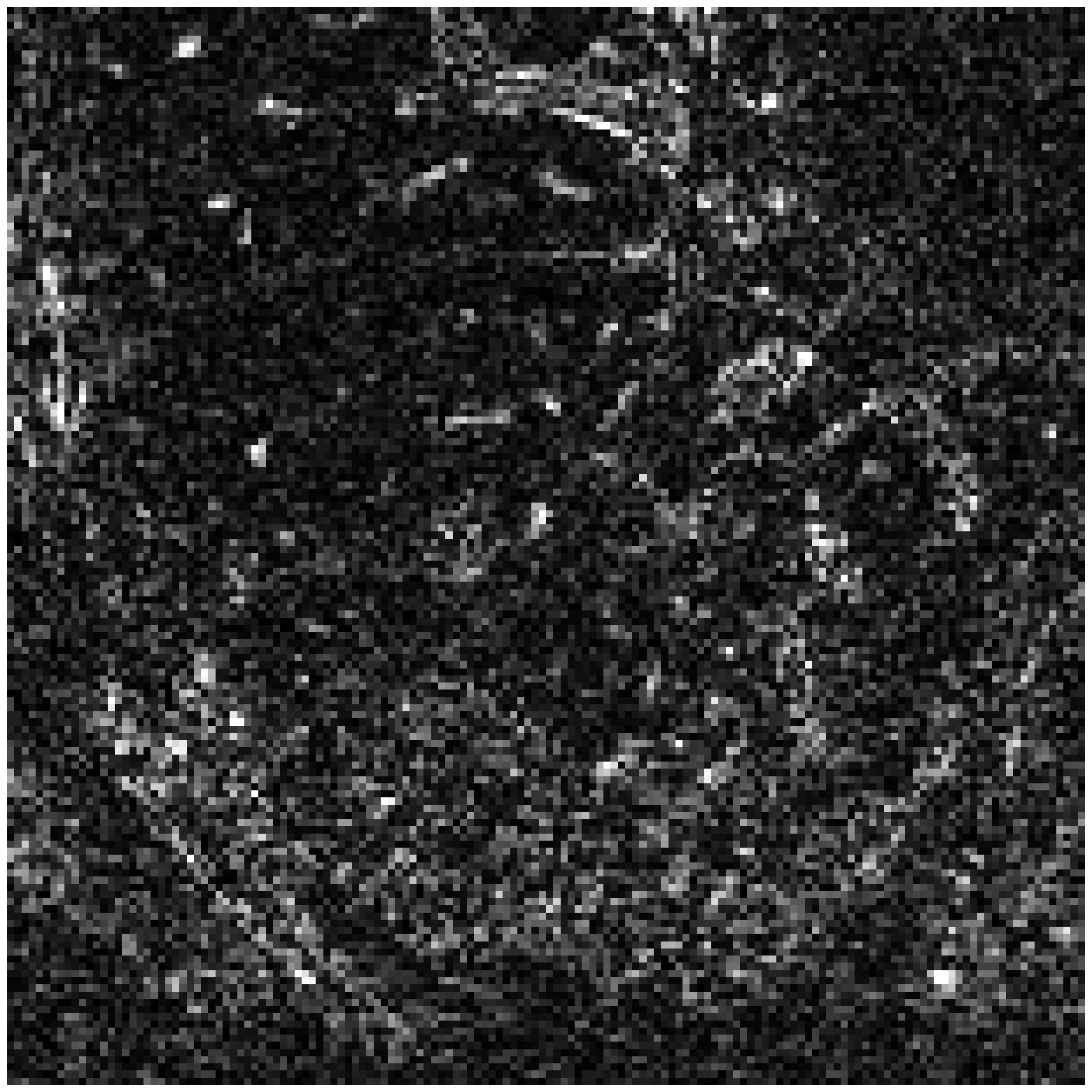}} 
    \end{varwidth}
    \\
    \begin{varwidth}{\x\linewidth}
      		\centering
      		\centerline{\includegraphics[width=\linewidth]{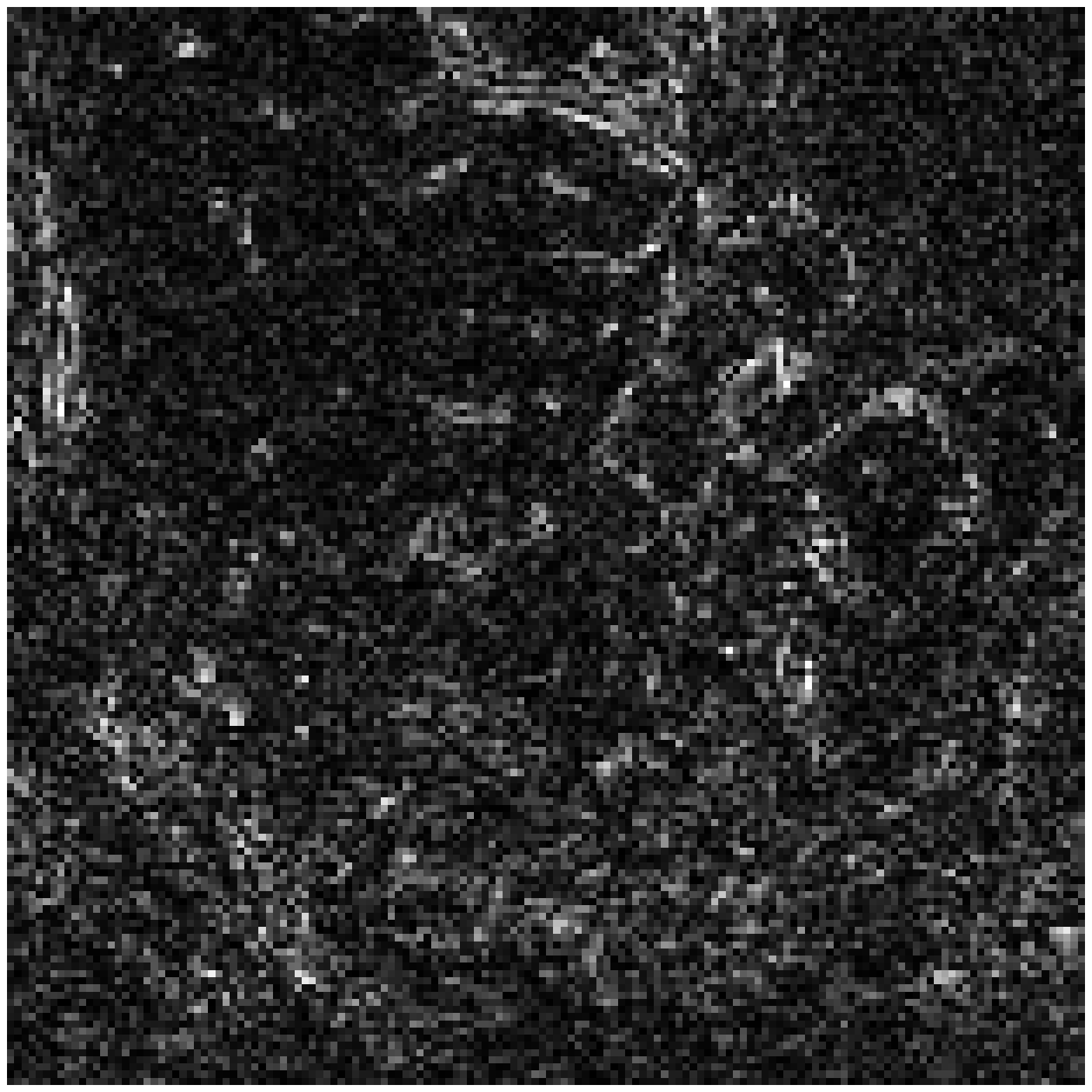}} 
      		\centerline{\footnotesize DD-Net}\smallskip
    \end{varwidth}
    \begin{varwidth}{\x\linewidth}
      		\centering
      		\centerline{\includegraphics[width=\linewidth]{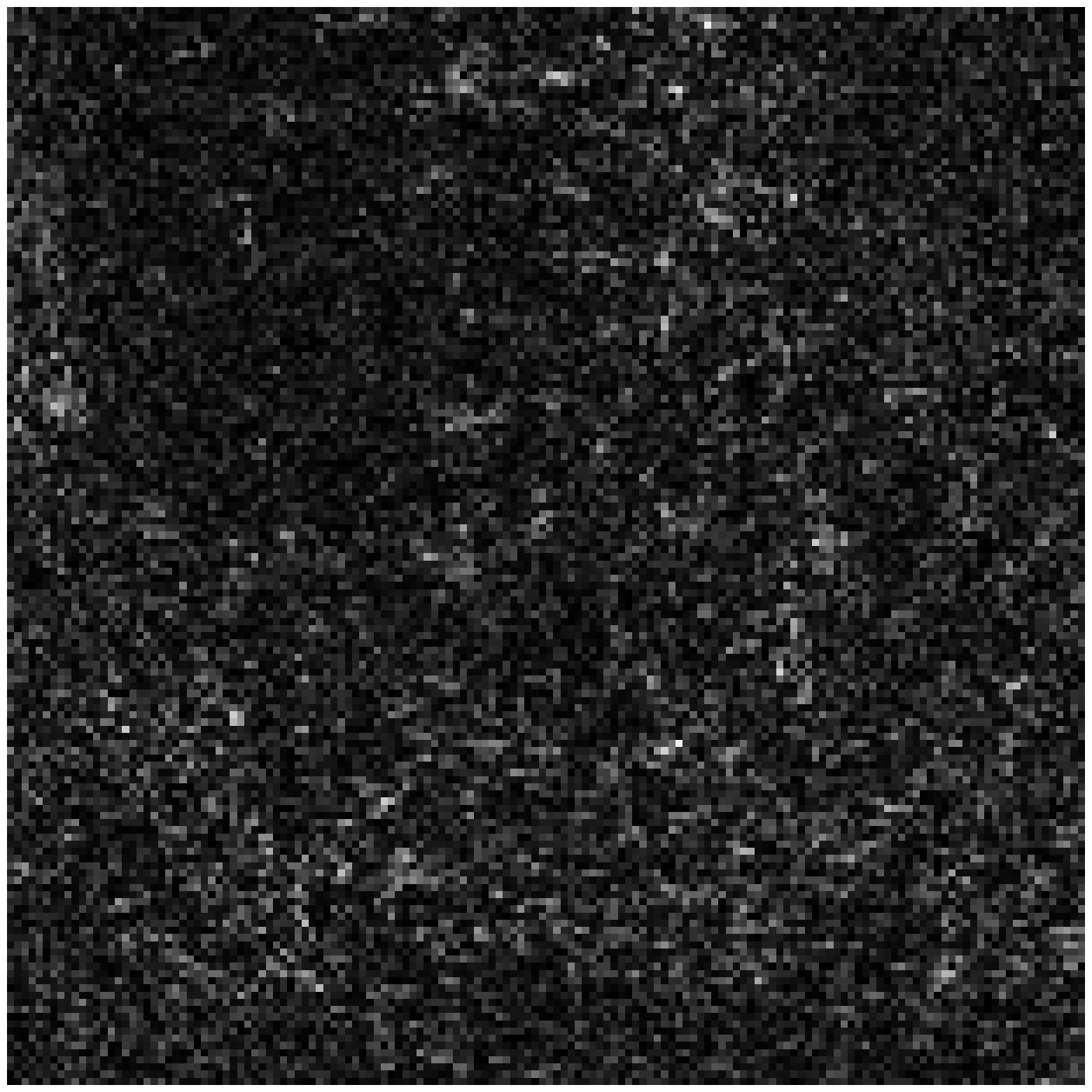}} 
      		\centerline{\footnotesize DRONE}\smallskip
    \end{varwidth}
    \begin{varwidth}{\x\linewidth}
      		\centering
      		\centerline{\includegraphics[width=\linewidth]{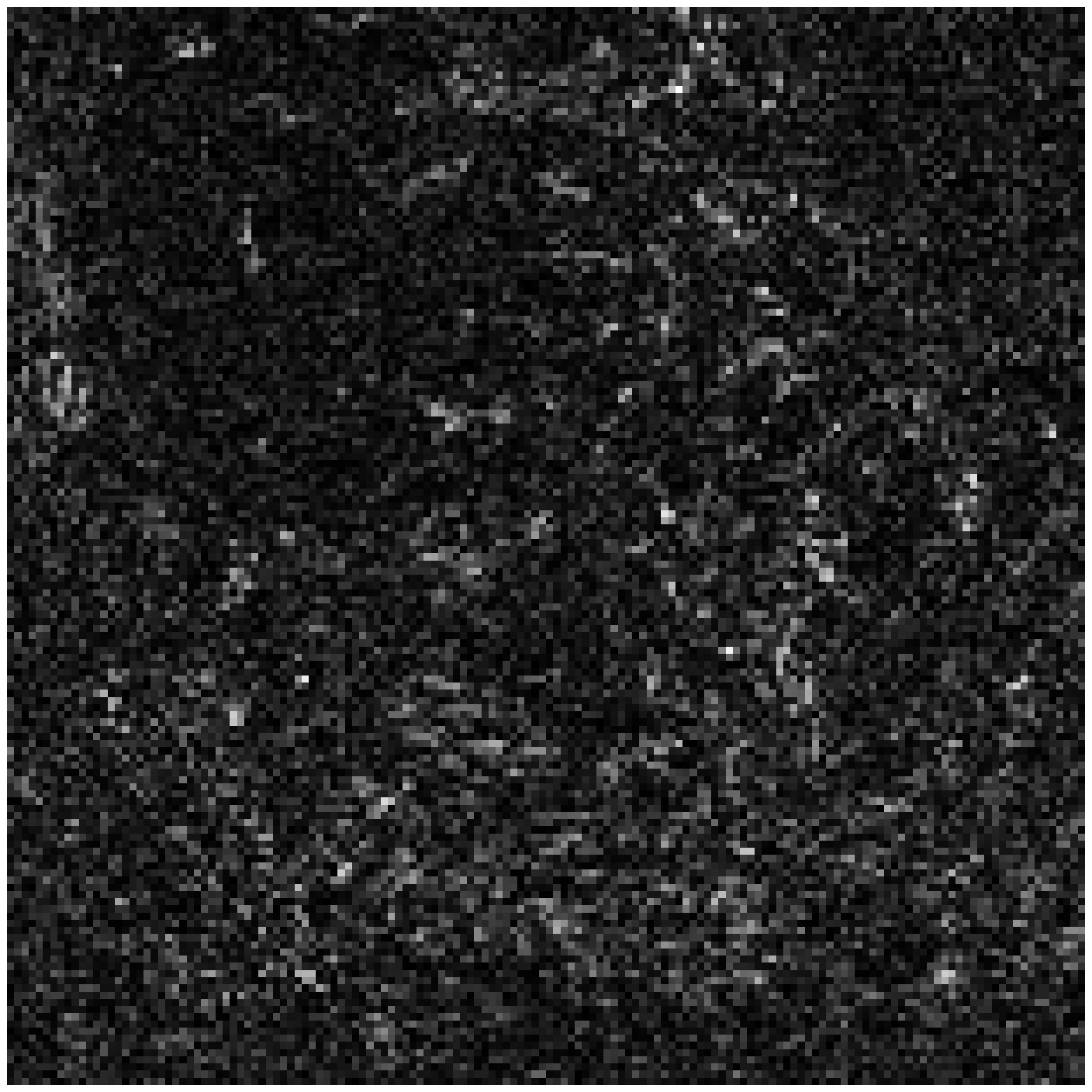}} 
      		\centerline{\footnotesize WNet (ours)}\smallskip
    \end{varwidth}
\end{varwidth}

\caption{Reconstruction results of a clinically relevant slice for determining COVID-19 pneumonia in a patient selected out of the test dataset. \textbf{(top set)} enlarged region-of-interest (ROI) for each individual reconstruction for a detailed view onto the local feature recovery performance of each algorithm investigated. Overall location of the ROI is shown as a red box in the full slice Ground Truth image which is the 2048-view FBP reconstruction $\tilde{x}_{2048}$ from Fig \ref{fig:main-results-1}. The display window for all reconstruction patches is [-1024,150] HU.\textbf{(bottom set)} Difference images to the ground truth corresponding to the output of each method. The display window for all generated difference images is [0,400] HU.}
\label{fig:main-results-2}
\end{figure}


\subsection{Data-driven FBP kernel weights} \label{section:methods,kernel}

An important part of the training process is the fine-tuning step performed on the ReM module filter $W$. At the start of the training procedure, this filter was initialized with the values of a Ram-Lak filter \cite{ramlak} and iteratively fine-tuned during training.

In an ideal continuous setting, inverting the tomographic reconstruction problem corresponds to finding the inverse of the Radon transform, which is given by the continuous FBP operation with a ramp filter \cite{precisionlearning}. However, in a real scenario, the discrete noisy measurements produce artifacts in the final reconstruction, for example cupping and \reviewerfoursecond{DC} shift artifacts \footnote{\reviewerfour{The discrete nature of the sinogram coupled with the zero-padding used for Fourier-based filtering in the FBP operation lead to a (\reviewerfoursecond{DC}) shift in the reconstructed values. For details on these types of artifacts see \cite{principles}.}} \cite{precisionlearning}, besides the sparse artifacts. Therefore, analytical solutions have been proposed, where the Ram-Lak filter has been replaced by custom made filters, like the Cosine or the Shepp-Logan filter, which penalize the frequency domain in different ways (the aforementioned three filters can be seen in Fig. \ref{fig:finetunedfilter}). We followed up to the study performed by W\"urfl et al. \cite{projdomweights} and enabled the fine-tuning of the FBP kernel during the training process. 

\subsection{Training Procedure}\label{section:training}

\begin{algorithm}
    \begin{algorithmic}
        \State Train $U_s$ in SdM for 5 \reviewertwo{\st{iterations}epochs} with $\mathcal{L}_s$
        \State Train $W$ in ReM for 5 \reviewertwo{\st{iterations}epochs} with $\mathcal{L}_r$
        \State Train $U_i$ in IdM for 5 \reviewertwo{\st{iterations}epochs} with $\mathcal{L}_i$
        \While {k in $0,1,\cdots$}
            \State Update $U_s$,$W$,$U_i$ with $\mathcal{L}_i$
        \EndWhile
    \end{algorithmic}
    \caption{Training pipeline}
    \label{alg:training}
\end{algorithm}

We train the three modules of WNet in a round-robin fashion as outlined in Alg. \ref{alg:training}. We first train $U_s$ in SdM for five \reviewertwo{\st{iterations}epochs} towards minimizing the loss function $\mathcal{L}_s$. \correction{\st{This ensures that the input of ReM is a properly $U_s$-denoised $K$-view sinogram $y_K^{U_s}$ and that training the ReM correctly fine-tunes its filter based on the denoised $K$-view sinogram}This ensures that ReM is trained from the beginning on a properly $U_s$-denoised $K$-view sinogram $y_K^{U_s}$}. The FBP filter is updated towards minimizing $\mathcal{L}_r$. The third and last layer, IdM, takes a reconstruction obtained from an FBP operation with the fine-tuned filter and is trained using $\mathcal{L}_i$. 
For both $\mathcal{L}_r$ and $\mathcal{L}_i$ we approximate $x$ with a full-view FBP reconstruction $\tilde{x}_{2048}$, as we will explain in more details in the next section.

\correction{\st{After the jump-start of all WNet modules the training continues iteratively, } The \enquote{jump-start} of all WNet modules is succeeded by a fine-tuning phase,} where during each iteration all three modules are concurrently updated by backpropagating the $\mathcal{L}_i$ loss (see \textbf{while}-loop in Alg. \ref{alg:training}):
\begin{equation*}
    \bm{\hat{U}_s}, \bm{\hat{W}}, \bm{\hat{U}_i} = \underset{U_i,W,U_s}{\text{arg min}} \, \mathcal{L}_i(U_i (A_{K}^T W U_s(y_{K}^{\mathcal{I}})),\tilde{x}_{2048})
\end{equation*}

\section{Experiments and Results} \label{section:results}

\subsection{Dataset} \label{section:dataset}
A total of 30 CT chest scans of patients suffering from COVID-19 pneumonia were acquired at Klinikum Rechts der Isar. Each scan was stripped of all the slices above and below the chest area which did not contain any lung region.
All 30 scans were obtained with a Philips iCT 256 based on a full-dose protocol in the lung window and reconstructed at a resolution of $512$ by $512$ pixels with a slice thickness of $0.9$mm and a pixel size of $\approx 0.72$mm. 
Data access was approved by the institutional ethics committee at Klinikum Rechts der Isar (Ethikvotum 87/18 S) and the data was anonymized.

To generate the required sparse sinograms and reconstructions we simulated an acquisition protocol based on a parallel-beam configuration with a detector resolution of $720$ pixels. In this study we chose $k=128$ uniformly sampled over a full-arc ($360\degree$) for the initial sparse-view sinogram $y_{128}$. This sparse-view protocol creates diagnostically challenging images for the radiologists. For the sinogram interpolation step, we set $C=4$, leading to $K=512$ angles in a quarter-view sinogram configuration. A quarter-view label sinogram $y_{512} $ was generated using the same parameters, as was a full-view label sinogram $y_{2048}$.

\reviewerthree{All slices were obtained in HU units and normalized to pixel values $[0,1]$. All subsequent operations are performed in this pixel value range.}

We generated a set of FBP reconstructions $\tilde{x}_{2048}$ from the simulated full-view sinograms $y_{2048}$ as labels for WNet. We chose not to use the slice data from the original scans, since this data has been processed and reconstructed in a blackbox clinical environment, with possibly additional post-reconstruction processing steps applied to the result. Therefore, generating matched forward- and filtered backprojections with a known reconstruction pipeline ensures compatibility of the input and label set and enables WNet to learn features related to the streaky nature of the SVCT images while remaining relatively transparent to the employed reconstruction parameters.

20 out of 30 scans were selected for the training set, while the other 10 were split in half and assigned to the validation and test set. We extend the training and validation set by applying an augmentation step to the individual slices which consists of a clockwise rotation of the original slice by $90\degree$, $180\degree$, and $270\degree$, thus expanding both sets by a factor of 4. In total we train WNet on $22800$ slices and we validate it on $4832$ slices. In the end, we test our trained network on the $5960$ slices in the test set.

For visualization purposes a physician chose a range of diagnostically relevant slices that display typical features of a COVID-19 pneumonia, which we show throughout this manuscript in Figs. \ref{fig:main-results-1}, \ref{fig:main-results-2},\ref{fig:sinograminterpolation}, \ref{fig:finetunedfilter}, \ref{fig:imagedenoising}. The COVID-19 related findings in the chosen slice are based on the ground glass opacity \footnote{\reviewerthree{refers to an area of increased attenuation in the lung with preserved inner structures \cite{ggo}}} of mixed consolidation in the dorsal region of the right lung, which is enlarged in the region-of-interest (ROI) in (red square in Fig. \ref{fig:main-results-2}-\enquote{Ground Truth} ). \correction{\st{We use the mentioned ROI to focus our analysis of the results in the next sections.}}

\subsection{Initialization and Training}\label{section:initialization}

WNet consists of around $17.3$M trainable parameters split up into $\approx 8.5$M for $U_s$, $\approx 8.5$M for $U_i$ and $1024$ for $W$. To initialize the weights of $U_s$ and $U_i$ we sample from the Xavier normal distribution \cite{xavier}, while the biases are generated with samples from the uniform distribution. The FBP operation in ReM is implemented with a Fourier-domain based filtering step. The filter weights are initialized with the Fourier-domain values of the Ram-lak filter, similar to the \textit{scikit} implementation of FBP (called \textit{iradon} in the \textit{scikit} framework) \cite{scikit-image}. The backprojection operation is implemented as a differentiable Pytorch layer, extending the \textit{torch.autograd} functionality with a custom backward gradient propagation operation \cite{pytorch}. The backend used for the matched forward and backward projectors is the C++17 tomographic reconstruction framework \textit{elsa} \cite{elsa}. The projector employed throughout the experiments is using Siddon's method in a parallel-beam configuration \cite{siddon}.

The Python library \textit{scikit-image} \cite{scikit-image} was used in the preprocessing stage for resizing and performing augmentation and in the training and inference stage to compute image comparison metrics.

To update the parameters of all three modules we use the Adam optimizer ($\beta_1=0.9$ and $\beta_2=0.999$) with a learning rate of $10^{-4}$ for each individual module. The batch size was set during the whole training process to 4. The number of epochs was 5 for the \enquote{jump-start} and then the whole network was fine-tuned for another 20 to ensure convergence, which was observed to be reached after 15 \reviewerfoursecond{epochs}.

\correction{The choice of the Huber function for $\mathcal{L}_s$, $\mathcal{L}_i$, and $\mathcal{L}_r$ is motivated empirically. We trained a stand-alone SdM and IdM module five times with three loss functions: L1 Norm, L2 Norm and Huber Norm and we concluded that the validation performance is on average approximately the same for all three types of loss function. As we trained our models against Ram-Lak filtered FBP reconstructions (Ground Truth), the models trained with Huber norm generated images with a good balance of smoothing and noise properties.}

\correction{We compare the impact of three different loss functions on the results of the corresponding trained network in the \enquote{Supplementary Materials} Section D and  Fig. A4.}

All networks and additional experiments were trained and run on a NVIDIA Quadro RTX 6000 GPU with 24GB pf VRAM. The underlying system was powered by an AMD EPYC 7452 32-Core CPU with 512GB RAM and running Ubuntu 20.04. The WNet was implemented in Pytorch 1.9 \cite{pytorch} with a CUDA 11.0 backend.

\renewcommand{\x}{0.23}
\begin{figure}[!t]
\begin{minipage}{0.01\linewidth}
        \vfill
  		\centerline{\begin{turn}{90}
  		\smallskip
  		\end{turn}}
\end{minipage}
\begin{minipage}{\x\linewidth}
  		\centering
  		\footnotesize 128-view \\ sparse \\ sinogram $y_{128}$
\end{minipage}
\begin{minipage}{\x\linewidth}
  		\centering
  		\footnotesize 512-view interpolated sinogram $y_{512}^{\text{int}}$
\end{minipage}
\begin{minipage}{\x\linewidth}
  		\centering
  		\footnotesize 512-view enhanced sinogram $y_{512}^{\mathcal{I}}$
\end{minipage}
\begin{minipage}{\x\linewidth}
  		\centering
  		\footnotesize 512-view \\ denoised sinogram $y_{512}^{\bm{\hat{U}_s}}$
\end{minipage}
\vspace{1em}
\\
\begin{minipage}{0.02\linewidth}
  		\begin{turn}{90}
  		\small Sino
  		\end{turn}
\end{minipage}
\begin{minipage}{\x\linewidth}
  		\centering
  		\centerline{\includegraphics[width=\linewidth,height=\linewidth]{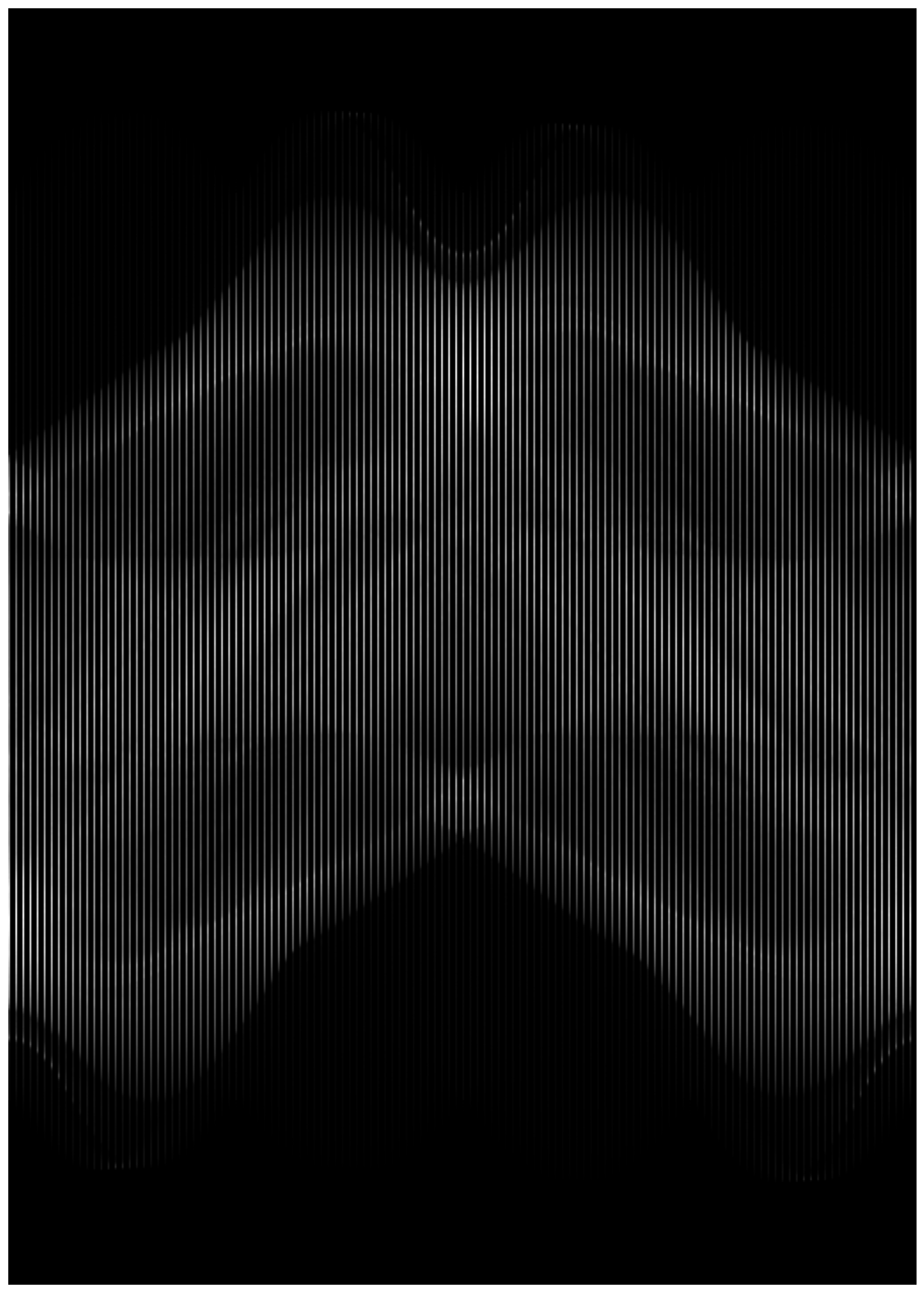}} 
\end{minipage}
\begin{minipage}{\x\linewidth}
  		\centering
  		\centerline{\includegraphics[width=\linewidth,height=\linewidth]{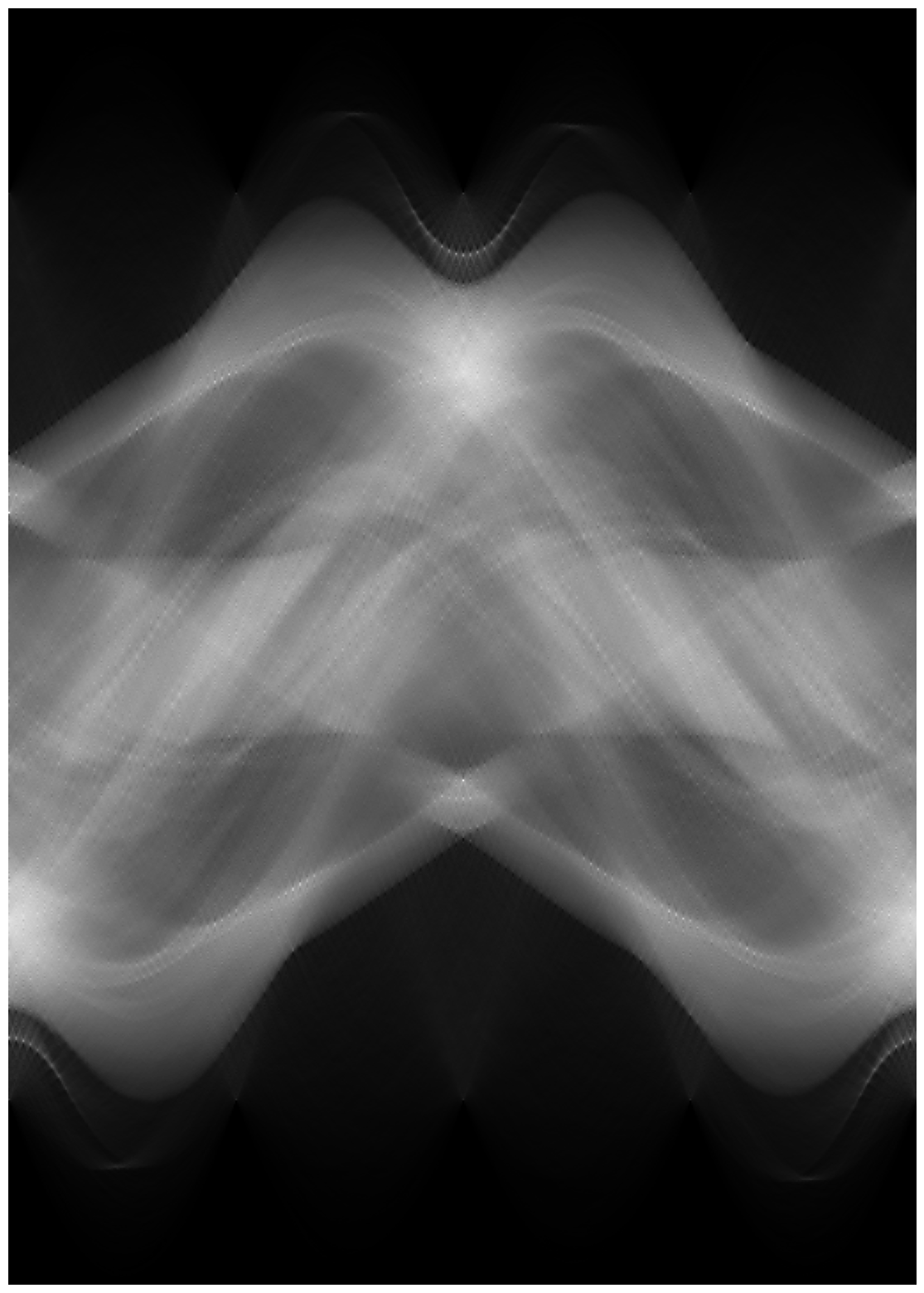}} 
\end{minipage}
\begin{minipage}{\x\linewidth}
  		\centering
  		\centerline{\includegraphics[width=\linewidth,height=\linewidth]{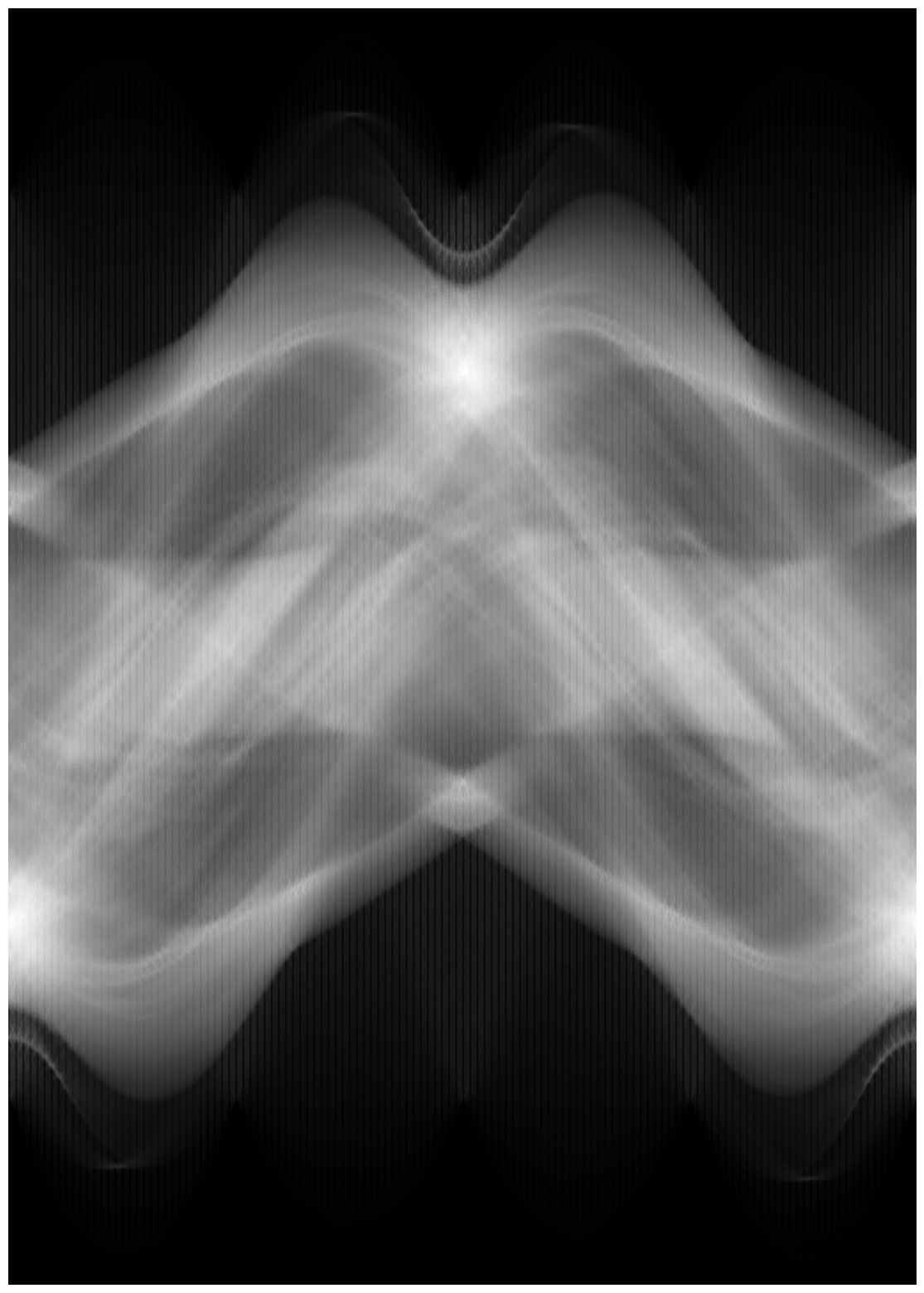}} 
\end{minipage}
\begin{minipage}{\x\linewidth}
  		\centering
  		\centerline{\includegraphics[width=\linewidth,height=\linewidth]{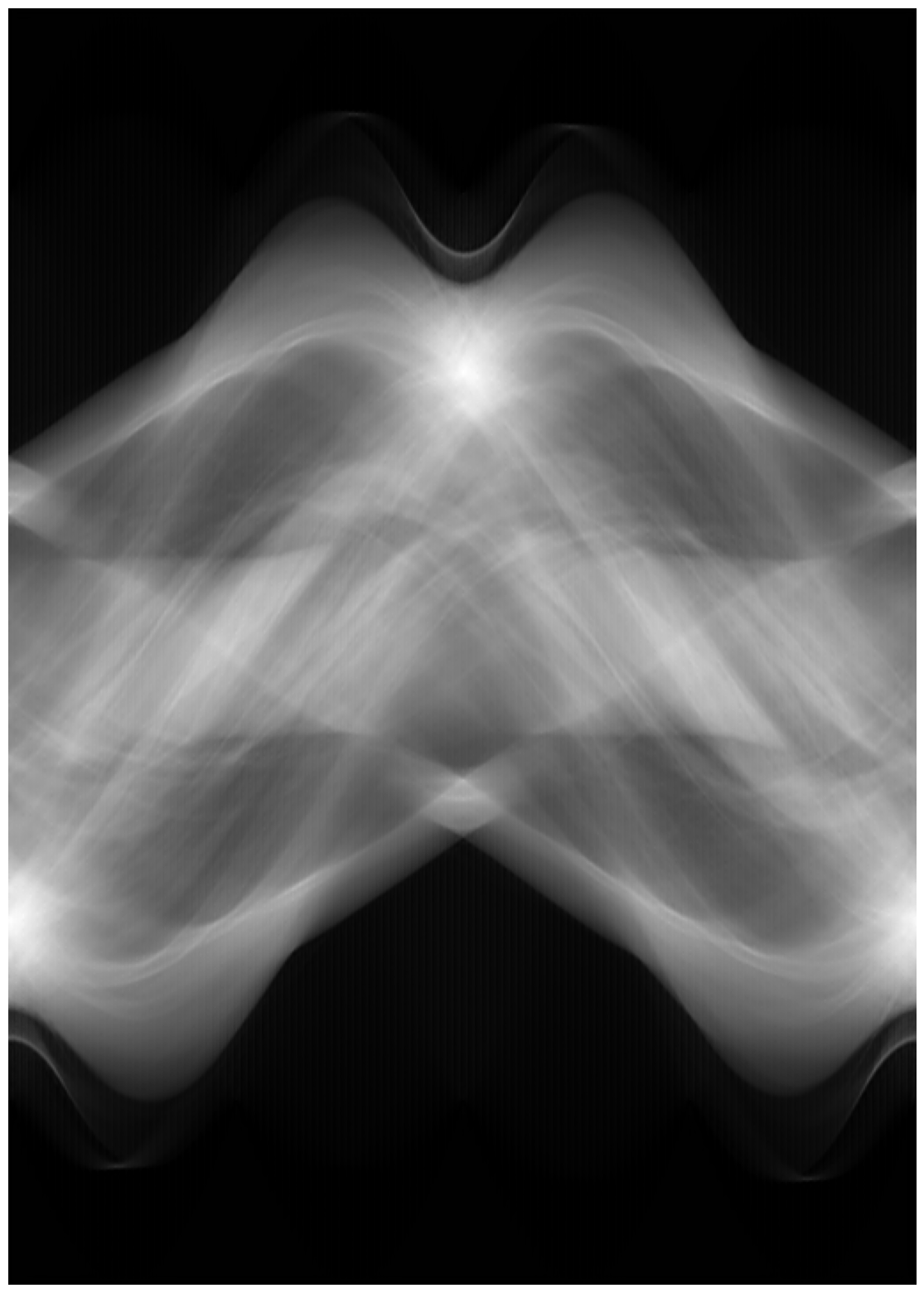}} 
\end{minipage}
\\
\begin{minipage}{0.02\linewidth}
  		\begin{turn}{90}
  		\footnotesize Diff. Sino
  		\end{turn}
\end{minipage}
\begin{minipage}{\x\linewidth}
  		\centering
  		\centerline{\includegraphics[width=\linewidth,height=\linewidth]{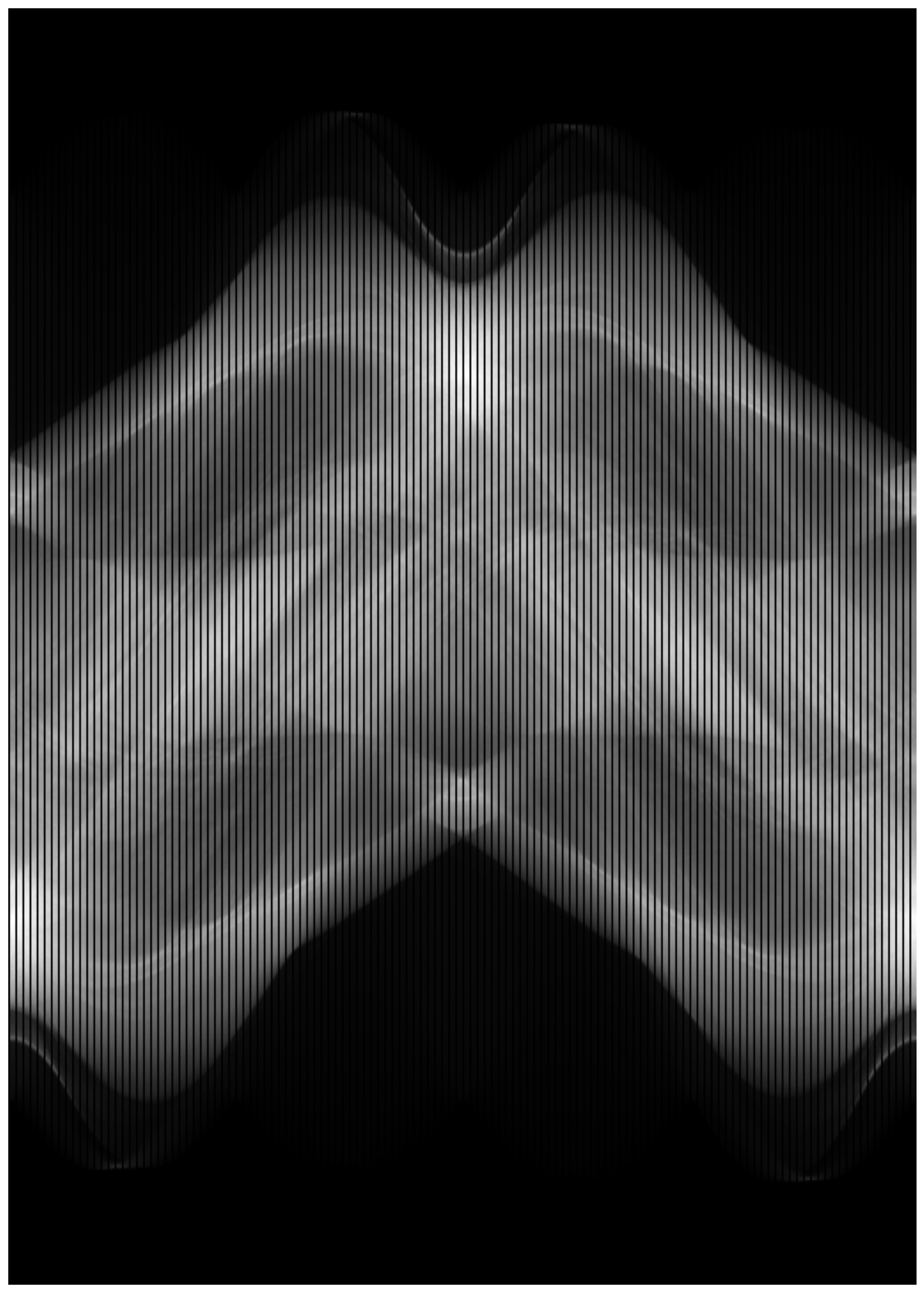}} 
\end{minipage}
\begin{minipage}{\x\linewidth}
  		\centering
  		\centerline{\includegraphics[width=\linewidth,height=\linewidth]{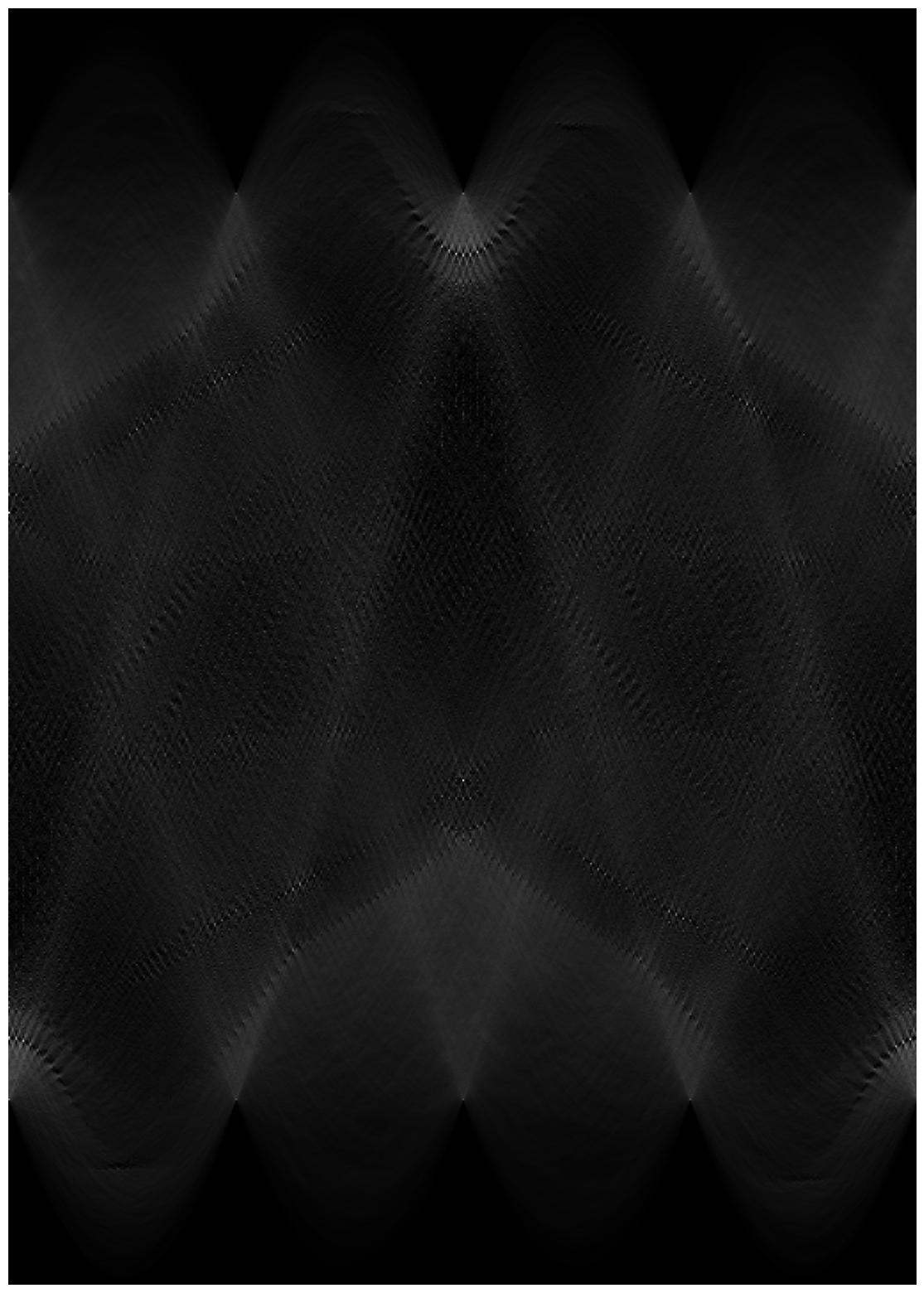}} 
\end{minipage}
\begin{minipage}{\x\linewidth}
  		\centering
  		\centerline{\includegraphics[width=\linewidth,height=\linewidth]{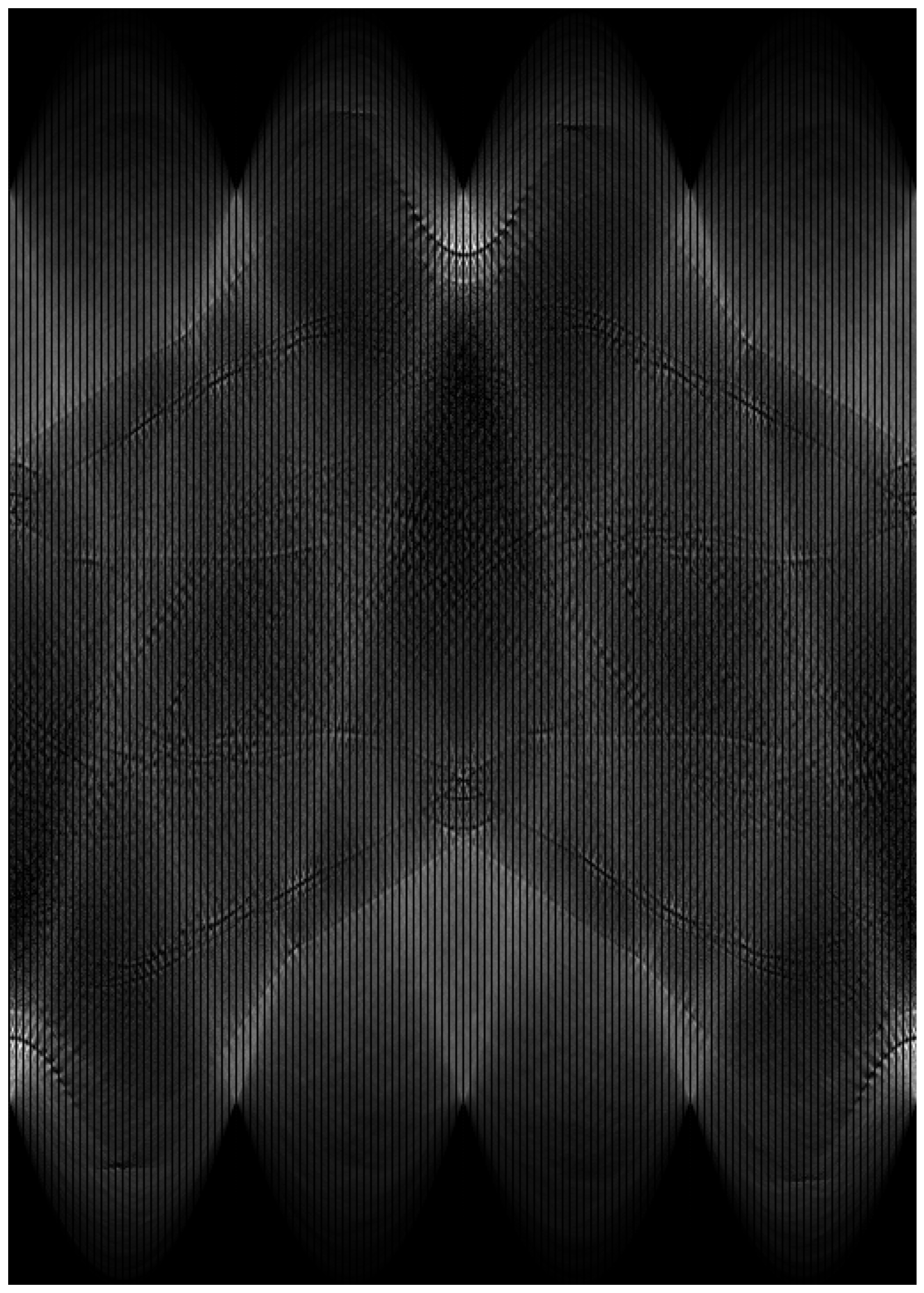}} 
\end{minipage}
\begin{minipage}{\x\linewidth}
  		\centering
  		\centerline{\includegraphics[width=\linewidth,height=\linewidth]{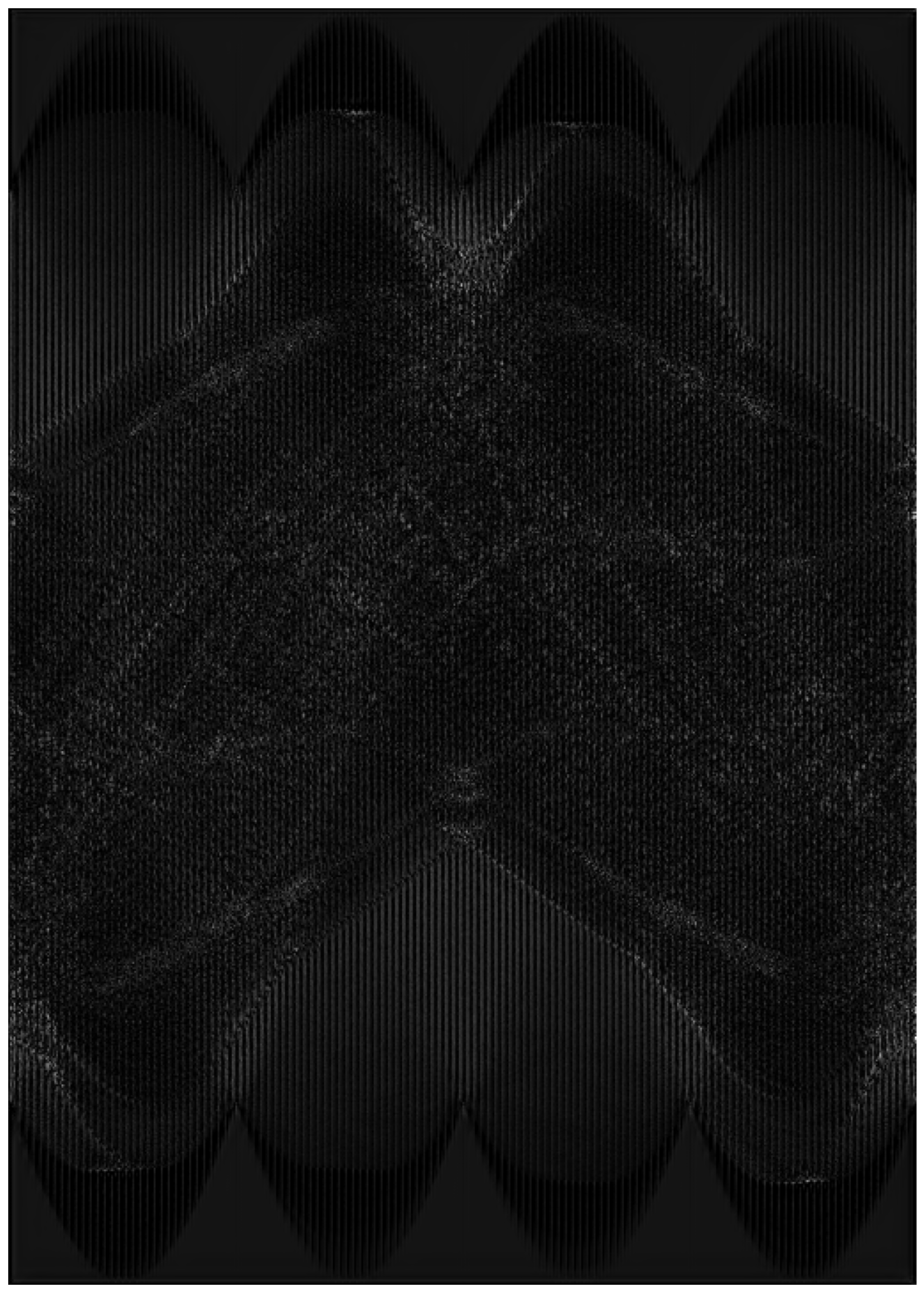}} 
\end{minipage}
\vspace{1em}
\\
\begin{minipage}{0.02\linewidth}
  		\begin{turn}{90}
  		\footnotesize Recon
  		\end{turn}
\end{minipage}
\begin{minipage}{\x\linewidth}
  		\centering
  		\centerline{\includegraphics[width=\linewidth]{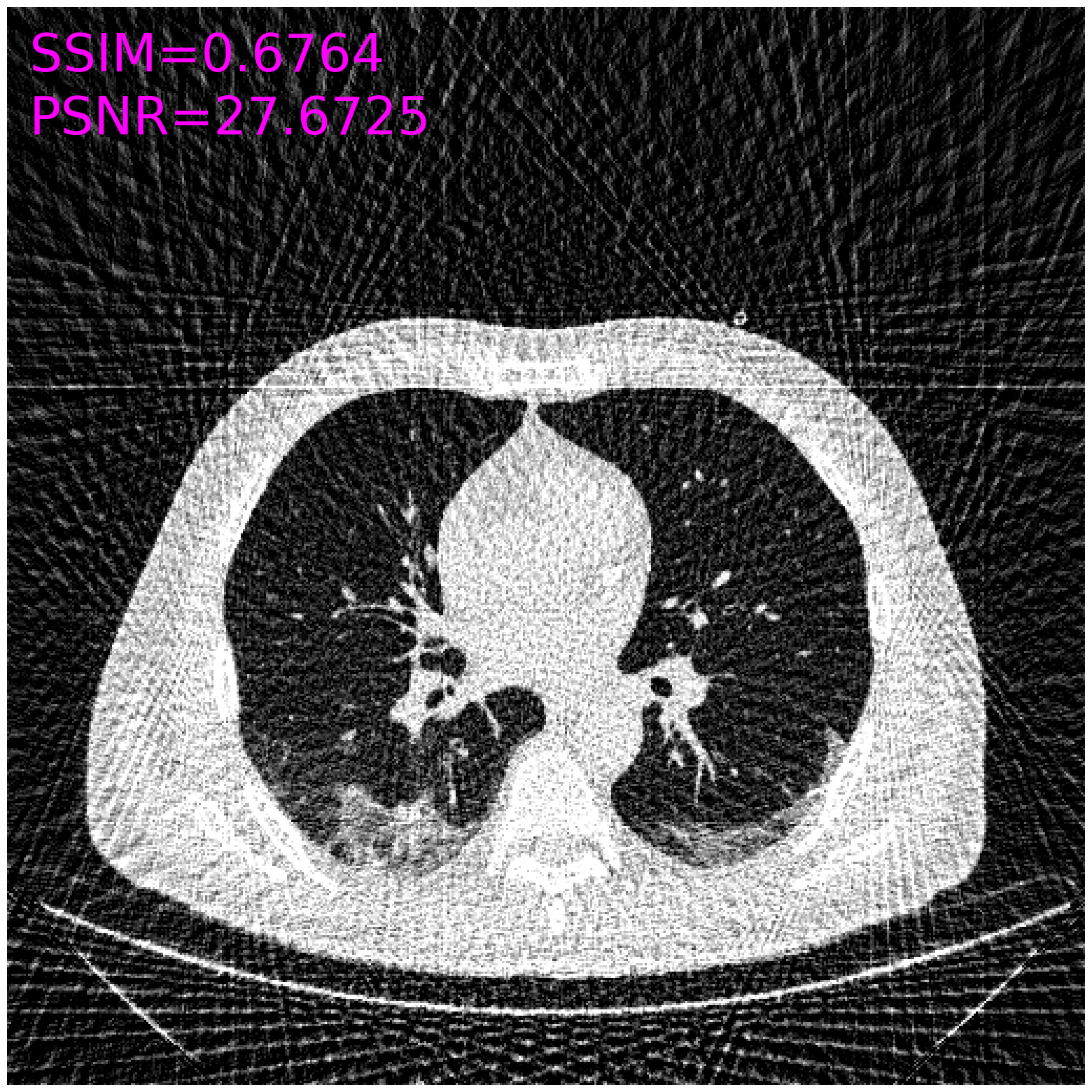}} 
\end{minipage}
\begin{minipage}{\x\linewidth}
  		\centering
  		\centerline{\includegraphics[width=\linewidth]{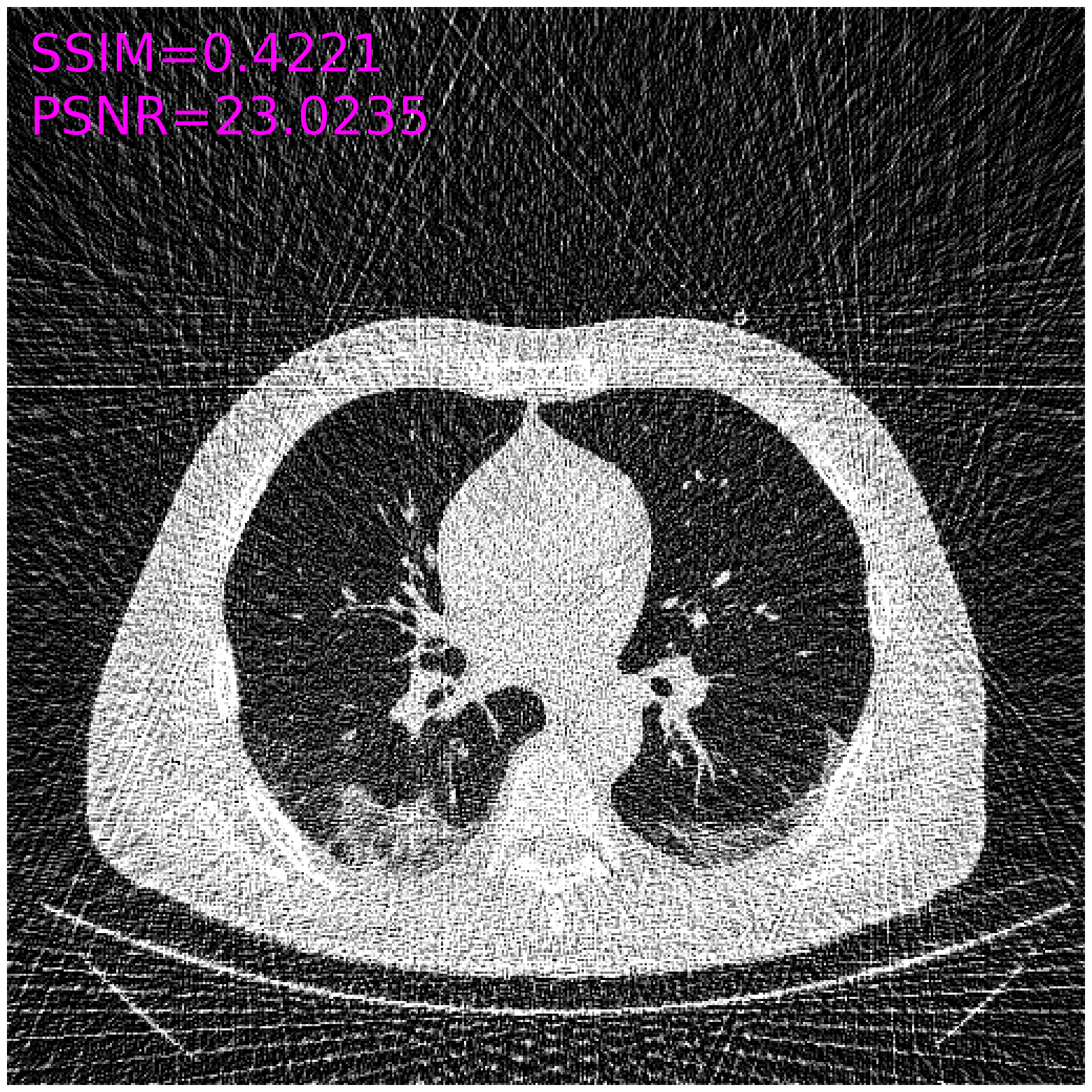}} 
\end{minipage}
\begin{minipage}{\x\linewidth}
  		\centering
  		\centerline{\includegraphics[width=\linewidth]{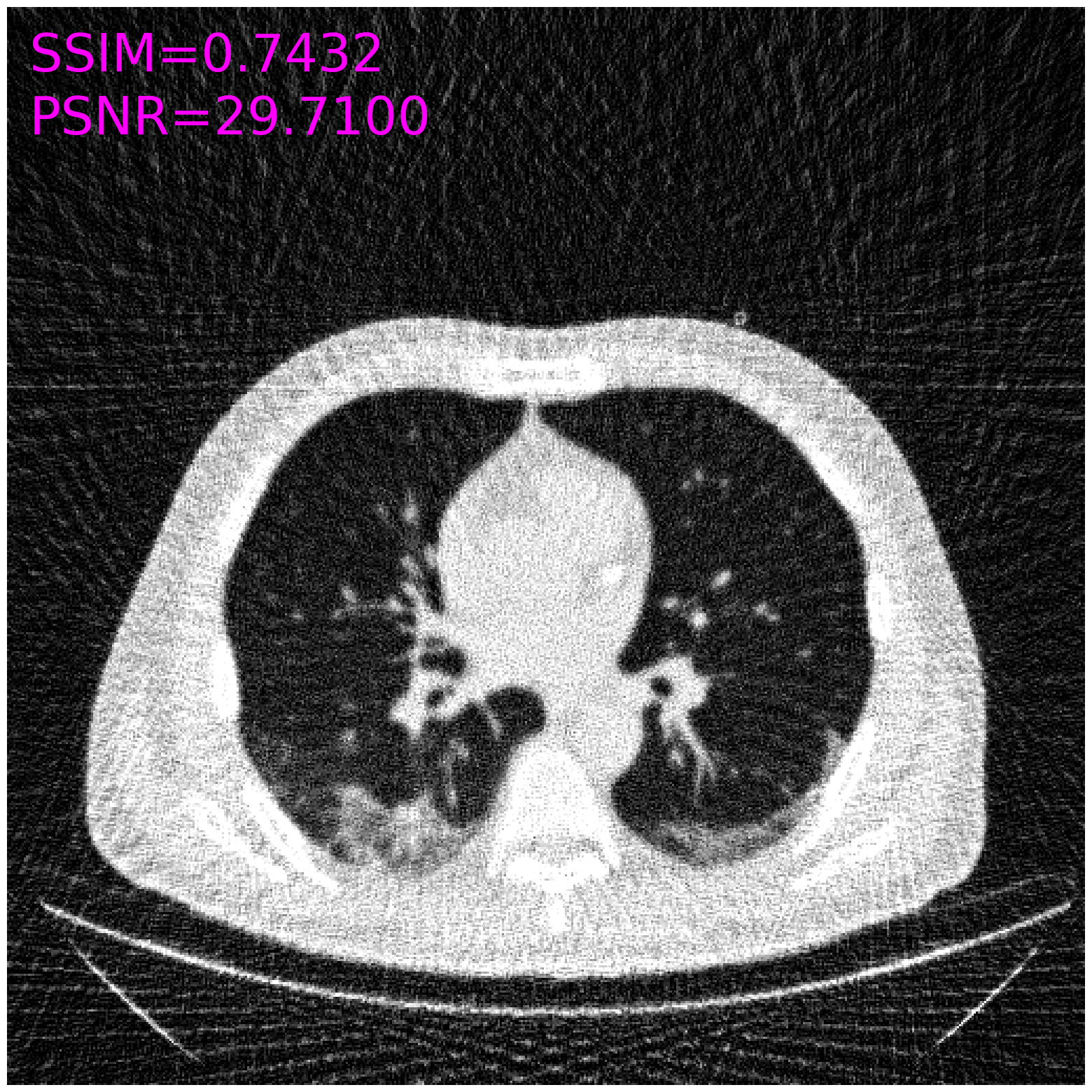}} 
\end{minipage}
\begin{minipage}{\x\linewidth}
  		\centering
  		\centerline{\includegraphics[width=\linewidth]{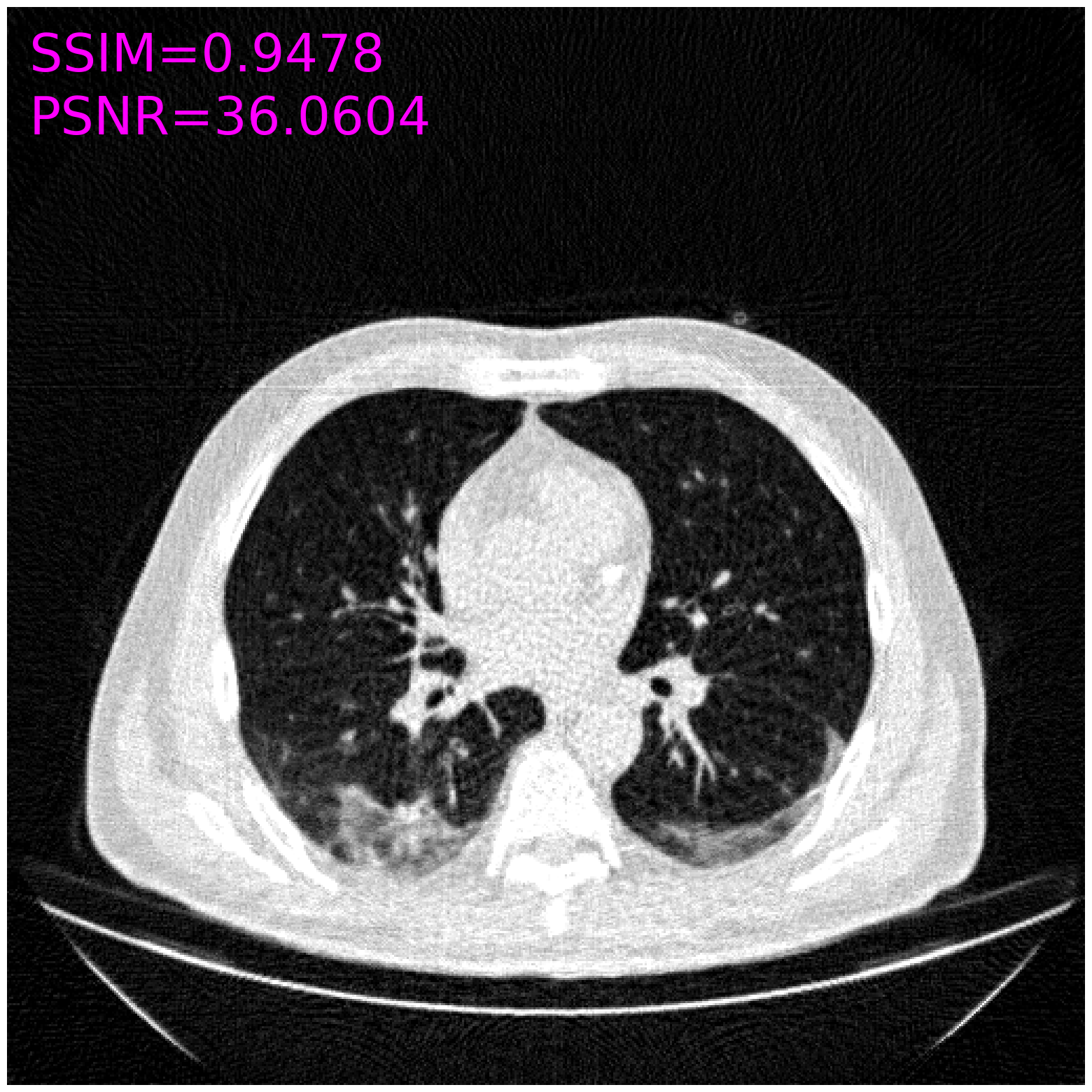}} 
\end{minipage}
\\
\begin{minipage}{0.02\linewidth}
  		\begin{turn}{90}
  		\footnotesize Diff. Recon
  		\end{turn}
\end{minipage}
\begin{minipage}{\x\linewidth}
  		\centering
  		\centerline{\includegraphics[width=\linewidth]{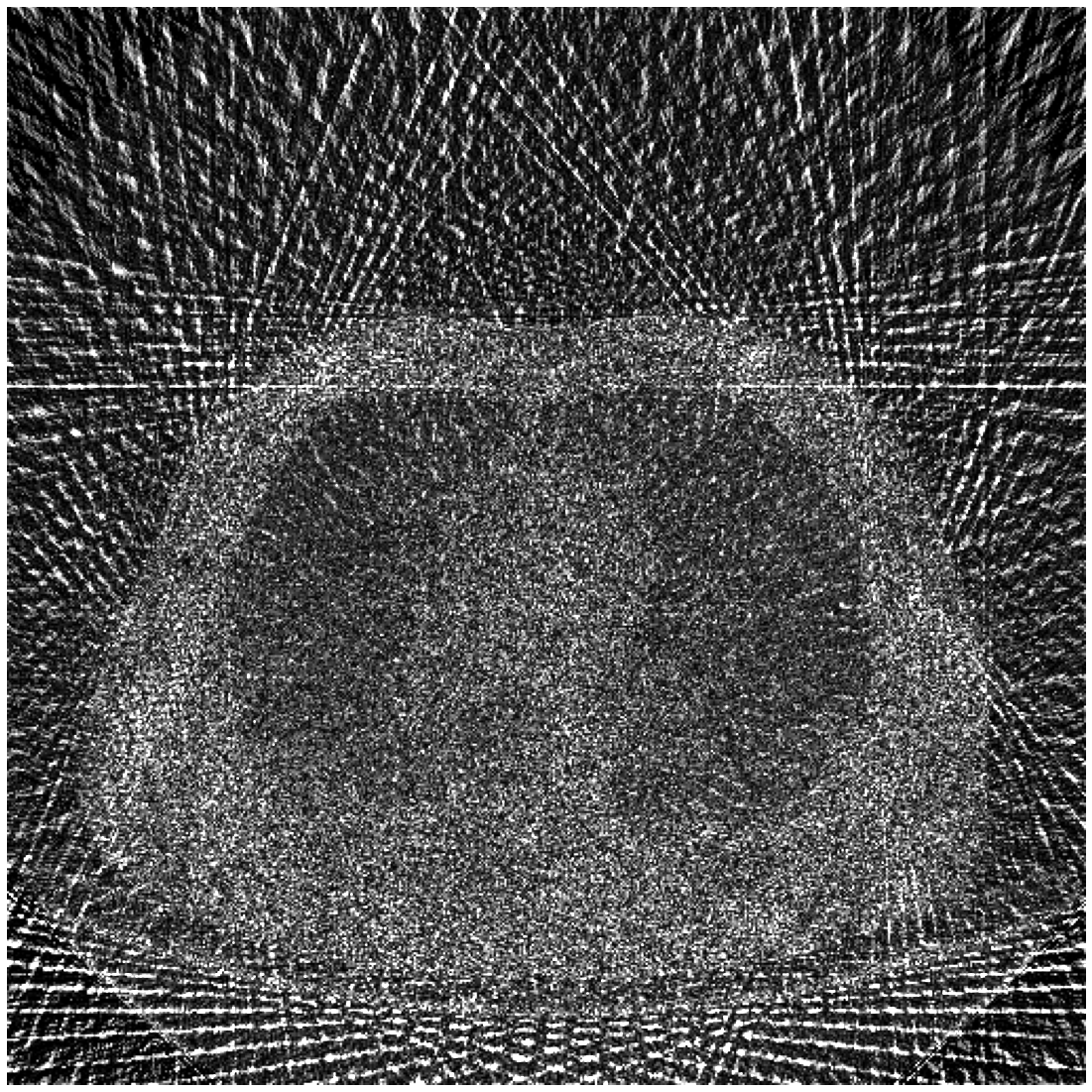}} 
\end{minipage}
\begin{minipage}{\x\linewidth}
  		\centering
  		\centerline{\includegraphics[width=\linewidth]{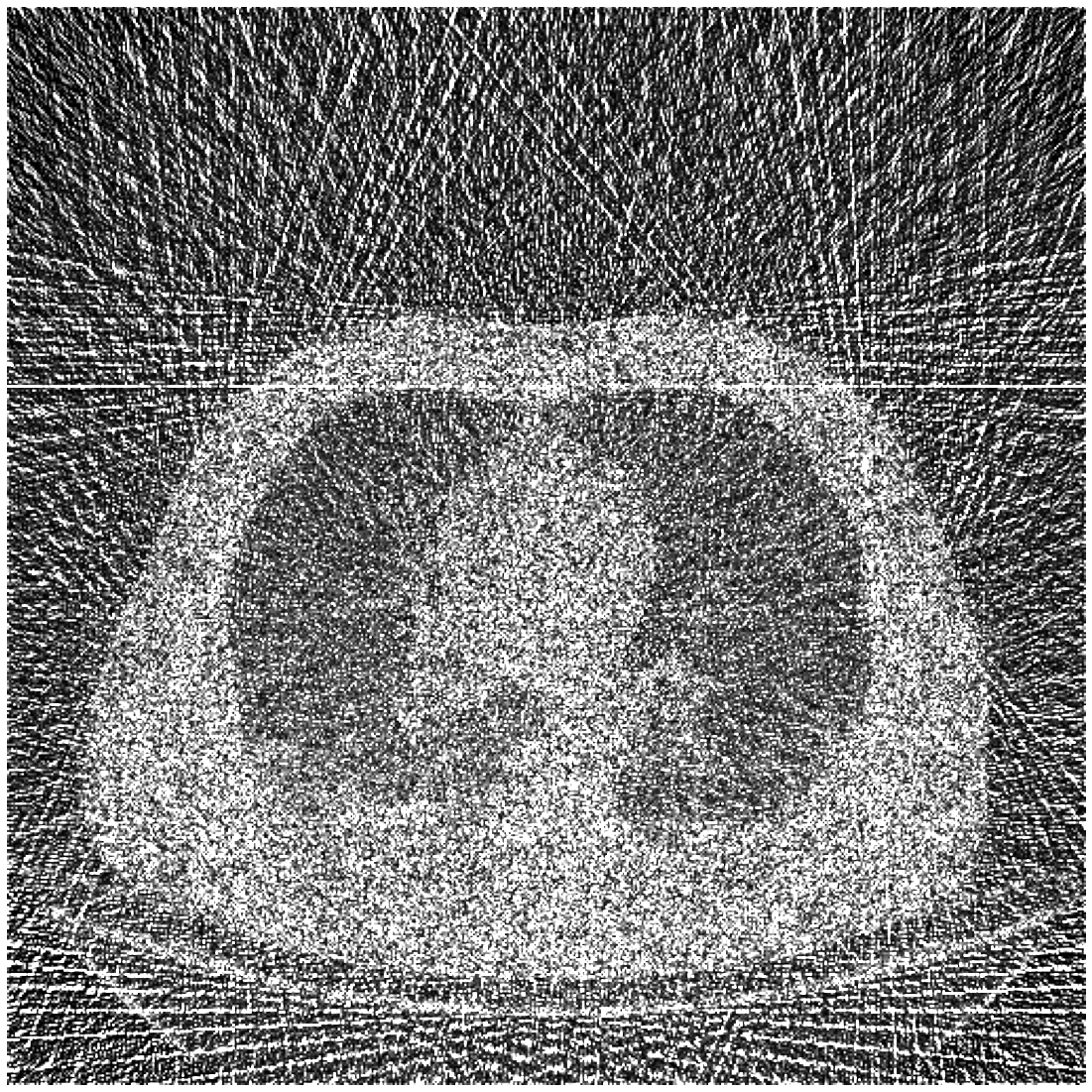}} 
\end{minipage}
\begin{minipage}{\x\linewidth}
  		\centering
  		\centerline{\includegraphics[width=\linewidth]{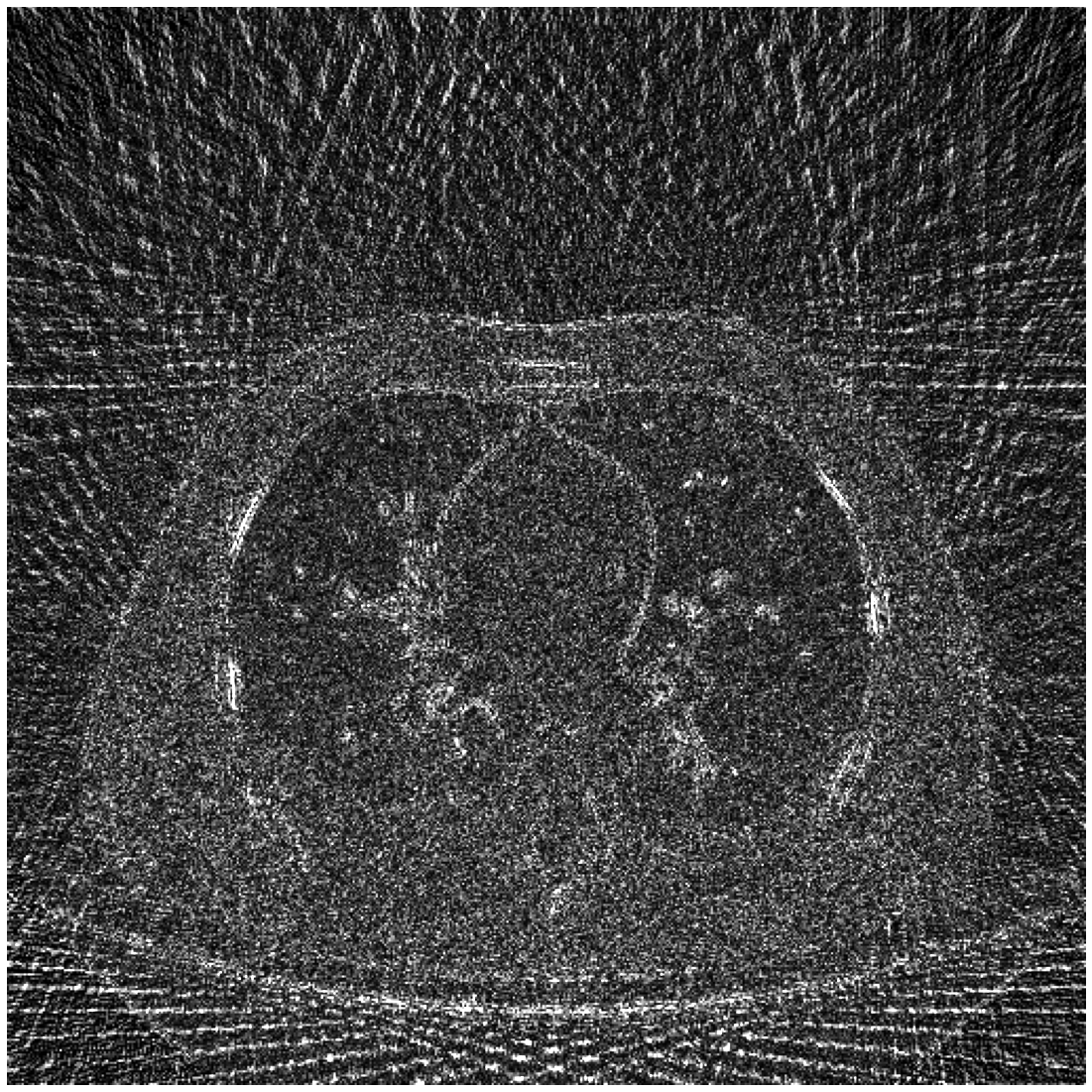}} 
\end{minipage}
\begin{minipage}{\x\linewidth}
  		\centering
  		\centerline{\includegraphics[width=\linewidth]{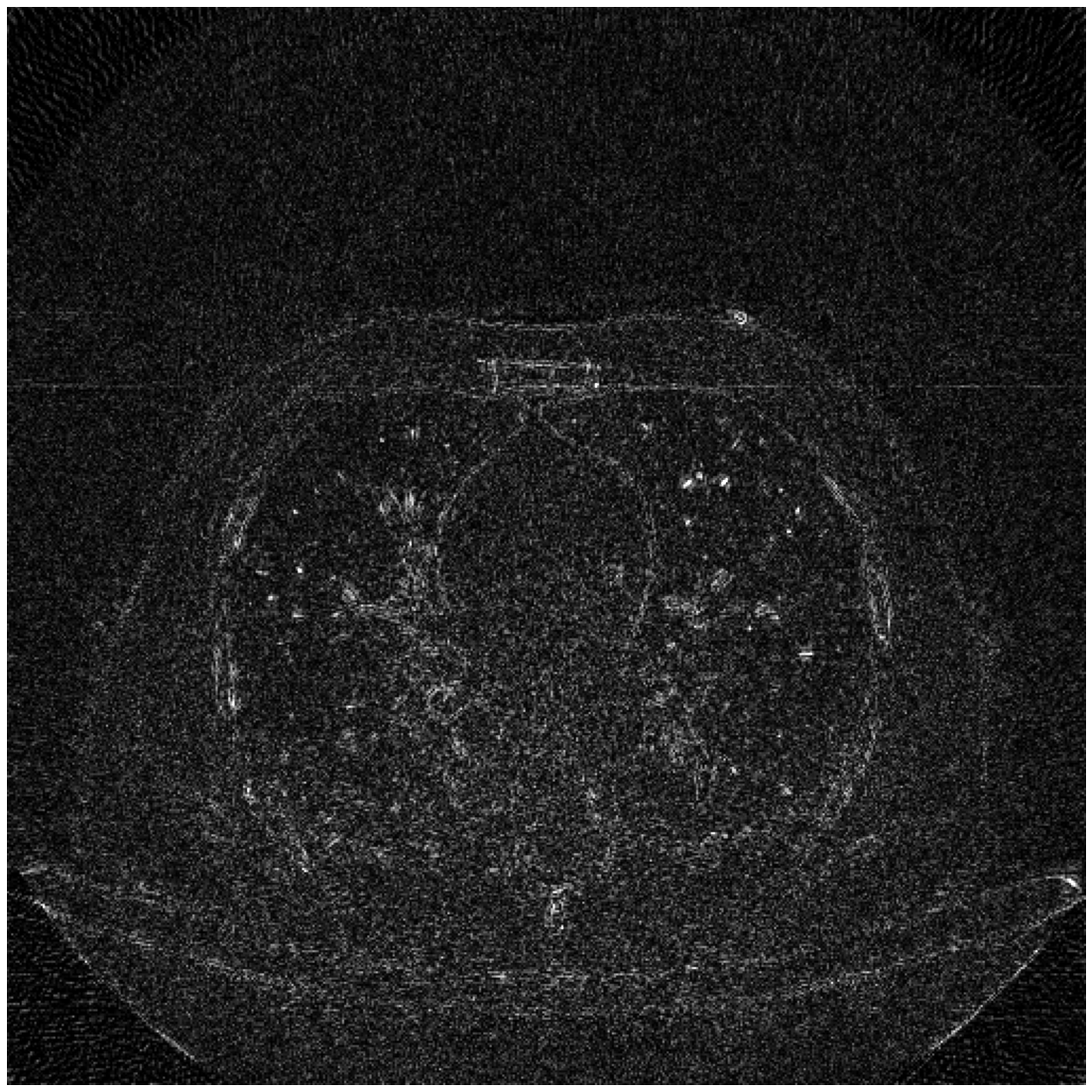}} 
\end{minipage}
\vspace{1em}
\\
\begin{minipage}{0.02\linewidth}
  		\begin{turn}{90}
  		\footnotesize ROI
  		\end{turn}
\end{minipage}
\begin{minipage}{\x\linewidth}
  		\centering
  		\centerline{\includegraphics[width=\linewidth]{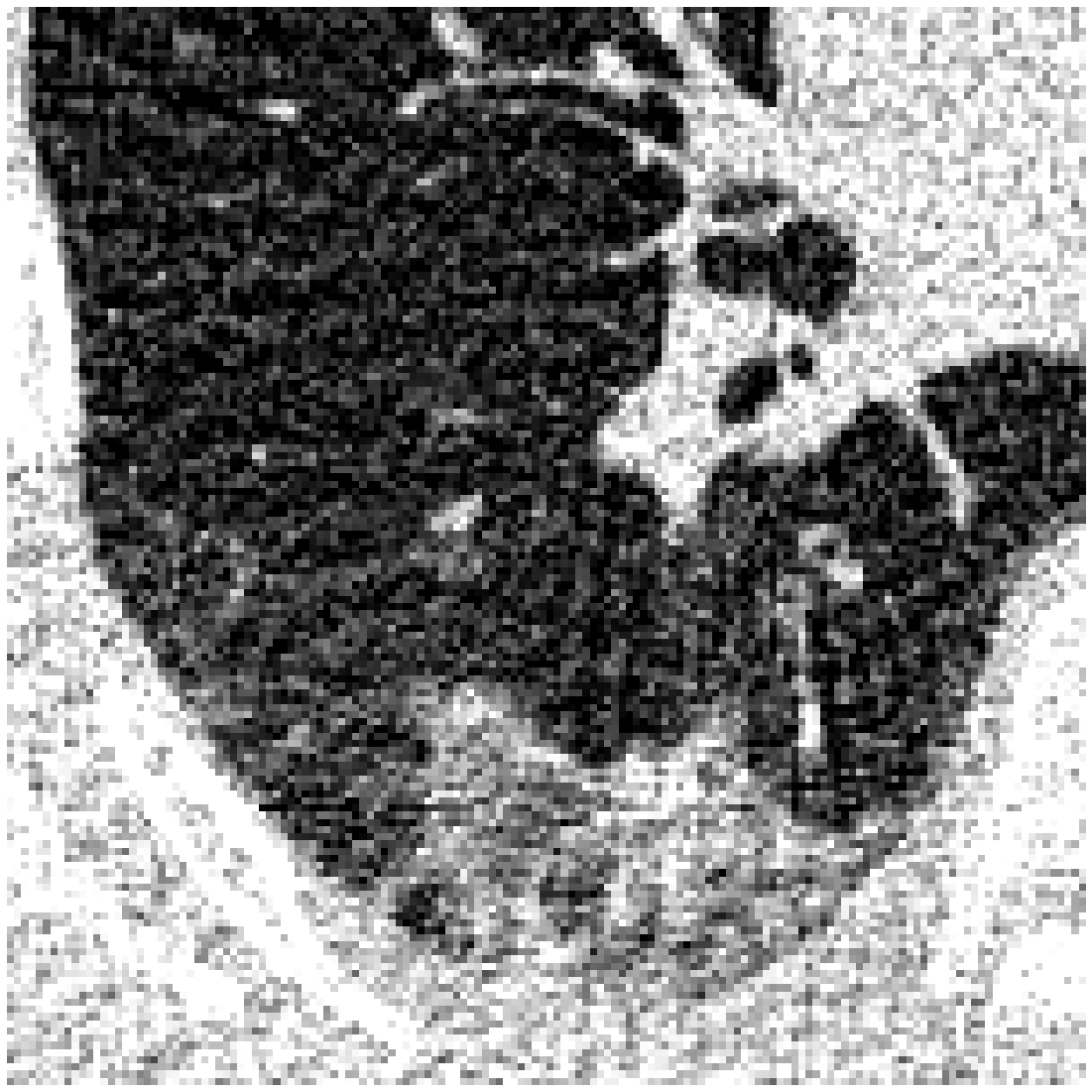}} 
\end{minipage}
\begin{minipage}{\x\linewidth}
  		\centering
  		\centerline{\includegraphics[width=\linewidth]{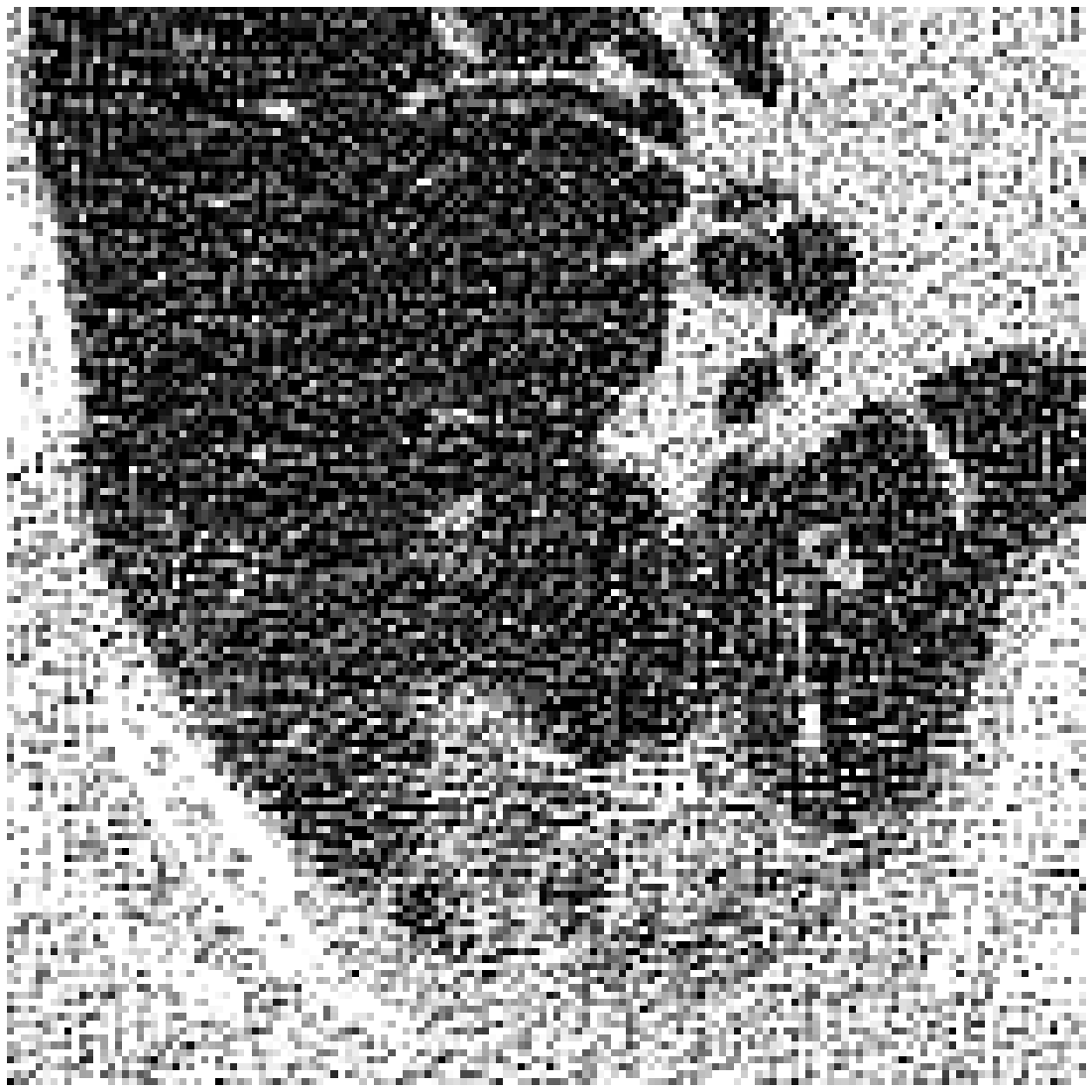}} 
\end{minipage}
\begin{minipage}{\x\linewidth}
  		\centering
  		\centerline{\includegraphics[width=\linewidth]{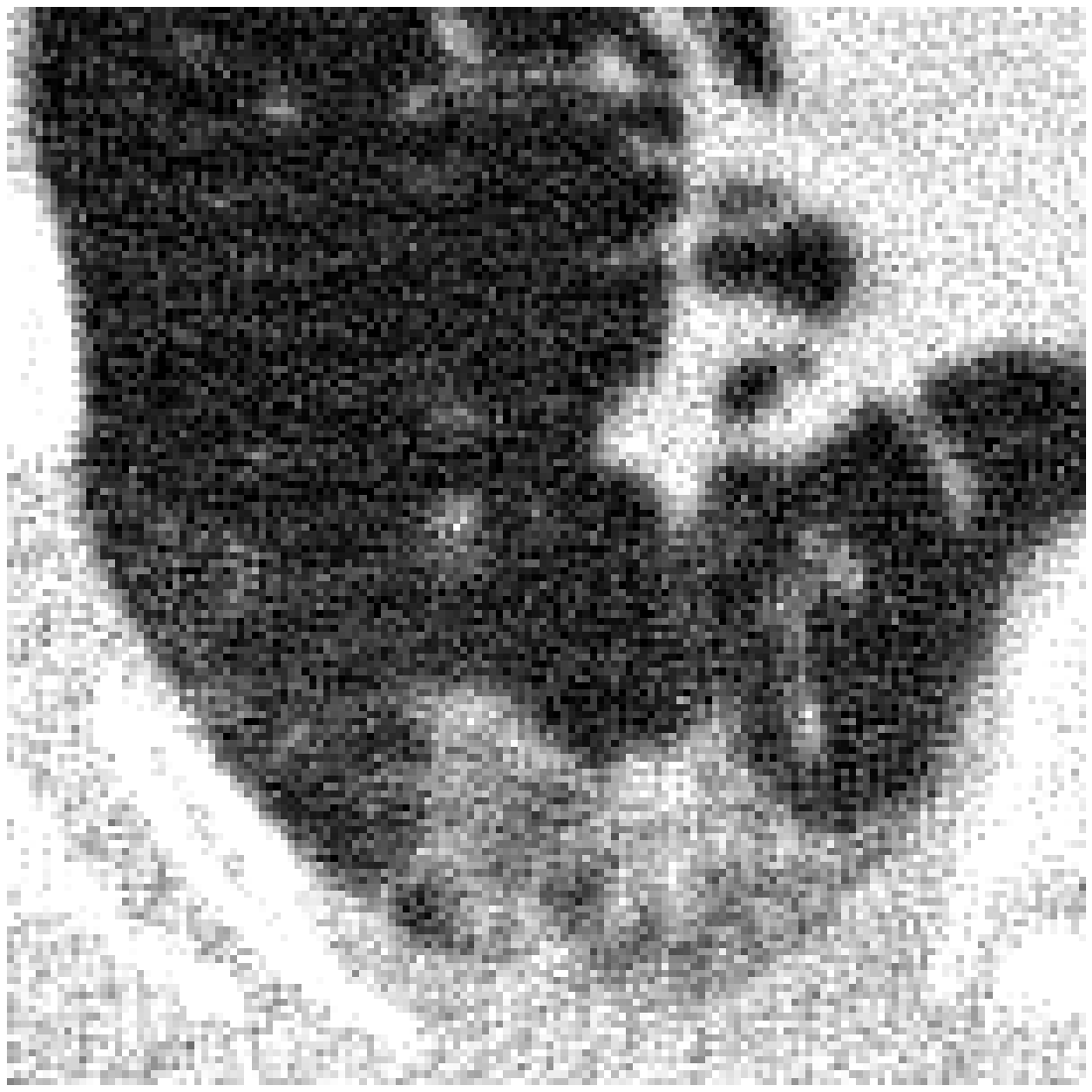}} 
\end{minipage}
\begin{minipage}{\x\linewidth}
  		\centering
  		\centerline{\includegraphics[width=\linewidth]{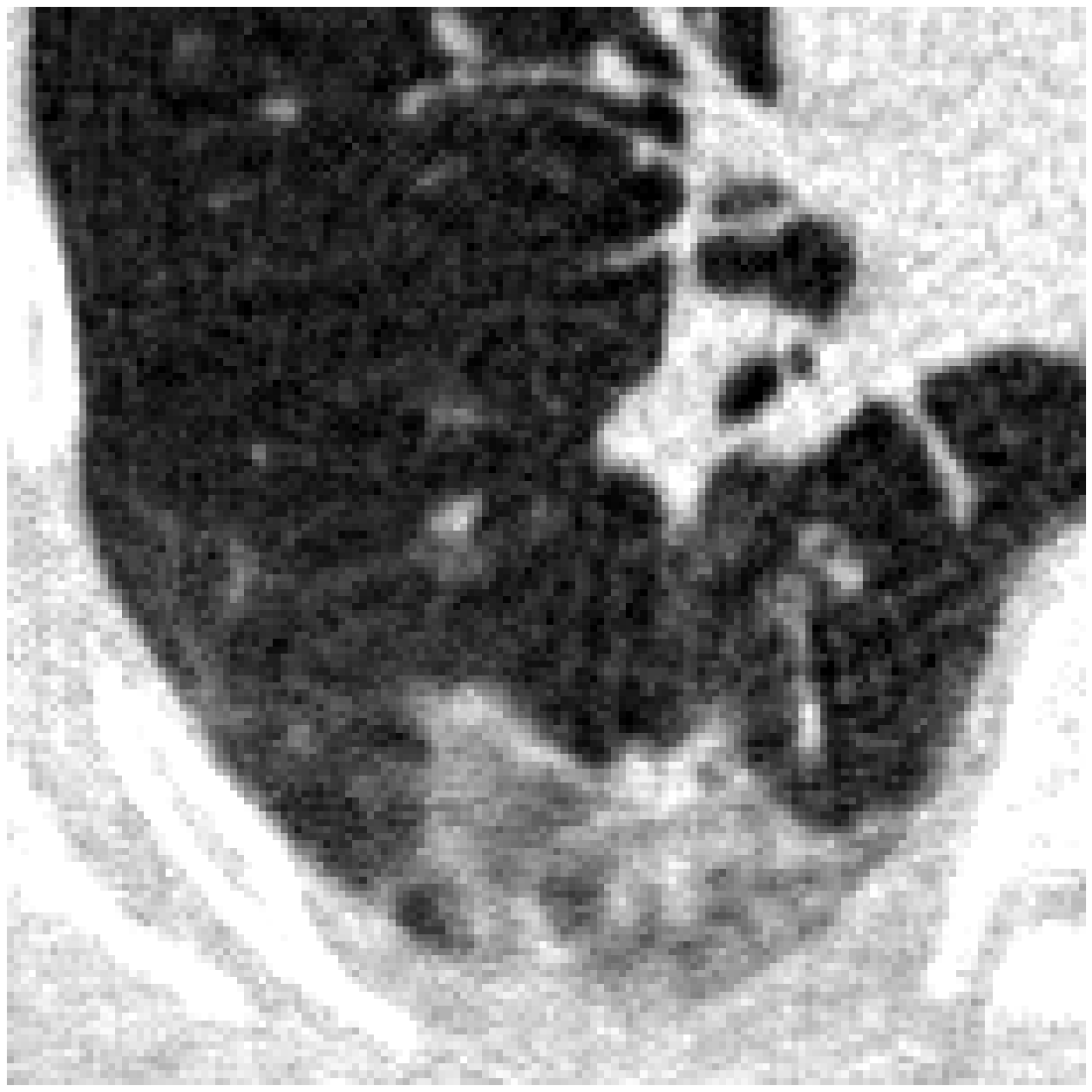}} 
\end{minipage}
\\
\begin{minipage}{0.02\linewidth}
  		\begin{turn}{90}
  		\footnotesize Diff. ROI
  		\end{turn}
\end{minipage}
\begin{minipage}{\x\linewidth}
  		\centering
  		\centerline{\includegraphics[width=\linewidth]{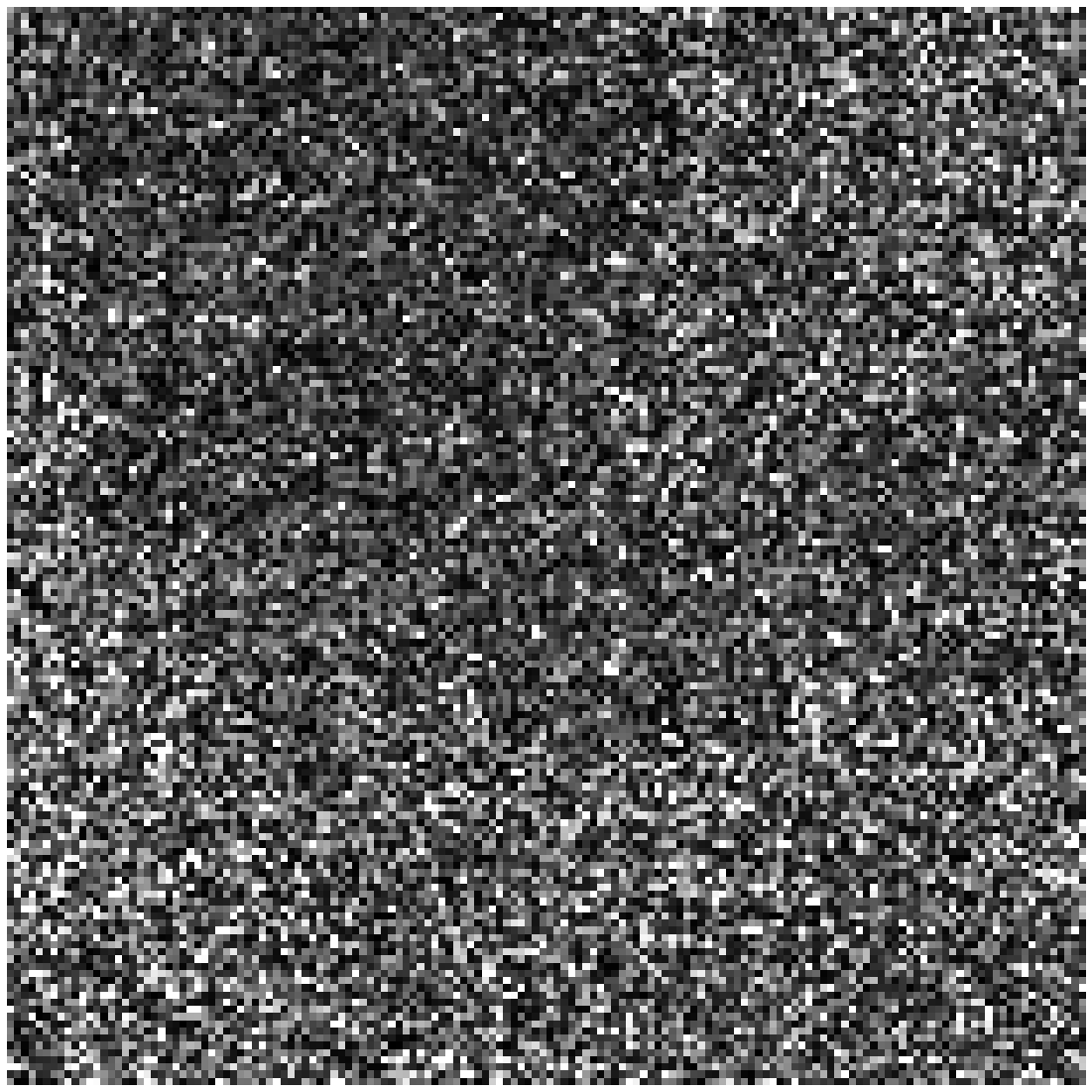}} 
\end{minipage}
\begin{minipage}{\x\linewidth}
  		\centering
  		\centerline{\includegraphics[width=\linewidth]{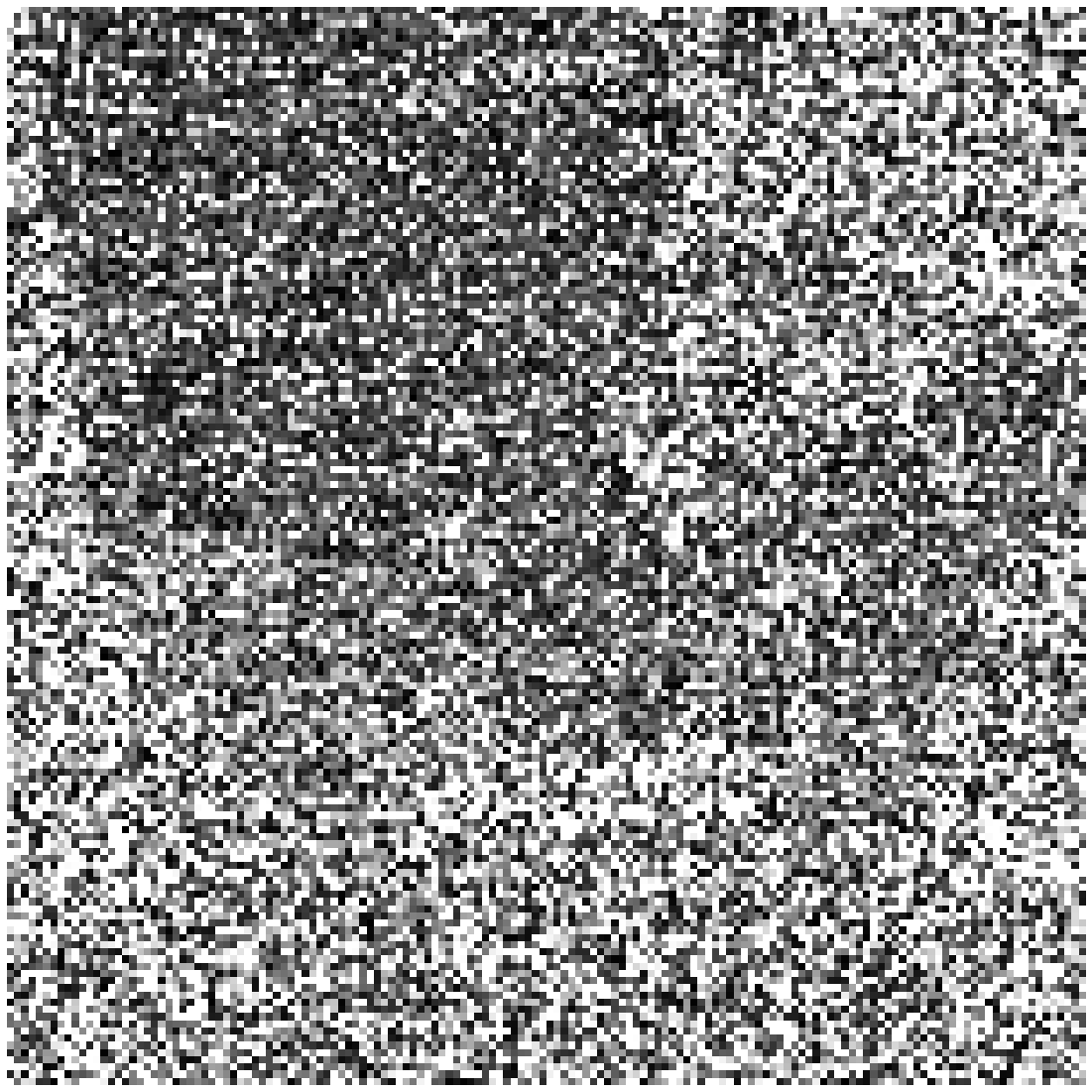}} 
\end{minipage}
\begin{minipage}{\x\linewidth}
  		\centering
  		\centerline{\includegraphics[width=\linewidth]{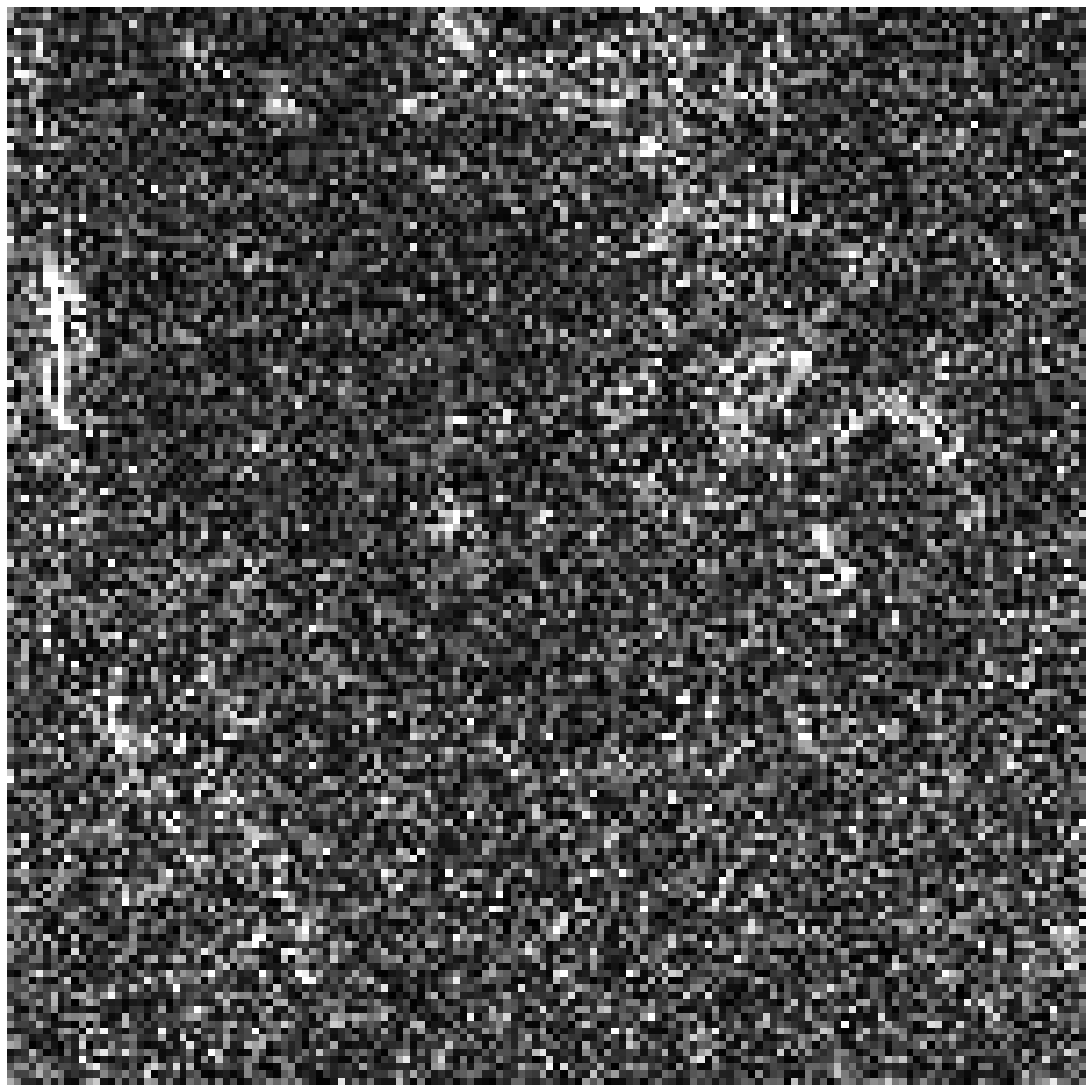}} 
\end{minipage}
\begin{minipage}{\x\linewidth}
  		\centering
  		\centerline{\includegraphics[width=\linewidth]{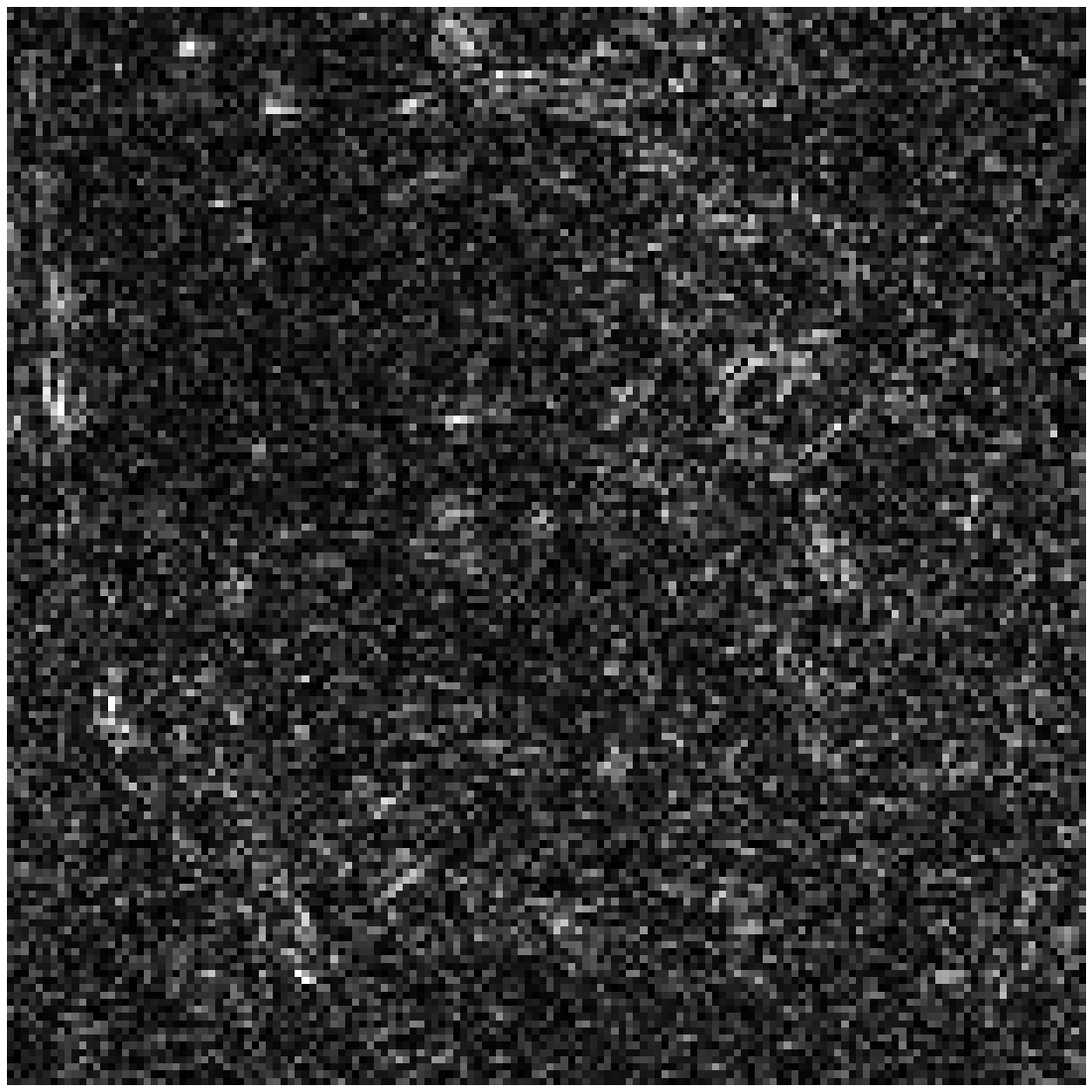}} 
\end{minipage}
\caption{FBP reconstructions corresponding to intermediate results in the preprocessing and sinogram-domain filtering stages. The respective equations for $\tilde{x}_{128}$, $\tilde{x}_{512}^{\text{int}}$, $\tilde{x}_{512}^{\mathcal{I}}$, and  $\tilde{x}_{512}^{\bm{\hat{U}_s}}$ can be found in Section \ref{section:methods} \associateed{\textbf{(Sino)} Respective sinogram used to generate the representative slice below. The 128-view sparse sinogram has been upsampled to a 512-view with missing 0's along the non-measured angles to match the size of the other sinograms for visualization purposes \textbf{(Diff. Sino)} The absolute difference to a 512-view ground truth sinogram} \textbf{(Recon)} Representative output slice in [-1024,150] HU window; \enquote{Ground Truth} is the 2048-view \correction{Ram-Lak} filtered FBP reconstruction $\tilde{x}_{2048}$; SSIM and PSNR metrics computed for the individual reconstruction with respect to the \enquote{Ground Truth} were added in the top-left corner (magenta); \textbf{(Diff. Recon)} Slice difference relative to Ground Truth in [0,400] HU window \textbf{(ROI)} Extracted patch containing the ROI, shown in a red box in Fig. \ref{fig:main-results-1}-\enquote{Ground Truth} \textbf{(Diff. ROI)} Patch difference to Ground Truth in [0,400] HU window.}
\label{fig:sinograminterpolation}
\end{figure}

\begin{table*}
\centering
\begin{tabular}{|c|c|c||c|c|c|c|}
\hline
  Metrics \textbackslash  Methods & FBP & WLS+L1 & FBPConvNet & DDNet & DRONE & WNet (ours) \\
\hline
PSNR &   $26.13 \pm 1.31$  & 
         $33.37 \pm 0.81$  & 
         $34.29 \pm 1.26$  & 
         $35.44 \pm 1.13$  &
         $36.27 \pm 1.28$  & 
         $\bm{37.10 \pm 1.35}$ \\
\hline
SSIM &  $0.6593 \pm 0.018$  &
        $0.9190 \pm 0.011$  & 
        $0.9192 \pm 0.011$  & 
        $0.9261 \pm 0.014$  & 
        $0.9530 \pm 0.016$  & 
        $\bm{0.9577 \pm 0.015}$ \\
\hline
\reviewerfour{Training duration} & $ - $ &
        $ -  $& 
        $ 40$h & 
        $ 31$h & 
        $ 55$h & 
        $ 52$h  \\
\hline
\reviewerfour{Reconstruction duration} &  $\approx 0.01$s  &
        $\approx 8.0$s  & 
        $\approx 0.5$s  & 
        $\approx 0.3$s  & 
        $\approx 73$s  & 
        $\approx 1.4$s \\
\hline
\# trainable parameters &  $0$  &
        $0$  & 
        $\approx 31$M  & 
        $\approx .5$M  & 
        $\approx 43$M  & 
        $\approx 17.5$M \\
\hline
\end{tabular}
\vspace{1em}
\caption{\textbf{(1st, 2nd row)} Quantitative metrics for our experiments (Average and Standard Deviation values over the test set)
\textbf{(3rd, 4th row)} Reconstruction duration and number of trainable parameters}
\label{table:results}
\end{table*}

\subsection{Comparison to state-of-the-art} \label{section:comparison}

We compare results obtained with methods from three different generations of reconstruction algorithms on our fixed test set. Fig. \ref{fig:main-results-1} contains the example slice reconstructions with each individual method while a corresponding zoomed-in region-of-interest view \correction{can be found in Fig. \ref{fig:main-results-2}. }The region-of-interest view is employed to shed more light on the localized performance of each algorithm. \correction{We also added Section A and Fig A1 in the \enquote{Supplementary Materials} showing results against FBP ground truths obtained with two other filters: Cosine and Shepp-Logan.}

The analytical reconstruction with Filtered Backprojection (FBP) was obtained from the simulated sparse-view sinogram $y_{128}$. Similarly, we ran 250 iterations of Weighted Least Squares with Total Variation regularization (WLS+TV) to obtain an iterative reconstruction from the same sparse-view set $y_{128}$. The regularization parameter was empirically set to $10^5$. 

Next, as FBPConvNet is one of the first networks which tackled the de-artifacting of SVCT, we reimplemented the network from \cite{fbpconvnet} in our framework and trained it for 100 \correctionsecond{epochs} on pairs of sparse-view FBP reconstructions (input) and full-view FBP reconstructions (label) following the procedures described in \cite{fbpconvnet}. 

The DD-Net was another network from the first generation of encoder-decoder models for SVCT that we investigated. We trained DD-Net for 50 \correctionsecond{epochs} on the same input-label pairs as FBPConvNet while employing the original architecture, code, and setup instructions from \cite{ddnet}. 

Finally, we compared our results to the results obtained by the most recent type of architecture, DRONE \cite{dronenet}. The architecture of DRONE consists of four networks which perform sequentially the tasks of projection-domain and image-domain denoising on both images and residual images (see Fig. 1 from \cite{dronenet} for more details about the architecture of DRONE). All four networks, similar to one another, implement the UNet architecture, each one of them containing approximately 10.9M parameters, for a gross total of 43.8M parameters. Every network is trained sequentially, its input representing the generated output \correction{\st{with}of} the previous one. 

The DRONE model employs custom forward and backward tomographic operators from the ASTRA Toolbox \cite{astra} both inside the network as well as for the final iterative denoising step, based on a half-arc two-dimensional fan-beam geometry. We adapted the original implementation of DRONE \cite{dronenet} with the forward-backward operators pair of \textit{elsa}-based operators and we changed the half-arc fan-beam geometry with the full-arc parallel-beam geometry employed by WNet, to ensure the compatibility of the results. We then followed the training procedures listed in \cite{dronenet} and performed in the end the required iterative TV denoising step. As for the denoising hyperparameters, we empirically set $\lambda_1,\alpha_1,\alpha_2,\beta_1$ and $\beta_2$ to $0.01,0.6,0.6,0.001$ and $0.001$ as the original values ($0.05,0.14,0.6,0.003$ and $0.003$) used by \cite{dronenet} did not fit our dataset.

To quantitatively evaluate the quality of the different results, we use the Peak Signal-to-Noise Ratio (PSNR) \cite{psnr} and the Structural Similarity Index Measure (SSIM) \cite{ssim} implementations from the \textit{scikit} library.

We listed the \correctionsecond{\st{average and the}mean and} standard deviation values for PNSR and SSIM over the test set in Table \ref{table:results} where we highlighted in boldface font the best values. We also added rows for the mean training and reconstruction durations and the number of trainable parameters, to support the method comparison.

\subsection{Module Analysis} \label{section:imagedenoising}

During the training procedure we noticed several interesting behaviors of the different submodules of WNet. Therefore, we performed an investigation into each individual WNet module to highlight aspects of the trained operators $\bm{\hat{U}_s}$, $\bm{\hat{U}_i}$ and $\bm{\hat{W}}$. 

Fig. \ref{fig:sinograminterpolation} contains the corresponding reconstructions \reviewertwo{(FBP with Ram-Lak filter)} to each individual sinogram preprocessing steps in the WNet pipeline shown in Fig. \ref{fig:wnet}b. To highlight the role of each step both in general and in local feature recovery we added a zoomed-in view into one of the regions of interest and the corresponding sinogram and sinogram difference images. \correction{The \enquote{Supplementary Materials} contain Section B and corresponding Fig. A2 discussing the difference between bilinear and geometry-aware interpolation for sinogram upsampling}

We illustrate the learned FBP filter from ReM and we plot it against three other widely used kernels in conventional CT applications. Reconstructions with matching colors to the corresponding filters (Fig. \ref{fig:finetunedfilter}a) used to generate them are shown \correction{\st{besides the plots}} in Fig. \ref{fig:finetunedfilter}b.

Last but not least, in Fig. \ref{fig:imagedenoising} we take a look at the performance of a trained $U_i$ given two different inputs, one obtained by an \reviewertwo{FBP with the Ram-Lak filter (SdM+Ram-Lak-IdM) and one generated by an FBP operation with a learned filter (SdM+ReM+IdM). The two models were trained separately, one with the fixed FBP kernel and one with the kernel optimization step. }We compare the results to a full-view FBP reconstruction with a Ram-Lak filter.

\FloatBarrier
\subsection{Generalization} \label{section:generalization}

One key aspect of DL based methods is their data-driven nature and high dependency on a training set representative of the whole modeled \correction{\st{population}distribution} of inputs. Having \correction{an insufficiently large} training set which does not contain all the representative information for the whole expected input domain, leads to poor performance of the neural network on samples with new features not represented in the training set. Therefore, understanding how a neural network \enquote{learns} to extract features \correction{\st{out of it}} becomes an important task in order to get insights into the training process and the generalizability performance of the trained model on new data.

We perform a short investigation into the performance of the trained WNet and DRONE on three out-of-distribution slices, \correction{which were part of neither the training nor the validation or the test set}.

Fig. \ref{fig:generalization} contains slices extracted from a CT scan belonging to an ex-vivo clinical study of deceased patients at Klinikum Rechts der Isar. The acquisition was performed with a high dose CT Protocol with an exposure time of 497ms at a nominal tube current of $512$mA and $120$ kV, and the images were reconstructed with a sharp reconstruction kernel in the lung window setting. A Philips iCT 256 scanner was used with a resolution of $(768,768)$ pixels at a voxel size of $(0.4362,0.4362,0.9)$ mm. Images were downsampled to $(512,512)$ pixels.

\correction{We attached generalization comparisons of the same resulting images to Ram-Lak filtered ground truths in the \enquote{Supplementary Materials} Section C and corresponding Fig. A3.}

\section{Discussion} \label{section:discussion}

\fboxsep=0pt
\fboxrule=2pt
\definecolor{green_custom}{rgb}{0.216, 0.494, 0.133}
\begin{figure*}[!t]
\begin{varwidth}{0.49\linewidth}
  		\centering
  		\centerline{\includegraphics[width=\linewidth]{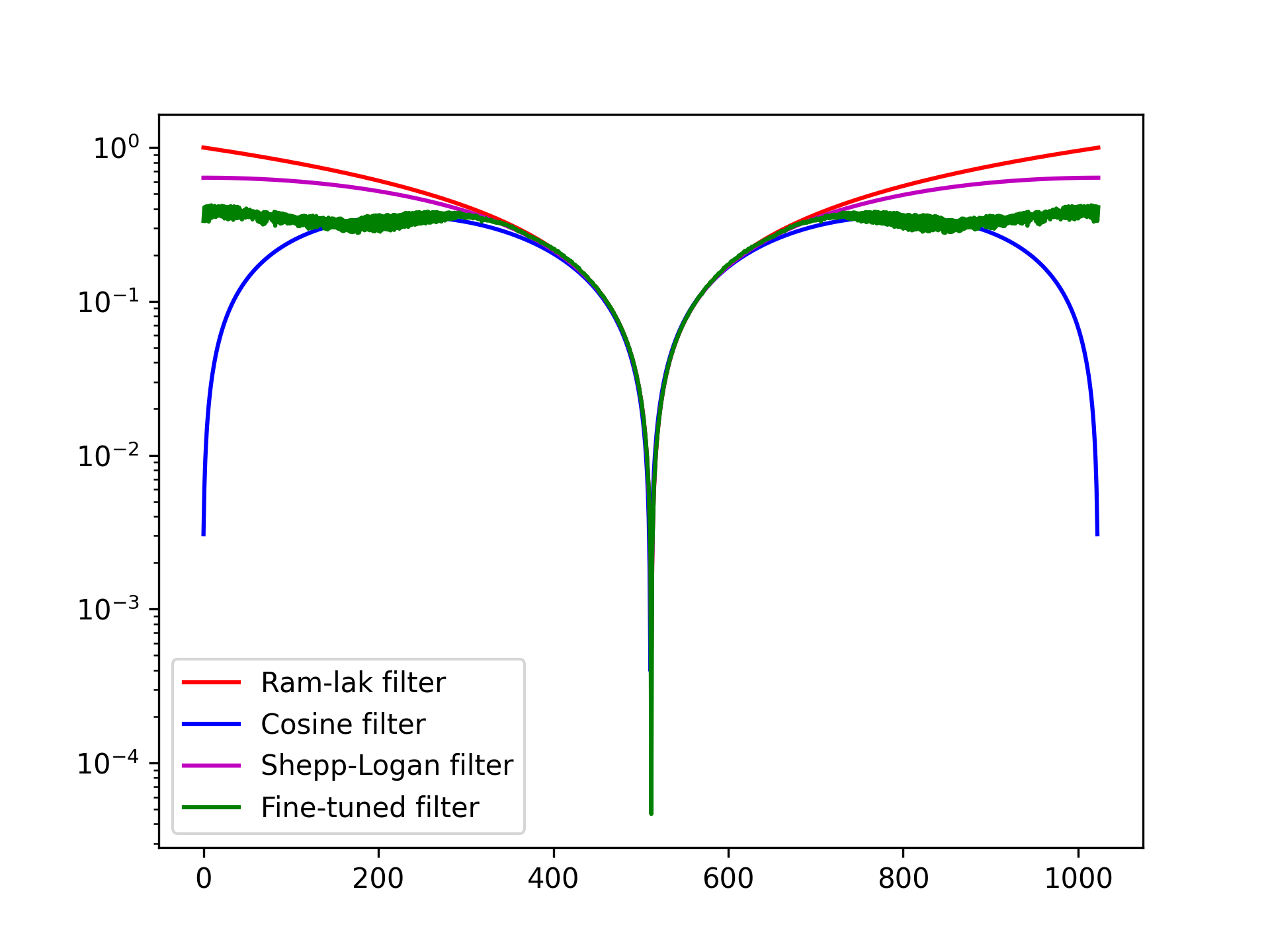}} 
\end{varwidth}
\begin{varwidth}{0.6\linewidth}
  		\centering
    		\begin{varwidth}{0.3\linewidth}
      		\centering
      		\hfill
      		\centerline{\footnotesize Recon}\medskip
      		\hfill
            \end{varwidth}
            \begin{varwidth}{0.3\linewidth}
      		\centering
      		\hfill
      		\centerline{\footnotesize Diff. Recon}\medskip
      		\hfill
            \end{varwidth}
            \begin{varwidth}{0.3\linewidth}
      		\centering
      		\hfill
    		\centerline{\footnotesize \reviewerfoursecond{Ground Truth}}\medskip
    		\hfill
            \end{varwidth}
            \vspace{3mm}
            \\
            \begin{varwidth}{0.05\linewidth}
      			\begin{turn}{90}
              		\footnotesize Ram-lak filter
          		\end{turn}
            \end{varwidth}
  		    \fcolorbox{red}{white}{
      		\begin{varwidth}{0.28\linewidth}
      		\centering
      		\centerline{\includegraphics[width=\linewidth]{figures/wnet_image_recon_512view_filter_patch_10.png}} 
            \end{varwidth}
            \begin{varwidth}{0.28\linewidth}
      		\centering
      		\centerline{\includegraphics[width=\linewidth]{figures/wnet_image_recon_512view_filter_difference_patch_10.png}} 
            \end{varwidth}
            \begin{varwidth}{0.28\linewidth}
      		\centering
      		\centerline{\includegraphics[width=\linewidth]{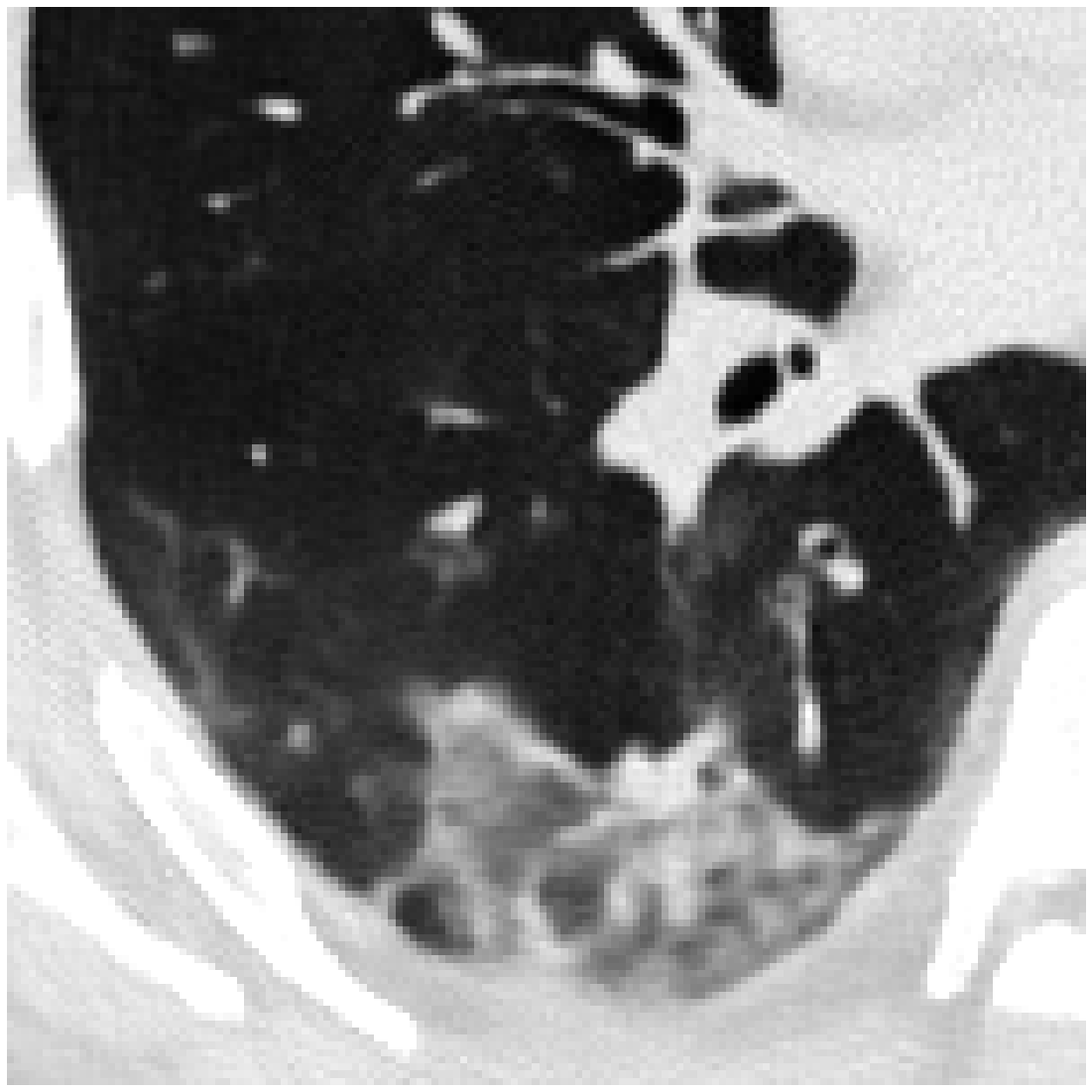}} 
            \end{varwidth}
            }
            \\
            \begin{varwidth}{0.05\linewidth}
      			\begin{turn}{90}
              		\footnotesize Fine-tuned filter
          		\end{turn}
            \end{varwidth}
            \fcolorbox{green_custom}{white}{
            \begin{varwidth}{0.28\linewidth}
      		\centering
      		\centerline{\includegraphics[width=\linewidth]{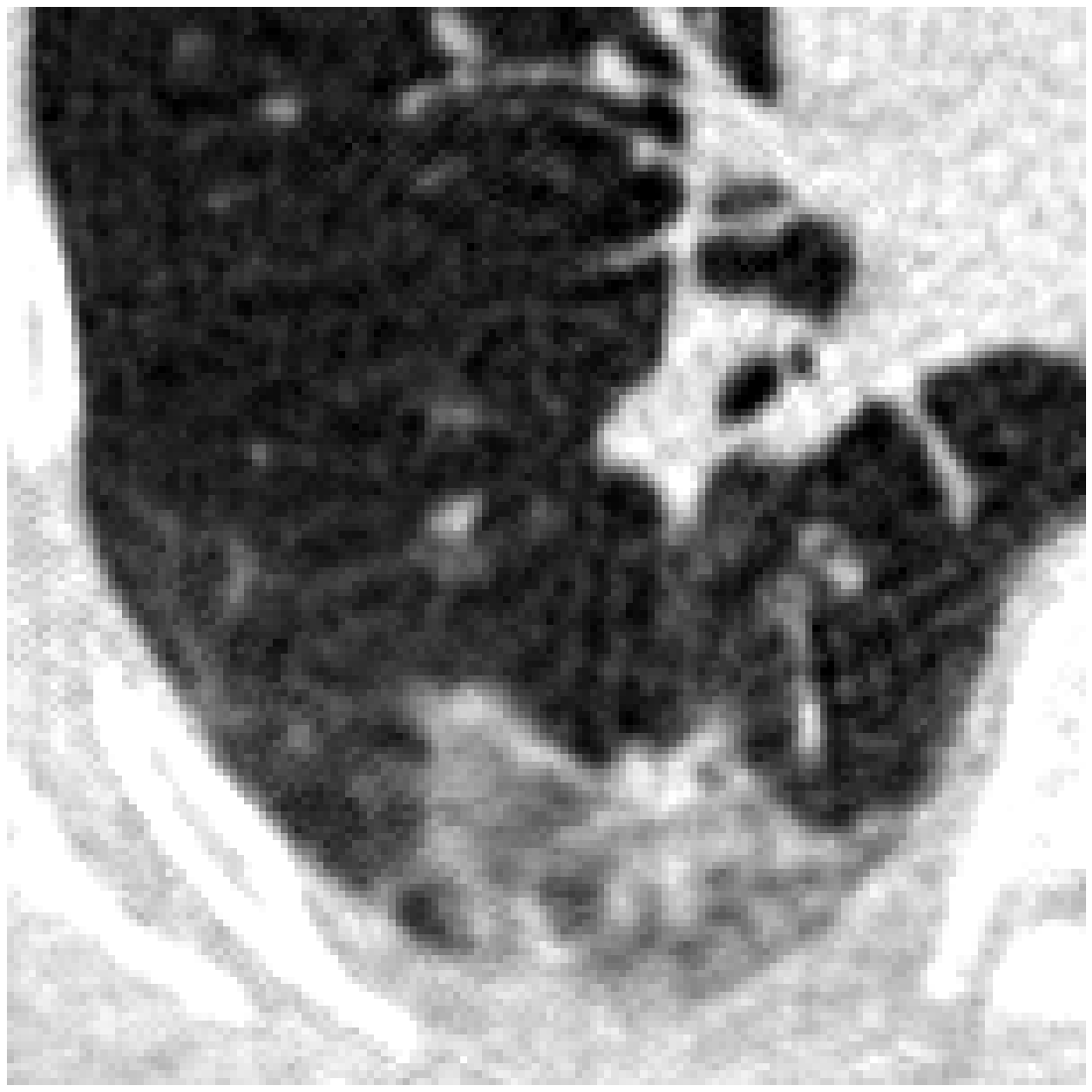}} 
            \end{varwidth}
            \begin{varwidth}{0.28\linewidth}
      		\centering
      		\centerline{\includegraphics[width=\linewidth]{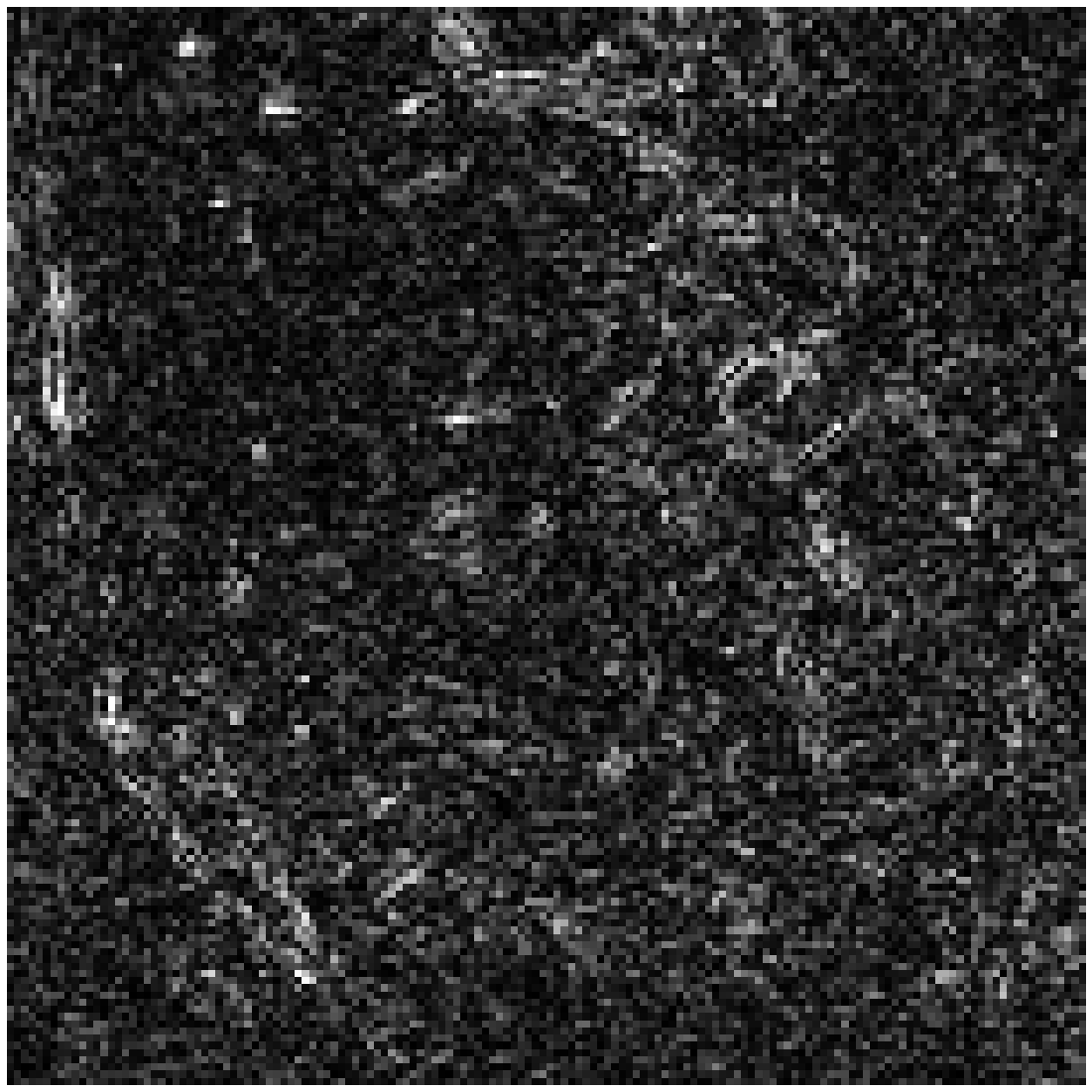}} 
            \end{varwidth}
            \begin{varwidth}{0.28\linewidth}
      		\centering
      		\centerline{\includegraphics[width=\linewidth]{figures/wnet_image_recon_512view_full_patch_10.png}} 
            \end{varwidth}
            }
\end{varwidth}
\vspace{1mm}
\\
\begin{varwidth}{0.49\linewidth}
  		\centering
  		\hfill
  		\centerline{(a) Filter outlines}\medskip
  		\hfill
\end{varwidth}
\begin{varwidth}{0.49\linewidth}
  		\centering
  		\hfill
  		\centerline{(b) Reconstructions}\medskip
  		\hfill
\end{varwidth}
\caption{\textbf{(a) Filter Outlines} The fine-tuned kernel plotted against other widely used FBP kernels (Ram-Lak, Cosine, Shepp-Logan). Details about the training of the kernel can be found in Section \ref{section:methods,kernel} and \ref{section:initialization}. The x-axis values represent detector pixel locations. \textbf{(b) Reconstructions,} Reconstruction, difference to ground truth and ground truth color-coded to the corresponding kernel plot color in \enquote{Filter outlines}. The denoised $y_K^{\bm{\hat{U}_s}}$ sinogram was used in both cases to generate the reconstructed patch via Ram-Lak filtered FBP. $\bm{\hat{U}_s}$ is the optimal SdM network obtained from eq. \ref{best-us}. The display window for the generated reconstructions is [-1024,150] HU and for the difference image [0,400] HU. The ground truth image is the 2048-view \correction{Ram-Lak filtered} FBP reconstruction $\tilde{x}_{2048}$.}
\label{fig:finetunedfilter}
\end{figure*}

\fboxsep=0pt
\fboxrule=1pt
\renewcommand{\x}{0.19}
\begin{figure*}[!h]
\begin{minipage}{0.01\linewidth}
        \vfill
  		\centerline{\begin{turn}{90}
  		\smallskip
  		\end{turn}}
\end{minipage}
\begin{minipage}{\x\linewidth}
  		\centering
  		\footnotesize $\tilde{x}_{512}^{\bm{\hat{U}_s}}$
\end{minipage}
\begin{minipage}{\x\linewidth}
  		\centering
  		\footnotesize $\tilde{x}_{512}^{\bm{\hat{W}}}$
\end{minipage}
\begin{minipage}{\x\linewidth}
  		\centering
  		\footnotesize (SdM+Ram-Lak+IdM only)
\end{minipage}
\begin{minipage}{\x\linewidth}
  		\centering
  		\footnotesize (SdM+ReM+IdM) $\bm{\hat{x}}$ 
\end{minipage}
\begin{minipage}{\x\linewidth}
	\centering
	\footnotesize \reviewerfoursecond{Ground Truth}
\end{minipage}
\vspace{1em}
\\
\begin{minipage}{0.02\linewidth}
  		\begin{turn}{90}
  		\footnotesize Recon
  		\end{turn}
		\vfill
	\end{minipage}
\begin{minipage}{\x\linewidth}
  		\centering
  		\centerline{\includegraphics[width=\linewidth]{figures/revision_results/wnet_image_recon_512view_filter_10.png}} 
\end{minipage}
\begin{minipage}{\x\linewidth}
  		\centering
  		\centerline{\includegraphics[width=\linewidth]{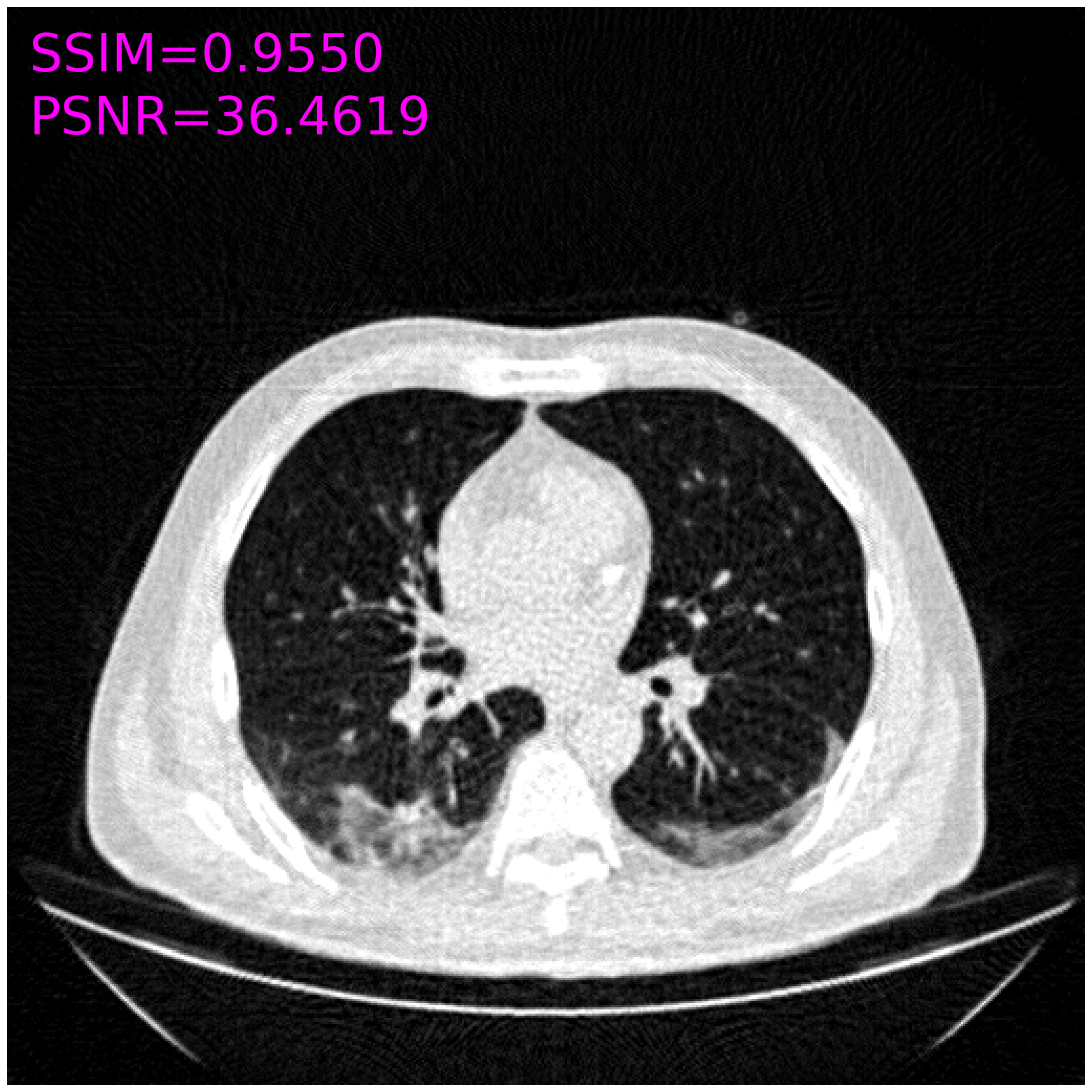}} 
\end{minipage}
\begin{minipage}{\x\linewidth}
  		\centering
  		\centerline{\includegraphics[width=\linewidth]{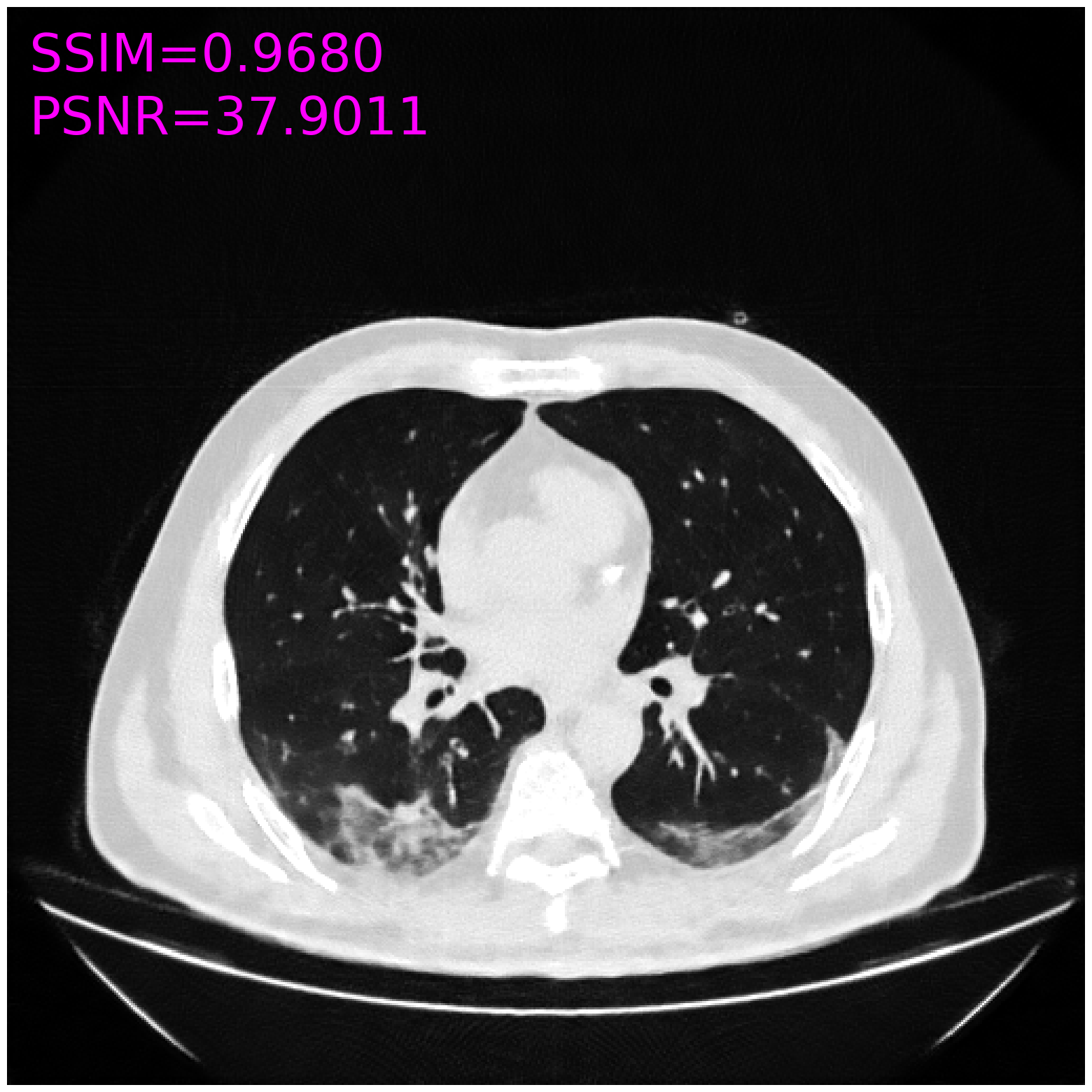}} 
\end{minipage}
\begin{minipage}{\x\linewidth}
  		\centering
  		\centerline{\includegraphics[width=\linewidth]{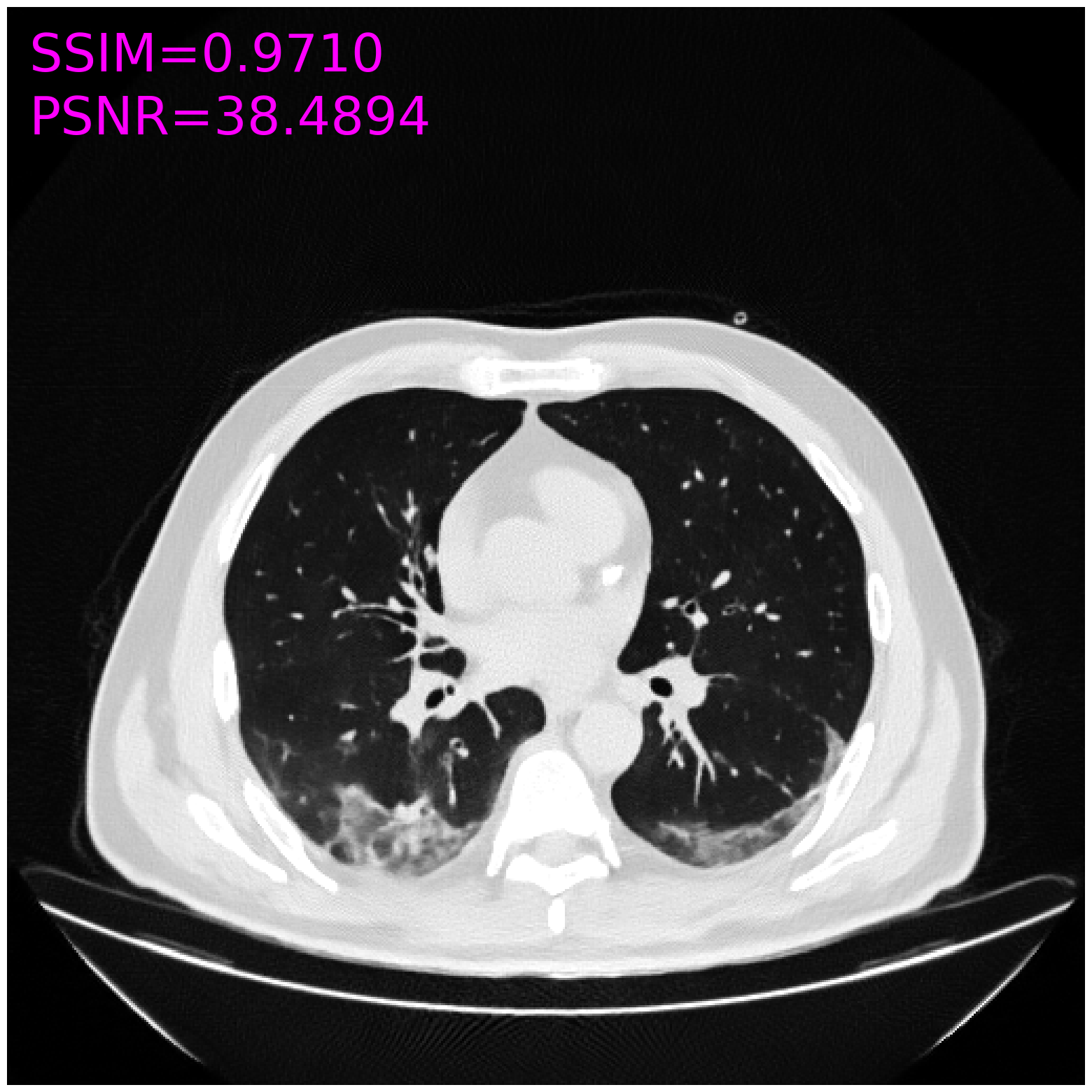}} 
\end{minipage}
\begin{minipage}{\x\linewidth}
	\centering
	\centerline{\includegraphics[width=\linewidth]{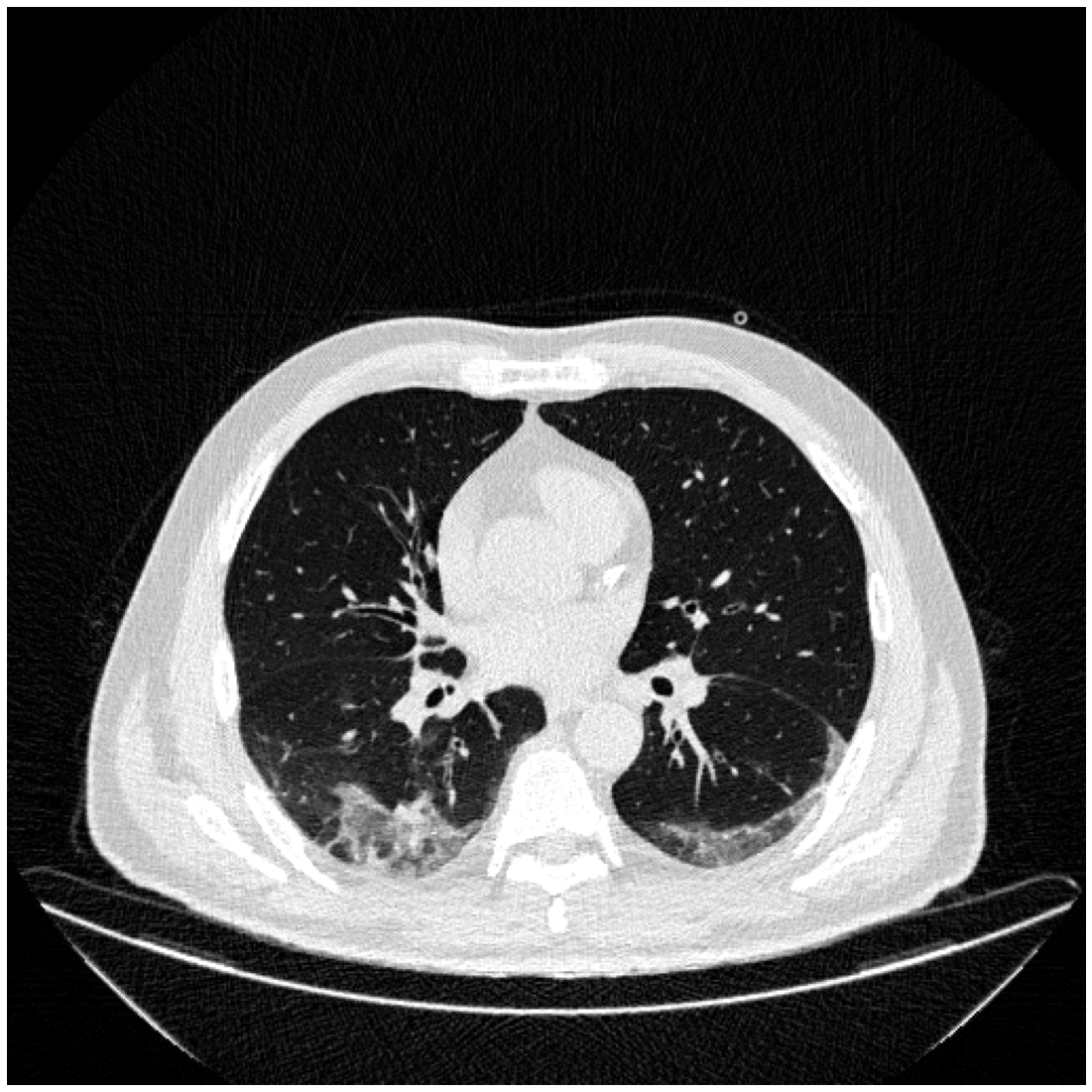}} 
\end{minipage}
\\
\begin{minipage}{0.02\linewidth}
		\begin{turn}{90}
  		\footnotesize Diff. Recon
  		\end{turn}
		\vfill
	\end{minipage}
\begin{minipage}{\x\linewidth}
  		\centering
  		\centerline{\includegraphics[width=\linewidth]{figures/wnet_image_recon_512view_filter_difference_10.png}} 
\end{minipage}
\begin{minipage}{\x\linewidth}
  		\centering
  		\centerline{\includegraphics[width=\linewidth]{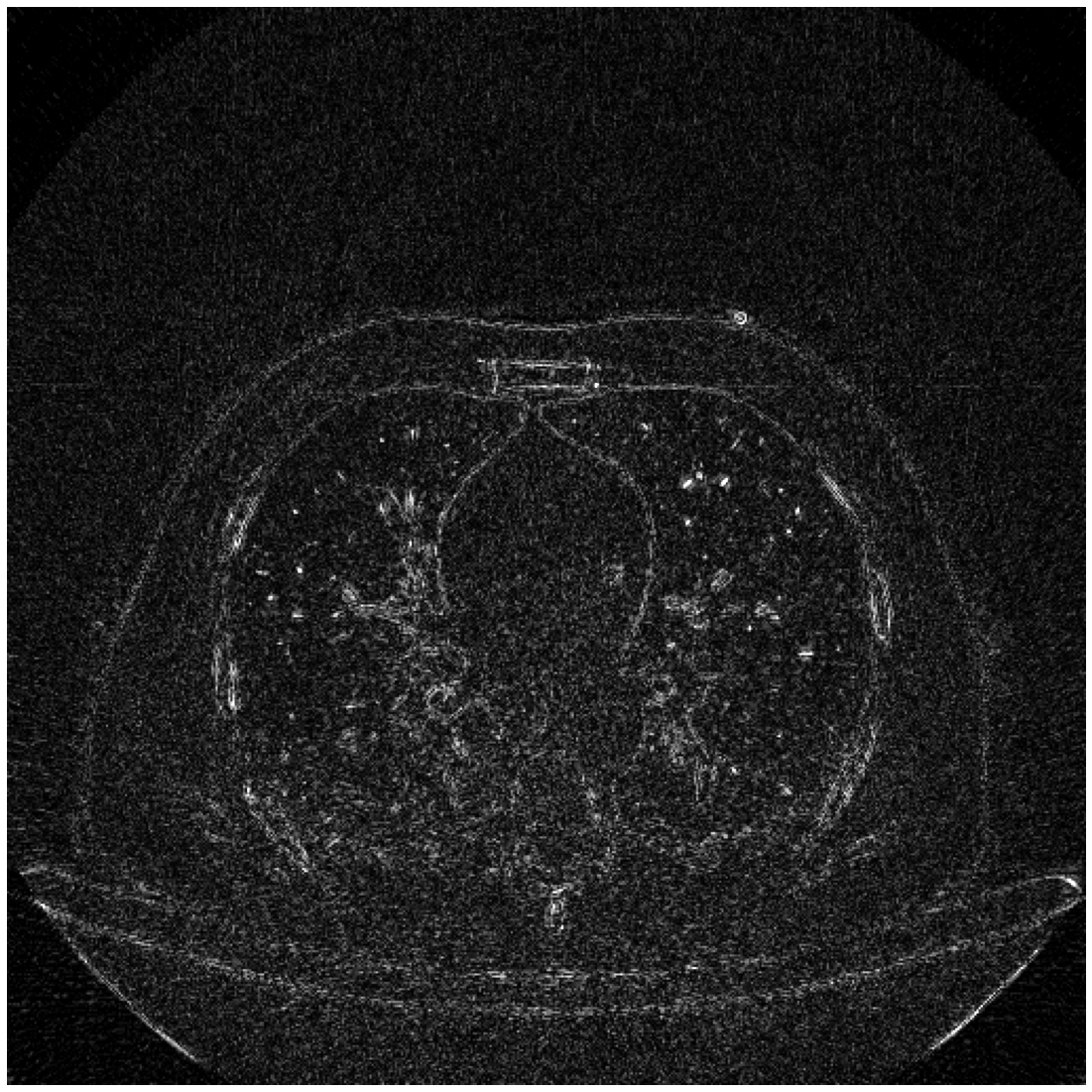}} 
\end{minipage}
\begin{minipage}{\x\linewidth}
  		\centering
  		\centerline{\includegraphics[width=\linewidth]{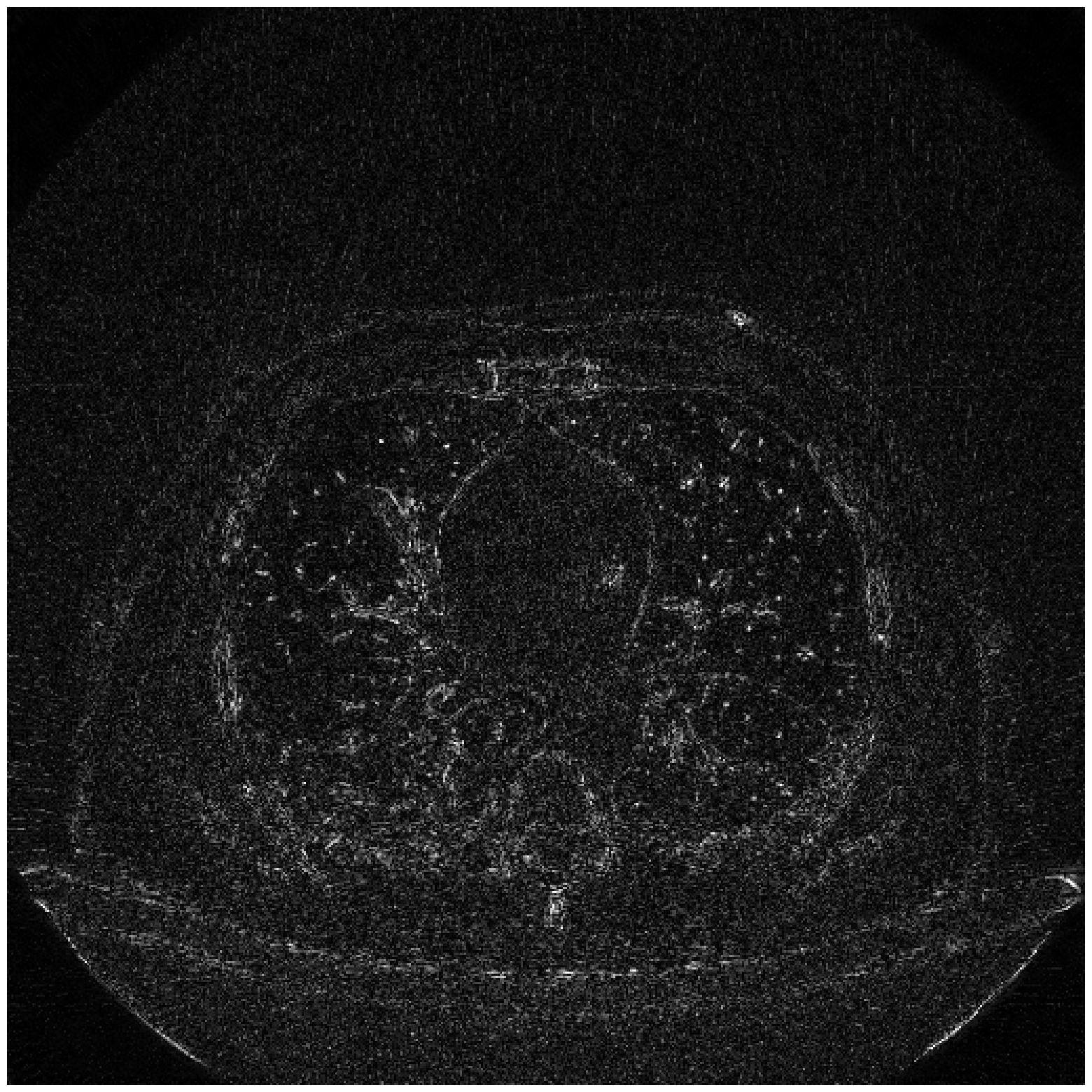}} 
\end{minipage}
\begin{minipage}{\x\linewidth}
  		\centering
  		\centerline{\includegraphics[width=\linewidth]{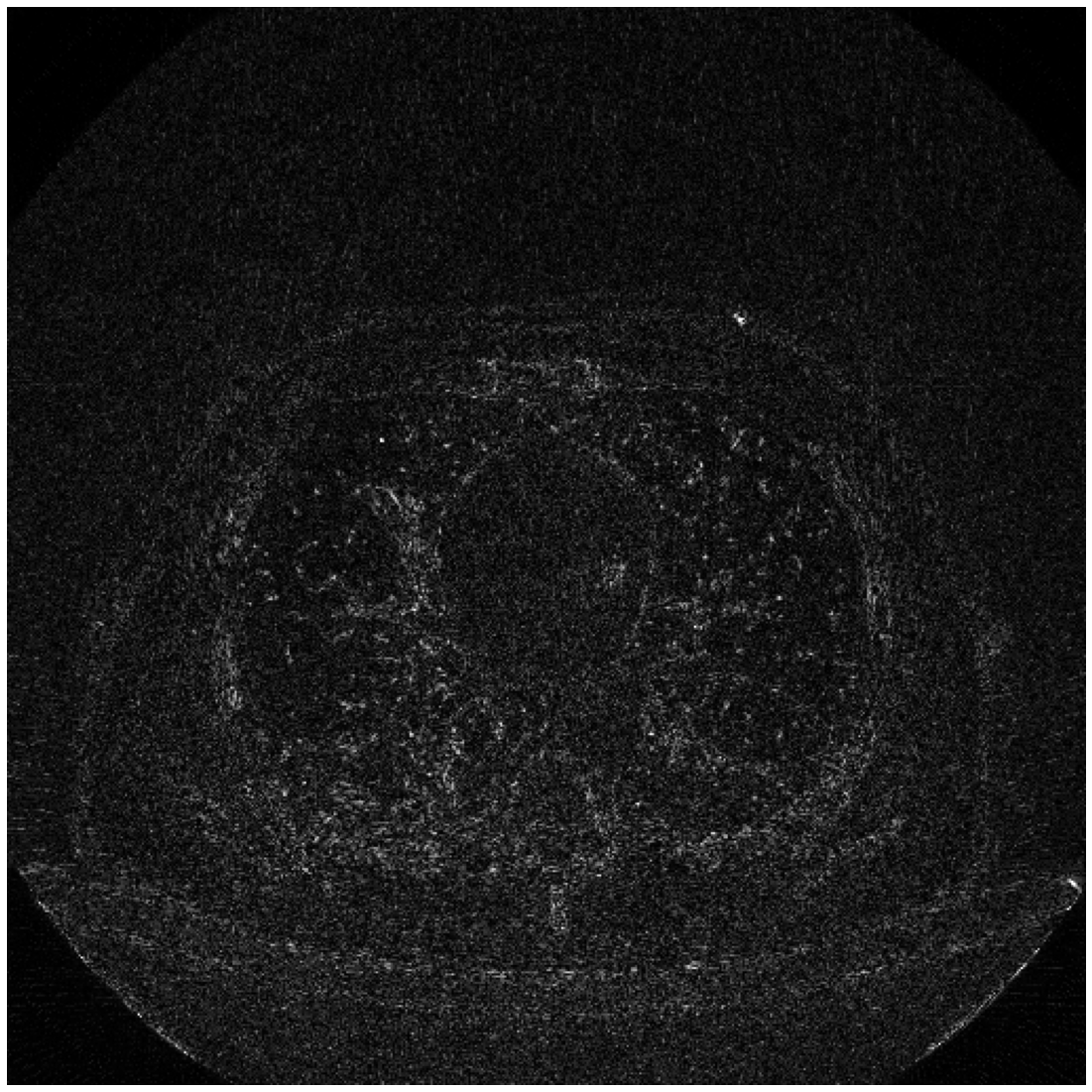}} 
\end{minipage}
\vspace{1em}
\\
\begin{minipage}{0.02\linewidth}
		\begin{turn}{90}
  		\footnotesize ROI
  		\end{turn}
		\vfill
	\end{minipage}
\begin{minipage}{\x\linewidth}
  		\centering
  		\centerline{\includegraphics[width=\linewidth]{figures/wnet_image_recon_512view_filter_patch_10.png}} 
\end{minipage}
\begin{minipage}{\x\linewidth}
  		\centering
  		\centerline{\includegraphics[width=\linewidth]{figures/wnet_image_recon_512view_finetu_patch_10.png}} 
\end{minipage}
\begin{minipage}{\x\linewidth}
  		\centering
  		\centerline{\includegraphics[width=\linewidth]{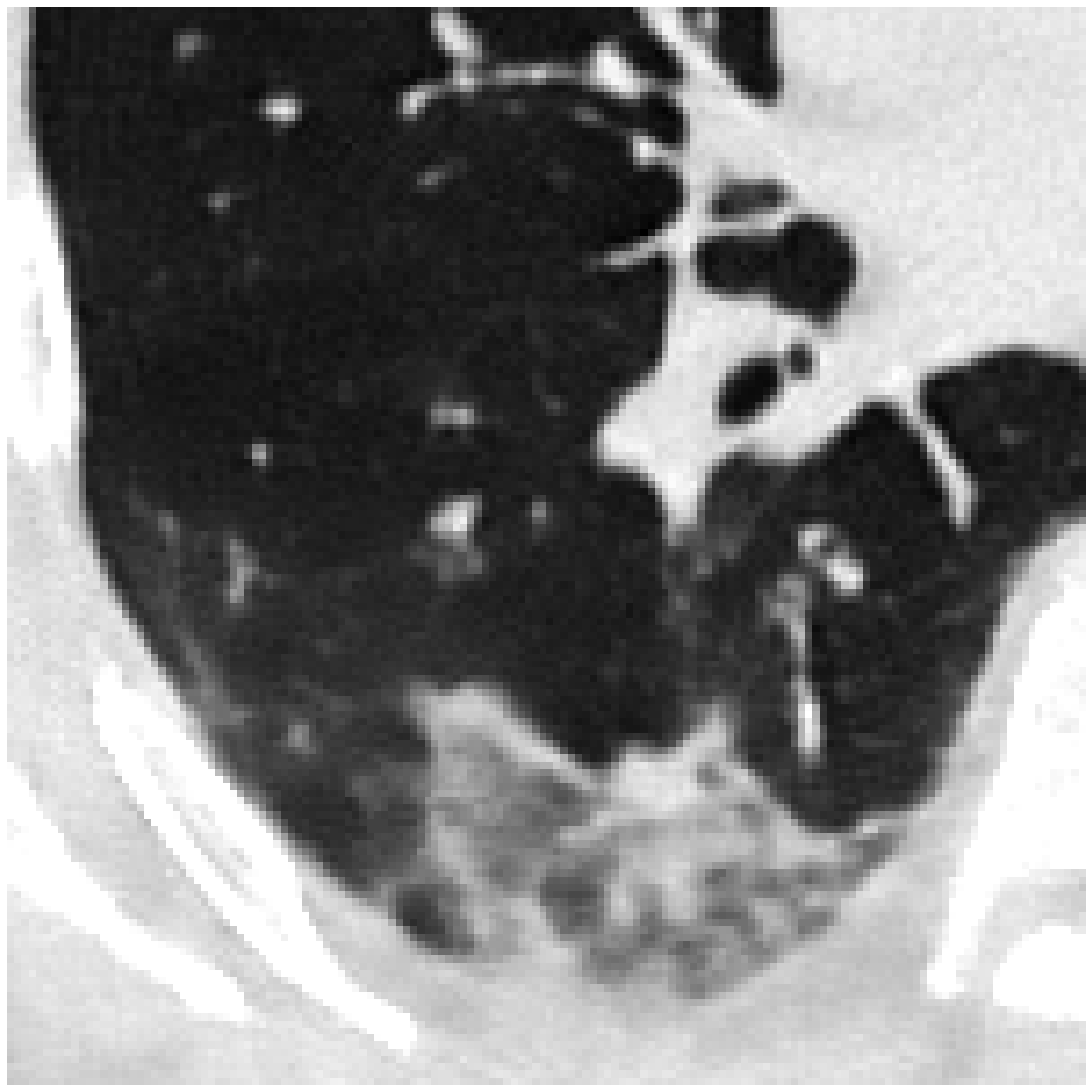}} 
\end{minipage}
\begin{minipage}{\x\linewidth}
  		\centering
  		\centerline{\includegraphics[width=\linewidth]{figures/wnet_image_recon_512view_full_patch_10.png}} 
\end{minipage}
\begin{minipage}{\x\linewidth}
 		\centering
 		\centerline{\includegraphics[width=\linewidth]{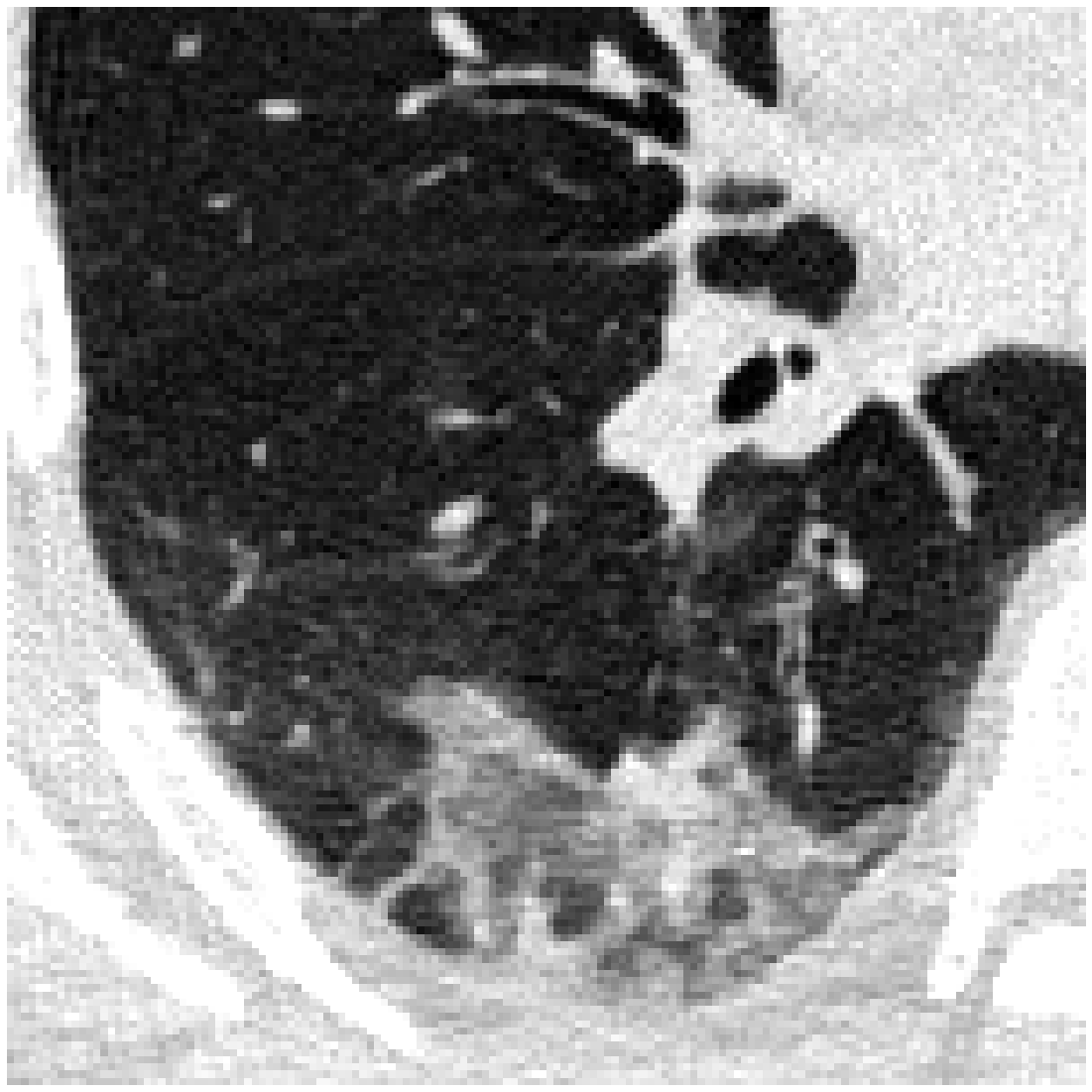}} 
\end{minipage}
\\
\begin{minipage}{0.02\linewidth}
		\begin{turn}{90}
  		\footnotesize Diff. ROI
  		\end{turn}
		\vfill
	\end{minipage}
\begin{minipage}{\x\linewidth}
  		\centering
  		\centerline{\includegraphics[width=\linewidth]{figures/wnet_image_recon_512view_filter_difference_patch_10.png}} 
\end{minipage}
\begin{minipage}{\x\linewidth}
  		\centering
  		\centerline{\includegraphics[width=\linewidth]{figures/wnet_image_recon_512view_finetu_difference_patch_10.png}} 
\end{minipage}
\begin{minipage}{\x\linewidth}
  		\centering
  		\centerline{\includegraphics[width=\linewidth]{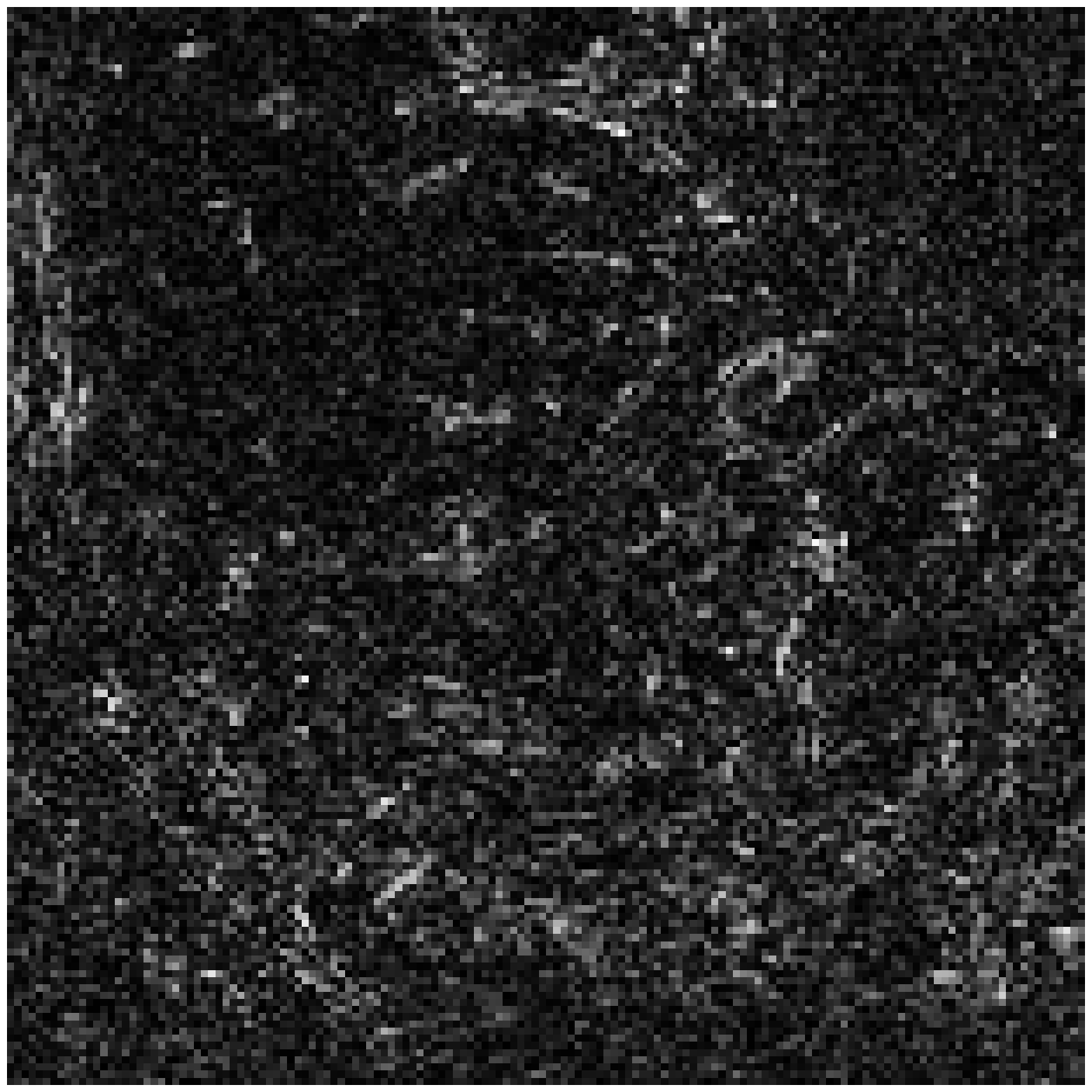}} 
\end{minipage}
\begin{minipage}{\x\linewidth}
  		\centering
  		\centerline{\includegraphics[width=\linewidth]{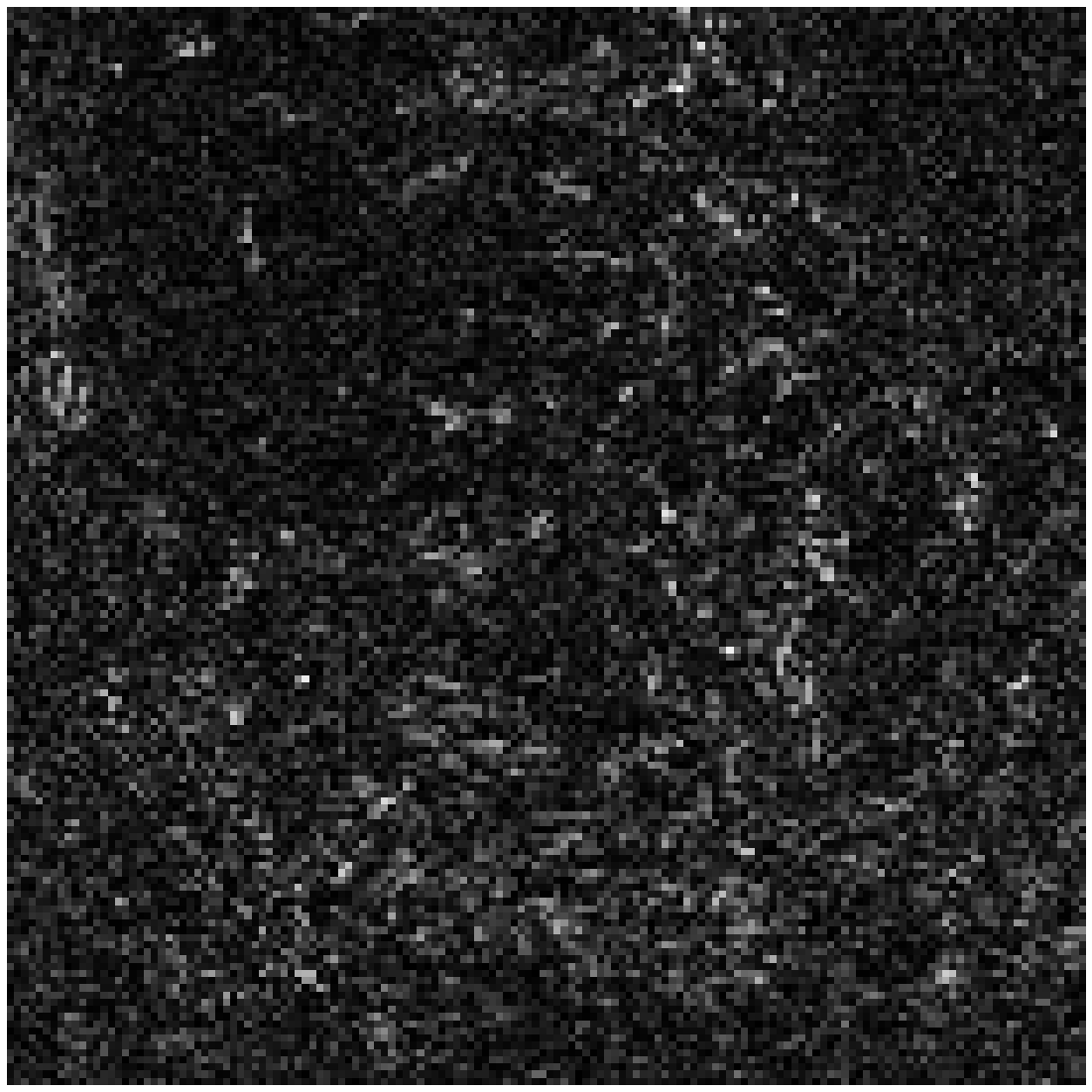}} 
\end{minipage}
\caption{Filter fine-tuning and image-domain denoising results. Details about the architecture of WNet can be found in Fig. \ref{fig:wnet}. Equations for $\tilde{x}_{512}^{\bm{\hat{U}_s}}$ and $\tilde{x}_{512}^{\bm{\hat{W}}}$ can be found in Section \ref{section:methods}. $\tilde{x}_{512}^{\bm{\hat{W}}}$ is an FBP reconstruction using the optimal filter $\bm{\hat{W}}$ \textbf{(Recon)} Representative output slice for the corresponding step in the WNet pipeline in [-1024,150] HU window; \enquote{Ground Truth} is the 2048-view \correction{Ram-Lak filtered} FBP reconstruction $\tilde{x}_{2048}$; The SSIM and PSNR metrics computed for the individual reconstruction with respect to the \enquote{Ground Truth} were added in the top-left corner (magenta); \textbf{(Diff. Recon)} Slice difference relative to ground truth in [0,400] HU window \textbf{(ROI)} Extracted patch containing the ROI in lung window \textbf{(Diff. ROI)} Patch difference relative to ground truth in [0,400] HU window.}
\label{fig:imagedenoising}
\end{figure*}

\subsection{Comparison to state-of-the-art} \label{section:comparison}

The results in Fig. \ref{fig:main-results-1} and \ref{fig:main-results-2} show that outputs of convolutional-based networks tend to share similarities in feature reconstruction performance like the over-smoothing of edges and loss of small and very-small details. The contrast of the image produced by the DD-Net is higher than the FBPConvNet result, with smaller details visible, which is also reflected in the increase in PSNR by $1$dB and in SSIM by $1\%$, with the major difference being that DD-Net requires $54\times$ less parameters than FBPConvNet and is $10\times$ faster, proving the potential of residual connections in such scenarios. However, DD-Net images are still plagued by low sharpness and the presence of oversmooth regions. 

The DRONE algorithm seems to have taken care of various drawbacks of previous convolutional-based methods, like oversmooth regions, blurry edges and loss of detail at the small and very-small scale, however, at the cost of network size and computational \correction{\st{burden}complexity}. For the reconstruction of the slice in Fig. \ref{fig:main-results-1} the DRONE pipeline requires around $73$s to complete one slice \footnote{\correction{The durations mentioned in this sections correspond to the average durations in Table \ref{table:results}}}. WNet needs less than $2$s to perform the corresponding inference step (on the same hardware in similar conditions). In a clinical setting, where time is of high importance, the DRONE network would require about $7$h to reconstruct a $300$ slice Chest CT volume, while our method would only need $\approx 6$min at a comparable image quality. \reviewerfour{The difference in visual quality of the reconstructions is mostly influenced by the display window recommended by radiologists for chest CT. As both DRONE and WNet are data-driven algorithms, their weight initialization plays a role in the final output of the network. As a consequence, some tiny details in the denoised reconstructions having values at the extremes of the display window spectrum, can appear to be missing from the final result. The quantitative metrics shown in Figs. \ref{fig:main-results-1}, \ref{fig:sinograminterpolation}, and \ref{fig:imagedenoising}, however, are a better measure of image quality and are not influenced by the windowing procedure.}

While the DRONE and WNet results are comparable in terms of contrast, resolution and overall image quality, WNet generates the image with less than half the number of parameters in DRONE and most notably at a fraction of the reconstruction time (see Table \ref{table:results}), which \correction{\st{is comparable to} lies in the same order of magnitude with} the inference times of other methods like FBPConvNet or DD-Net. Most notably, WNet does not require any hyperparameter tuning for the iterative denoising or reconstruction process like WLS+L1 or DRONE do, thus making it more efficient for similar use-cases where parameter tweaking might be necessary for a DRONE-like pipeline.

Last but not least, we improved on the factor of 3 increase in interpolated sinogram resolution mentioned in \cite{dronenet} \associateed{(which is limited by the nearest neighbor interpolation method employed by the authors)} by showing that sinogram upsampling by a factor of 4 (from $128$-view to $512$-view sinograms) is indeed feasible in the context of SVCT.

\subsection{Network Interpretability} \label{section:interpretability}
Progress in the development of new deep learning methods for image processing is often a slow process based on small incremental steps from previous model iterations. Therefore, a key to understanding the way WNet "works" is to analyze the outputs of its individual modules.

\subsubsection{Sinogram-domain Denoising} \label{section:sinogramdenoising}

In Fig. \ref{fig:sinograminterpolation} it is noticeable that the geometry-aware interpolation $\mathcal{I}$ coupled with the measurement consensus step $\mathcal{C}$, reduces some of the streaking artifacts seen in the original $128$-view sparse reconstruction $\tilde{x}_{128}$. This quality improvement is also reflected in the quantitative measures of the two images: an increase of approx. $2$dB in PSNR and $7\%$ in SSIM for $\tilde{x}_{512}^{\mathcal{I}}$ over $\tilde{x}_{128}$ ($\tilde{x}_{512}^{\mathcal{I}}$ is the \correction{Ram-Lak filtered} FBP reconstruction of $y_{512}^{\mathcal{I}}$ from eq. \ref{eq:enhanced_sinogram}). However, there is no new information inferred in the enhanced sinogram $y_{512}^{\mathcal{I}}$ (eq. \ref{eq:enhanced_sinogram}). Only an optimal sinogram-denoising operator $\bm{\hat{U}_s}$ significantly reduces the amount of streaking and interpolation artifacts in the sinogram, such that in its corresponding reconstruction $\tilde{x}_{512}^{\bm{\hat{U}_s}}$ contrast and features are remarkably restored to some extent (see Sec. \ref{section:artifact_suppression} for details on how $\tilde{x}_{512}^{\bm{\hat{U}_s}}$ is computed). The PSNR and SSIM metrics reflect the increase in information recovered in the output with an increase of another 7dB in PSNR and 20\% in SSIM over $\tilde{x}_{512}^{\mathcal{I}}$.

\subsubsection{Reconstruction module fine-tuning} \label{section:finetuning}

As shown in Fig. \ref{fig:finetunedfilter}, the learned filter reveals similarities to other filters in the low-frequency domain. It starts to diverge from the Ram-Lak filter once it reaches the medium-frequency sector, its values lying close to the Cosine filter. In the high-frequency domain, coefficients diverge from the Cosine filter values towards the Shepp-Logan kernel. Note that, the learned filter is optimized for the specific sparse-view reconstruction problem ($k=128$ and $K=512$) and might require re-training for a different \correction{\st{measurement} sinogram} sparsity factor.

\subsubsection{Image-domain Denoising}
The results in Fig. \ref{fig:imagedenoising} show that $U_i$ is a necessary step to handle the \enquote{left-over} sparse-artifacts from the previous modules. The natural features present in the image, compared to a sinogram, enable this module to learn more meaningful representations and to recover small and very-small details in the image, thus creating a better result \correction{both qualitatively and quantitatively}.

\subsection{Generalization} \label{section:discussiongeneralization}

\renewcommand{\x}{0.32}
\begin{figure}[!h]
\centering
\begin{varwidth}{0.02\linewidth}
  		\begin{turn}{90}
  		\footnotesize Ground Truth
  		\end{turn}
\end{varwidth}
\begin{varwidth}{0.97\linewidth}
	\begin{varwidth}{\x\linewidth}
	  		\centering
	  		\centerline{\footnotesize Abdomen}\smallskip
	  		\centerline{\includegraphics[width=\linewidth]{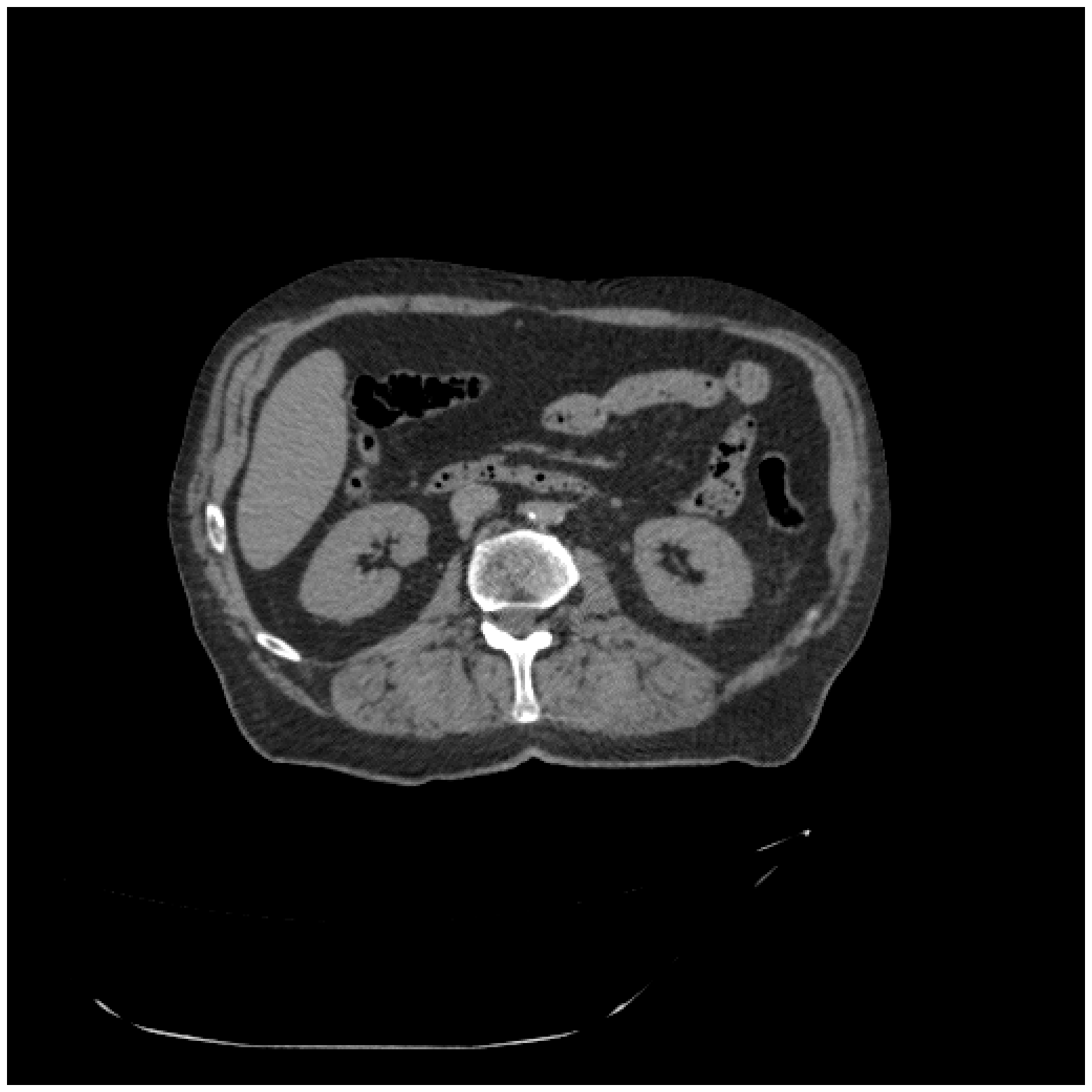}} 
	\end{varwidth}
		\begin{varwidth}{\x\linewidth}
	  		\centering
	  		\centerline{\footnotesize Chest}\smallskip
	  		\centerline{\includegraphics[width=\linewidth]{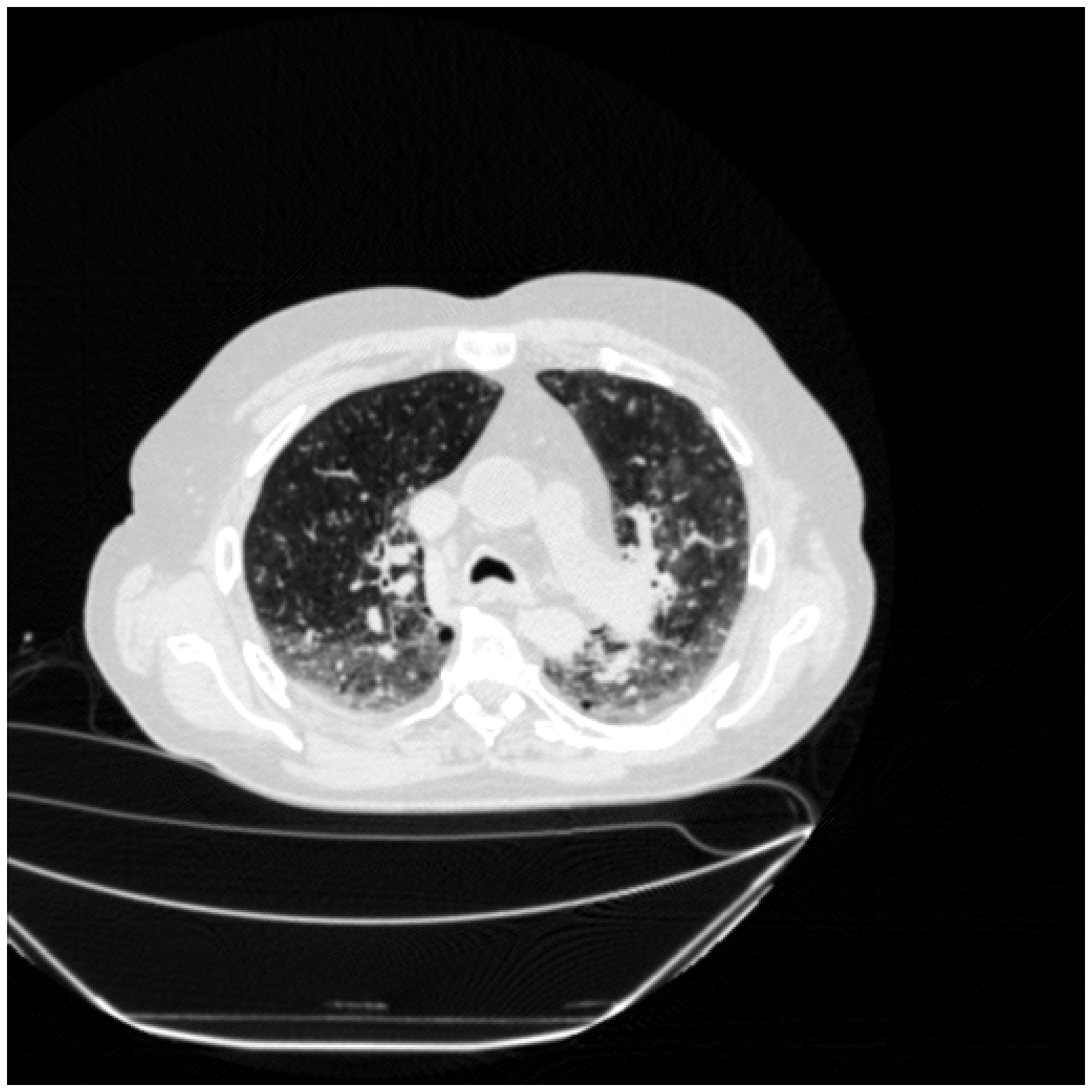}} 
	\end{varwidth}
	\begin{varwidth}{\x\linewidth}
	  		\centering
	  		\centerline{\footnotesize Head}\smallskip
	  		\centerline{\includegraphics[width=\linewidth]{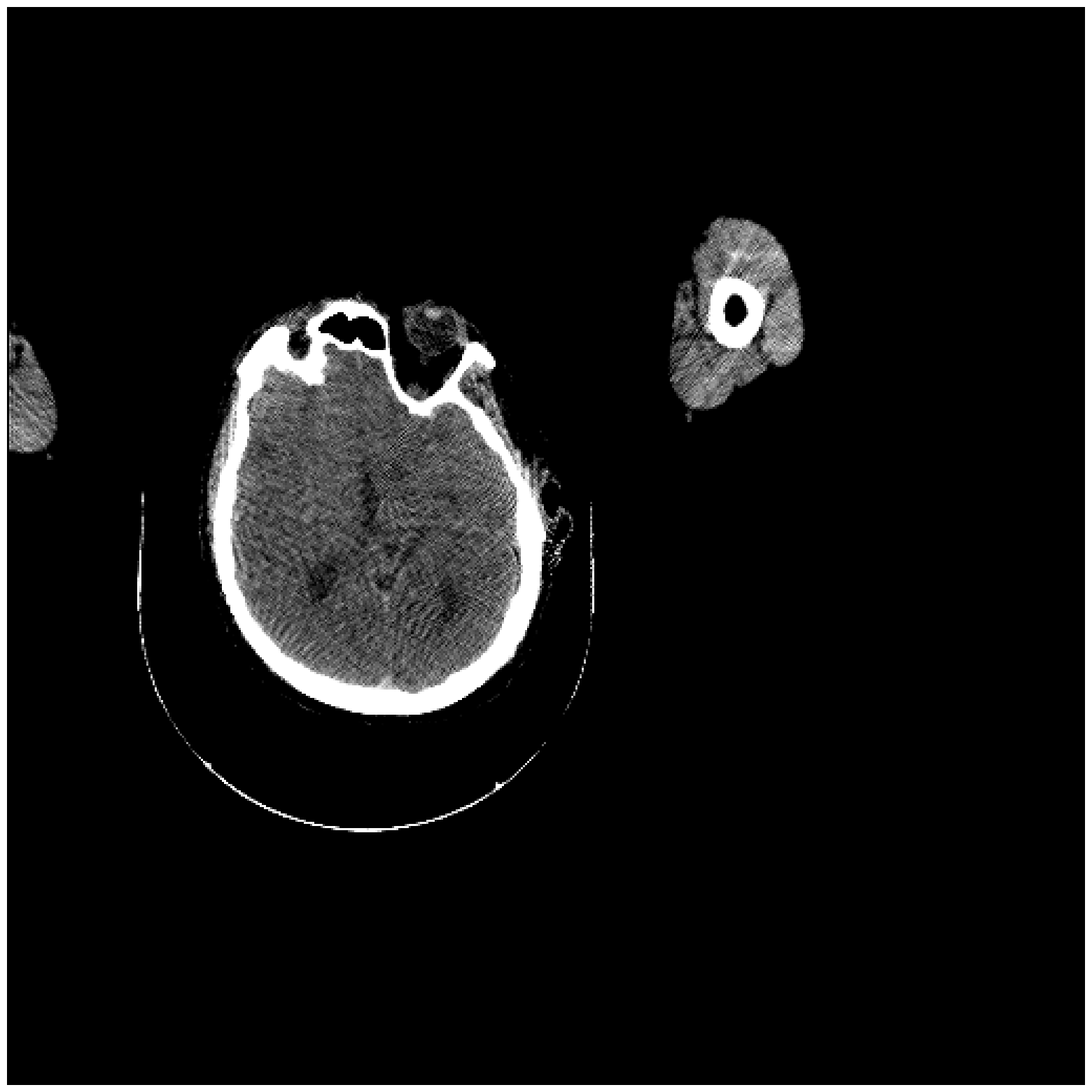}} 
	\end{varwidth}
\end{varwidth}
\\
\vspace{1em}
\begin{varwidth}{0.02\linewidth}
  		\begin{turn}{90}
  		\footnotesize DRONE
  		\end{turn}
\end{varwidth}
\begin{varwidth}{0.97\linewidth}
	\begin{varwidth}{\x\linewidth}
	  		\centering
	  		\centerline{\includegraphics[width=\linewidth]{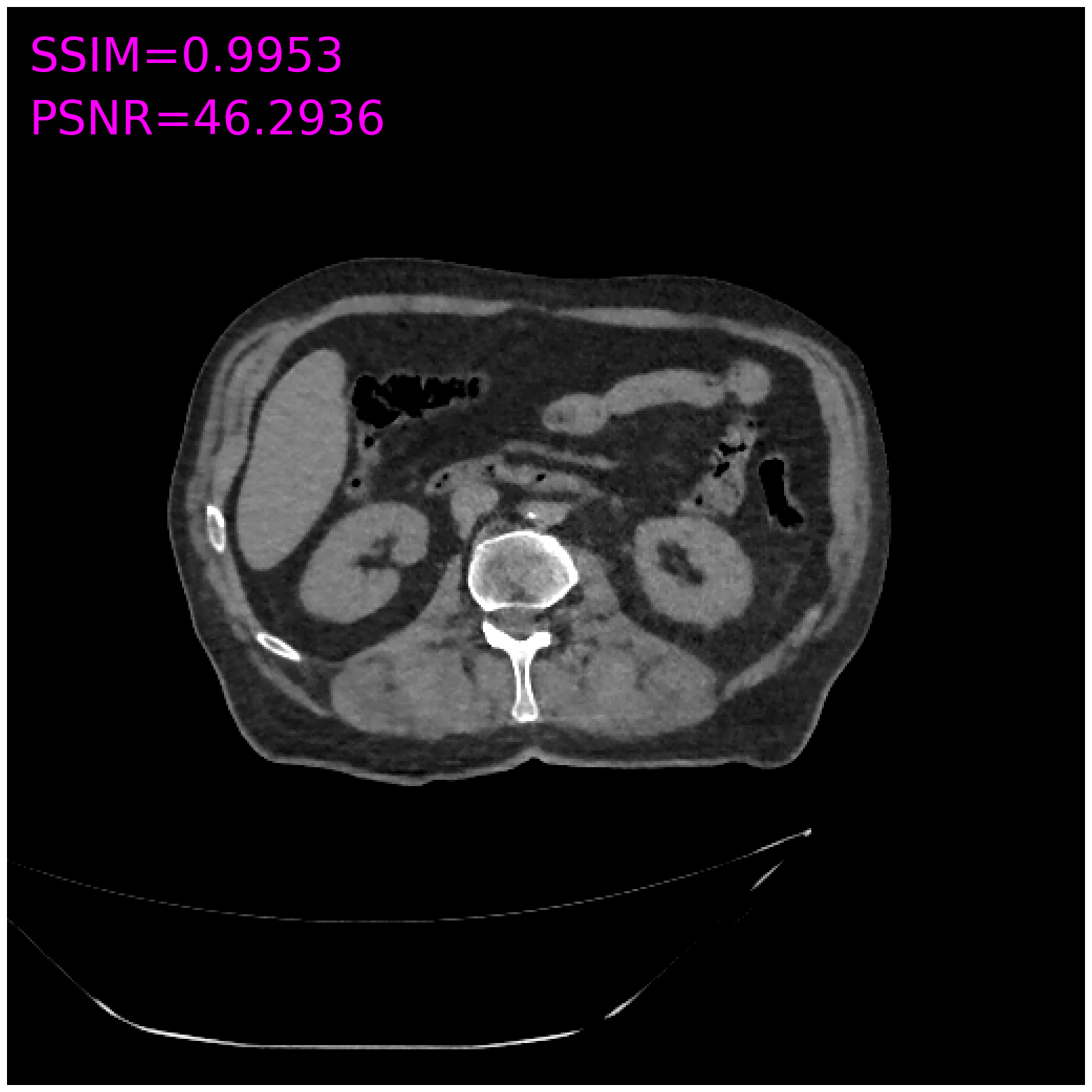}} 
	\end{varwidth}
		\begin{varwidth}{\x\linewidth}
	  		\centering
	  		\centerline{\includegraphics[width=\linewidth]{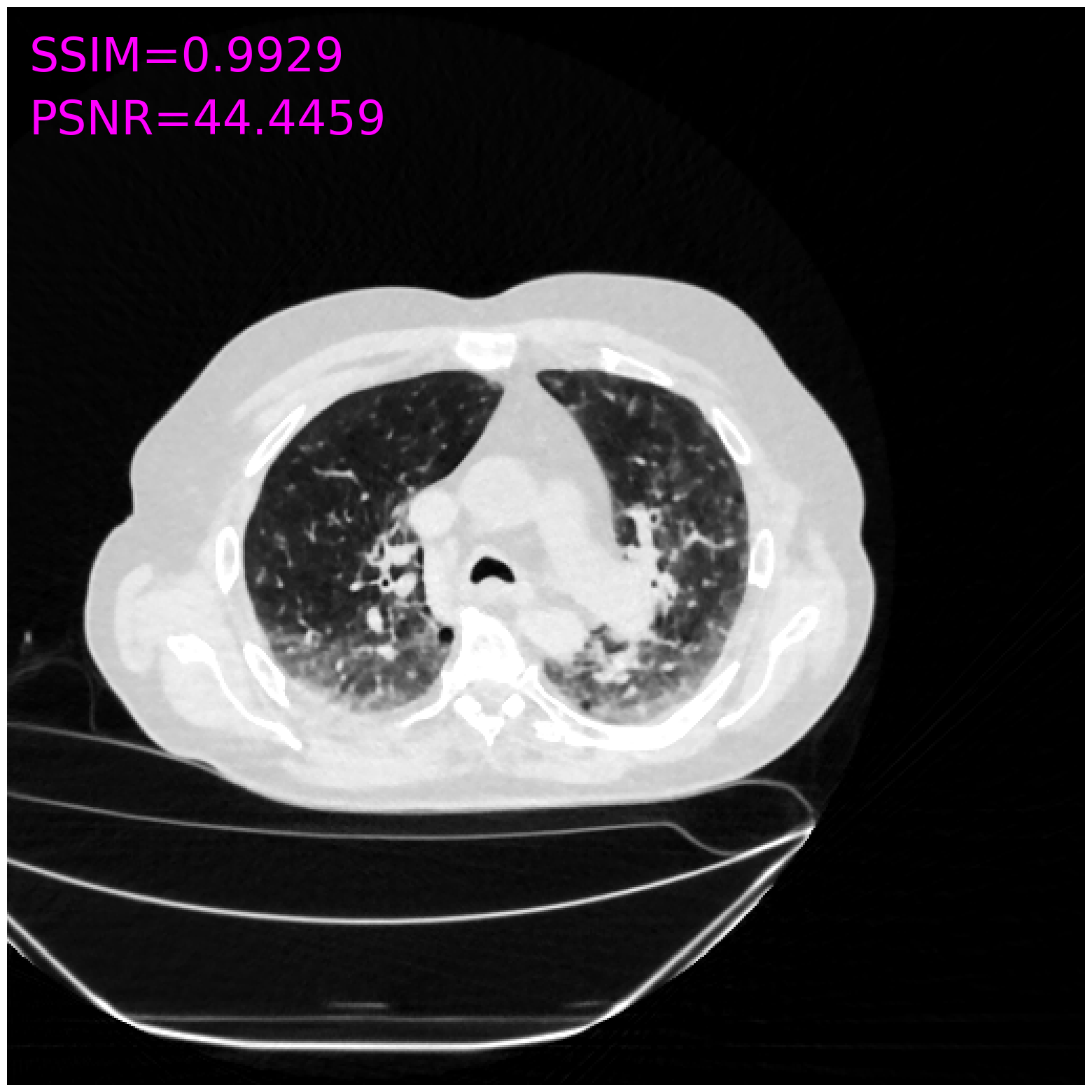}} 
	\end{varwidth}
	\begin{varwidth}{\x\linewidth}
	  		\centering
	  		\centerline{\includegraphics[width=\linewidth]{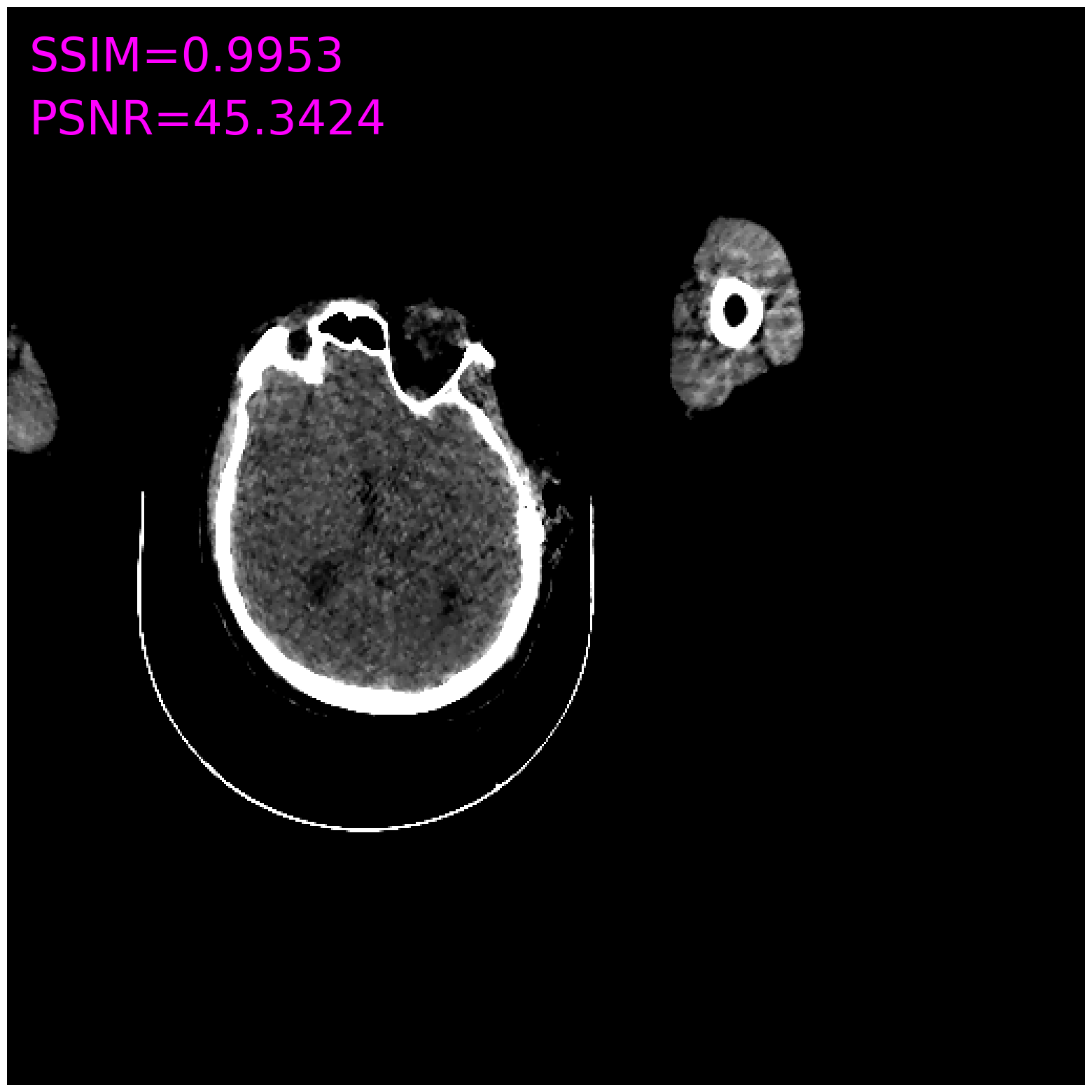}} 
	\end{varwidth}
\end{varwidth}
\\
\begin{varwidth}{0.02\linewidth}
  		\begin{turn}{90}
  		\footnotesize DRONE Diff.
  		\end{turn}
\end{varwidth}
\begin{varwidth}{0.97\linewidth}
	\begin{varwidth}{\x\linewidth}
	  		\centering
	  		\centerline{\includegraphics[width=\linewidth]{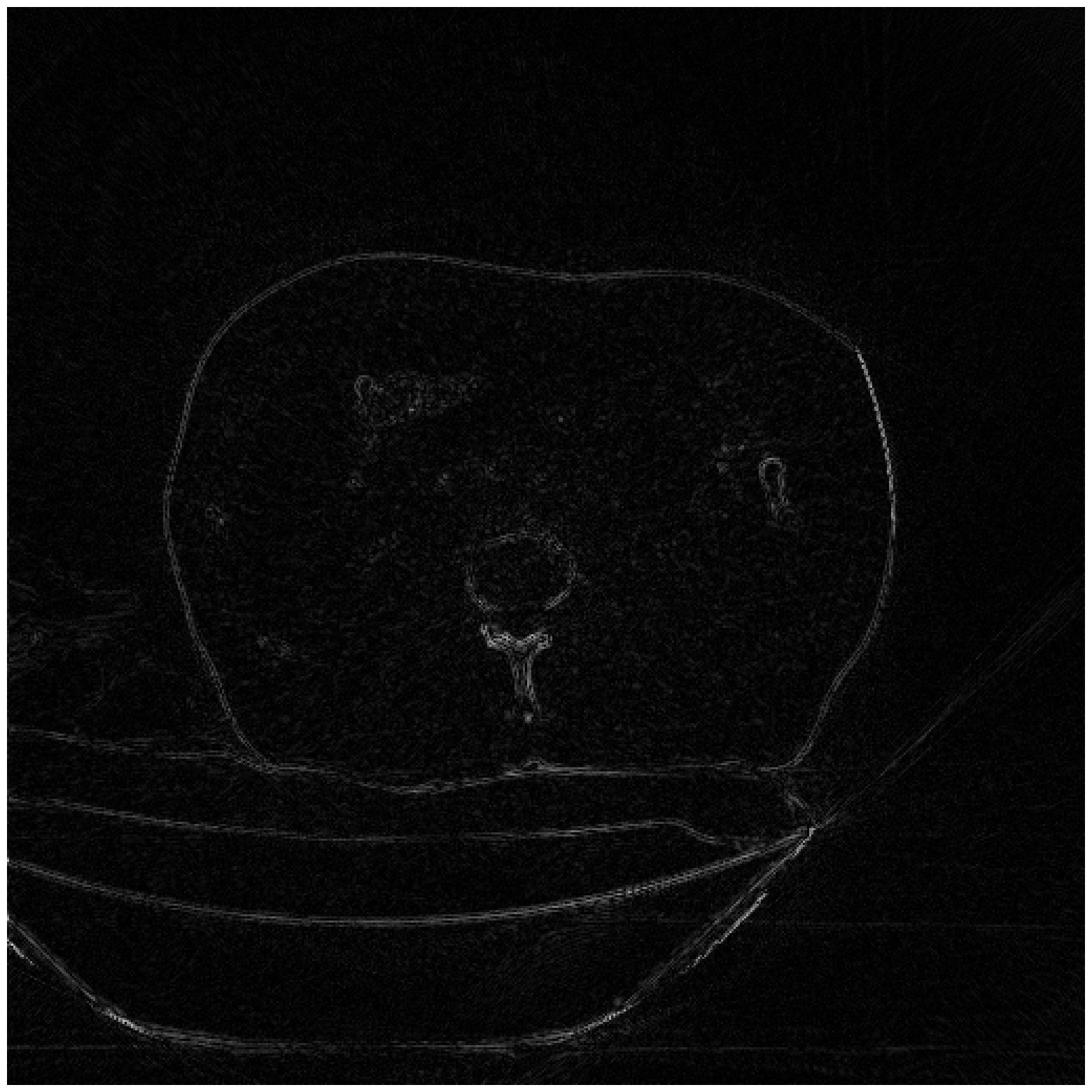}} 
	\end{varwidth}
		\begin{varwidth}{\x\linewidth}
	  		\centering
	  		\centerline{\includegraphics[width=\linewidth]{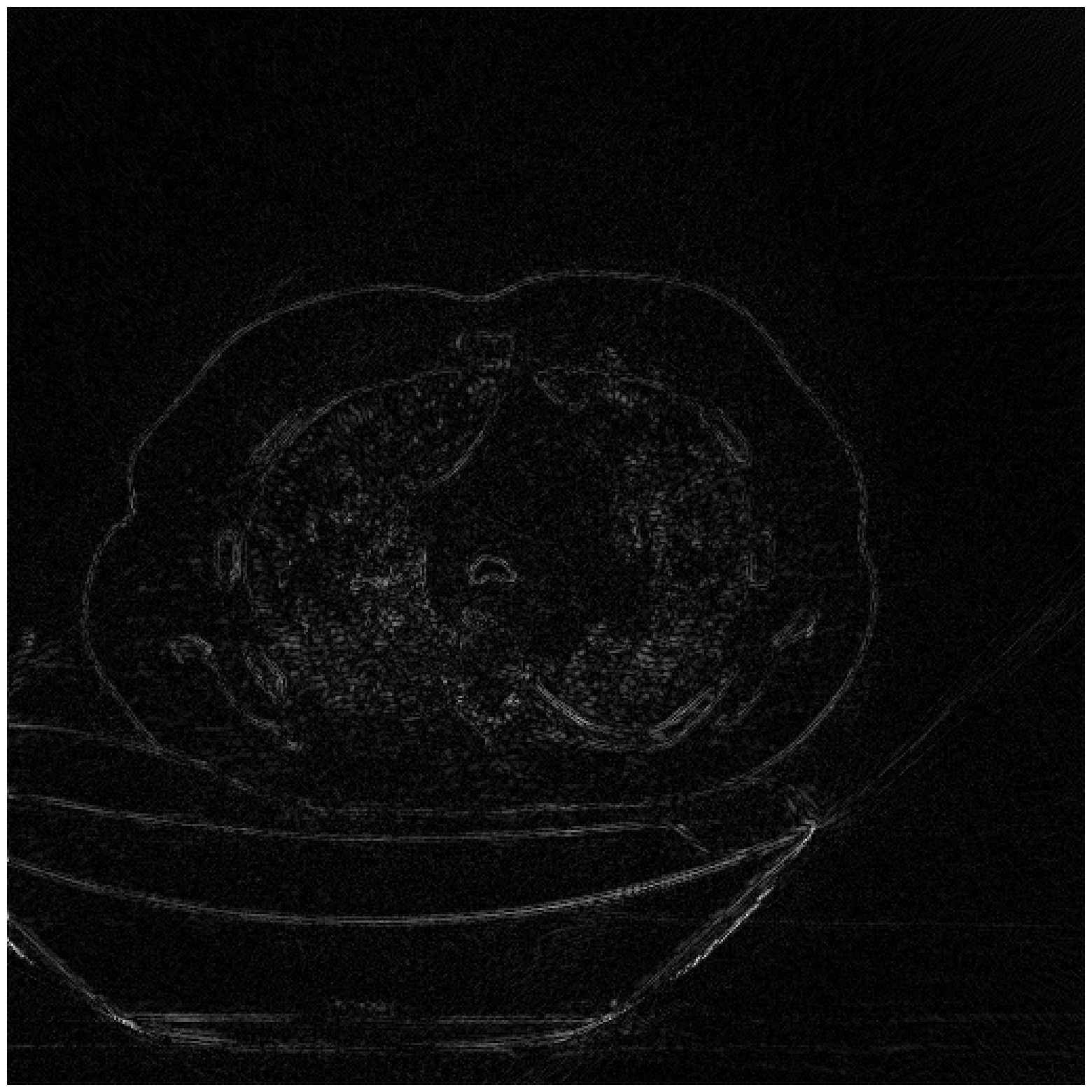}} 
	\end{varwidth}
	\begin{varwidth}{\x\linewidth}
	  		\centering
	  		\centerline{\includegraphics[width=\linewidth]{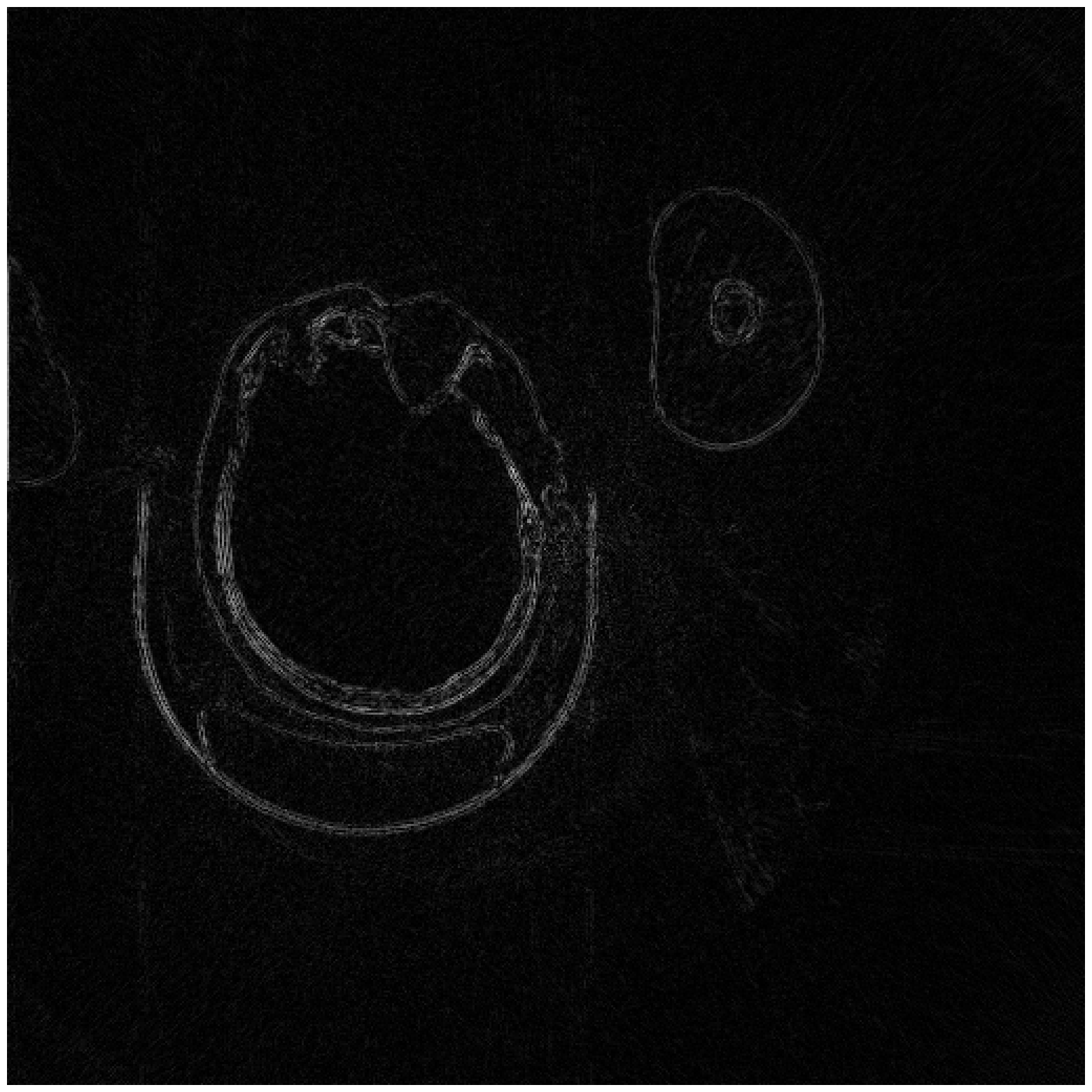}} 
	\end{varwidth}
\end{varwidth}
\\
\vspace{1em}
\begin{varwidth}{0.02\linewidth}
  		\begin{turn}{90}
  		\footnotesize WNet
  		\end{turn}
\end{varwidth}
\begin{varwidth}{0.97\linewidth}
	\begin{varwidth}{\x\linewidth}
	  		\centering
	  		\centerline{\includegraphics[width=\linewidth]{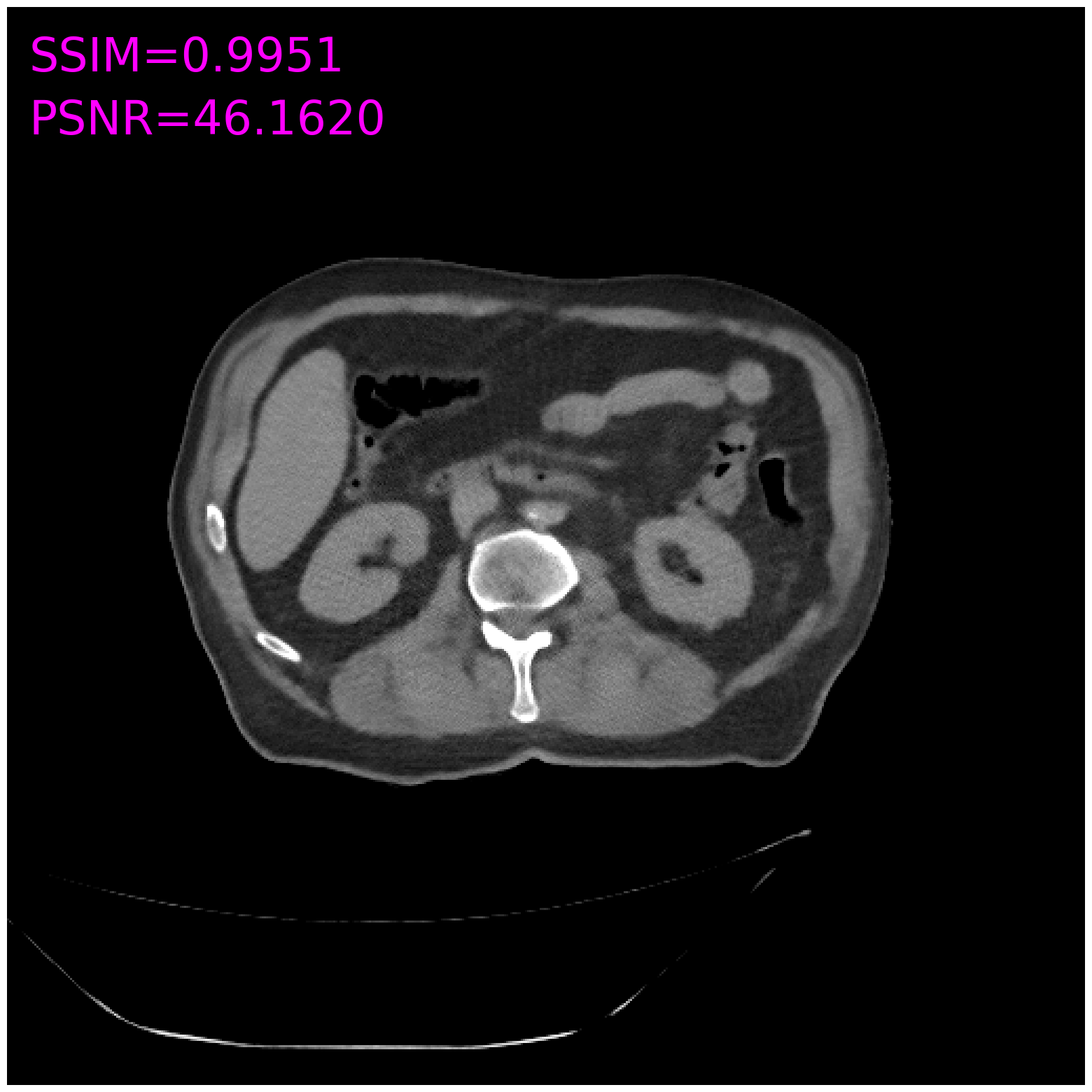}} 
	\end{varwidth}
		\begin{varwidth}{\x\linewidth}
	  		\centering
	  		\centerline{\includegraphics[width=\linewidth]{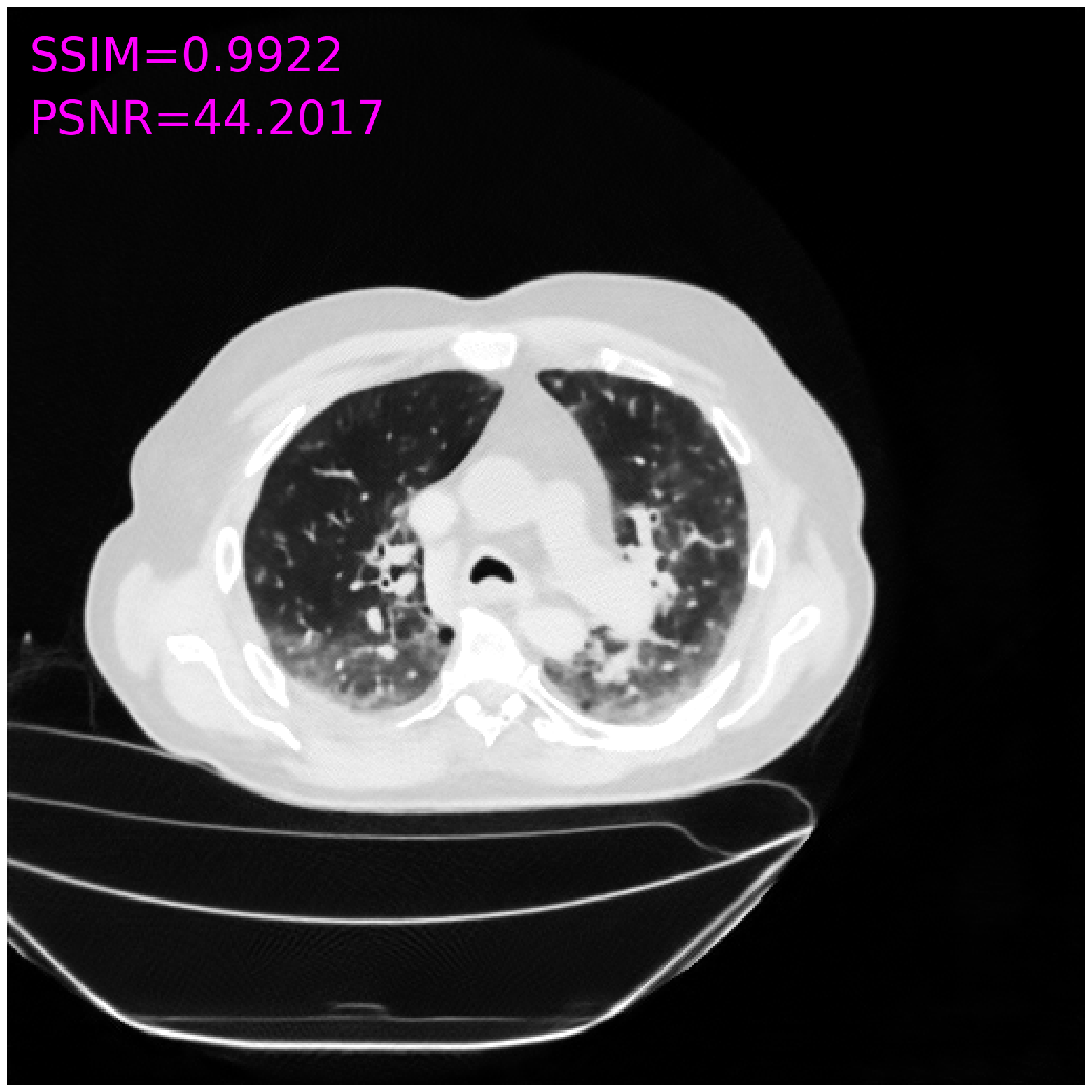}} 
	\end{varwidth}
	\begin{varwidth}{\x\linewidth}
	  		\centering
	  		\centerline{\includegraphics[width=\linewidth]{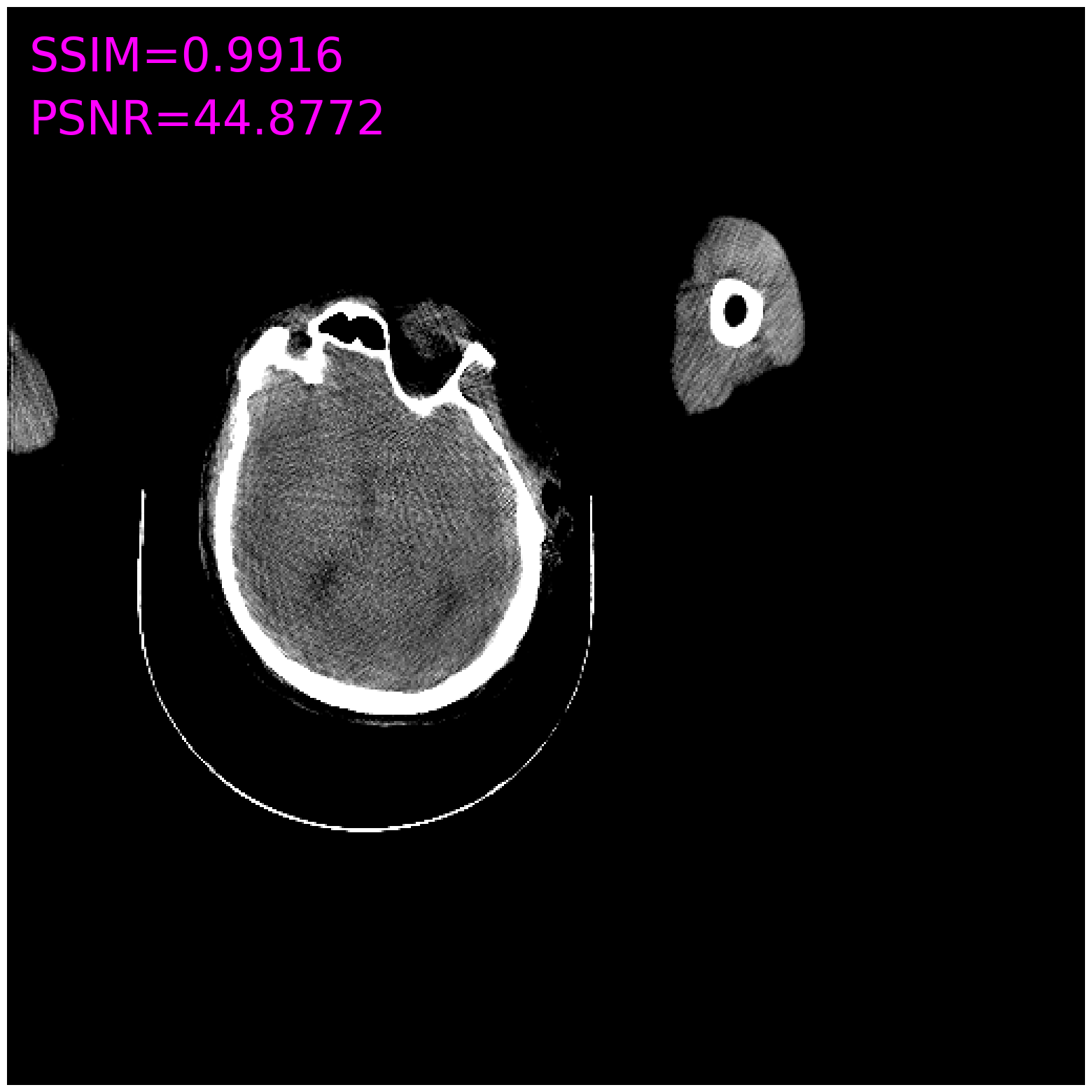}} 
	\end{varwidth}
\end{varwidth}
\\
\begin{varwidth}{0.02\linewidth}
  		\begin{turn}{90}
  		\footnotesize WNet Diff.
  		\end{turn}
\end{varwidth}
\begin{varwidth}{0.97\linewidth}
	\begin{varwidth}{\x\linewidth}
	  		\centering
	  		\centerline{\includegraphics[width=\linewidth]{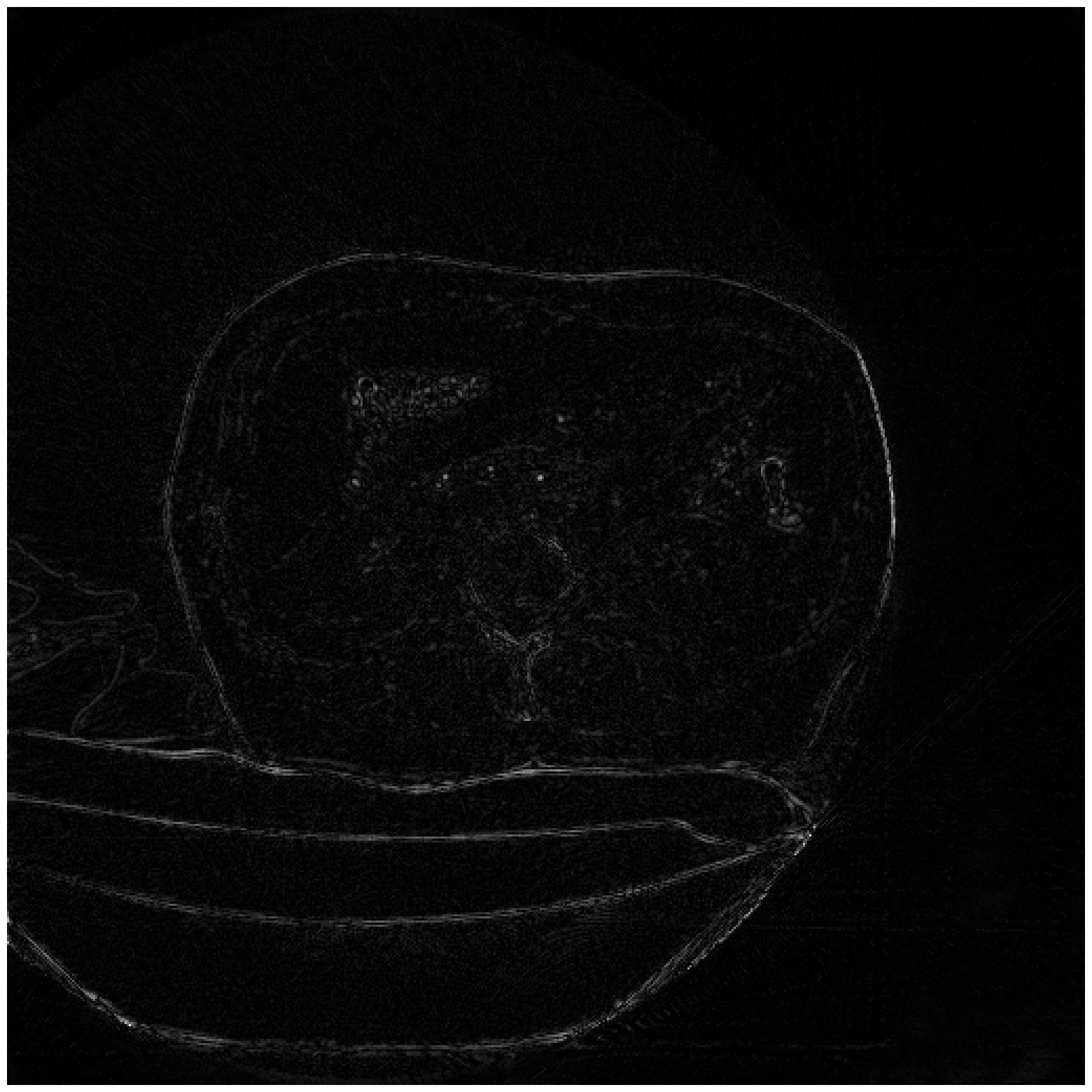}} 
	\end{varwidth}
		\begin{varwidth}{\x\linewidth}
	  		\centering
	  		\centerline{\includegraphics[width=\linewidth]{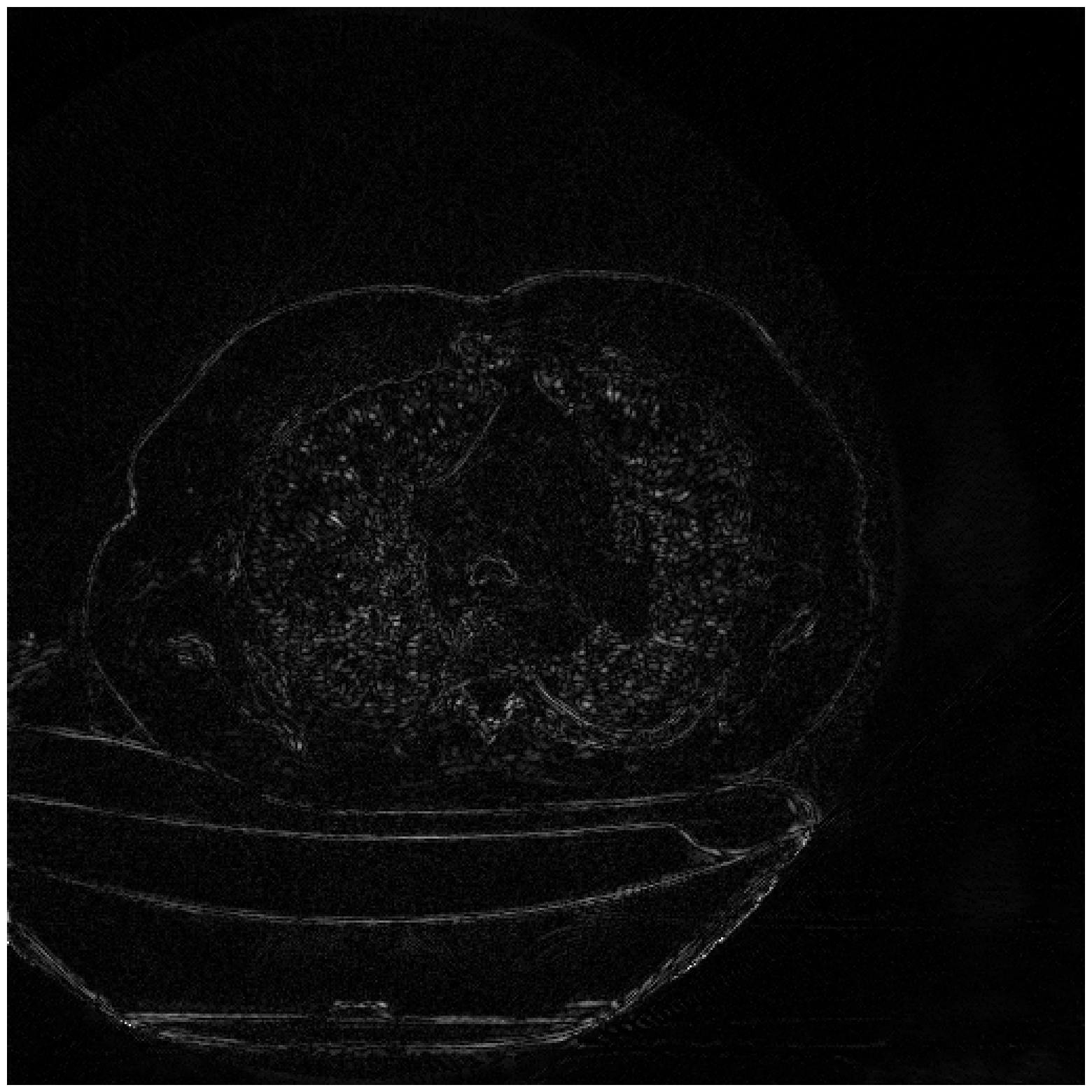}} 
	\end{varwidth}
	\begin{varwidth}{\x\linewidth}
	  		\centering
	  		\centerline{\includegraphics[width=\linewidth]{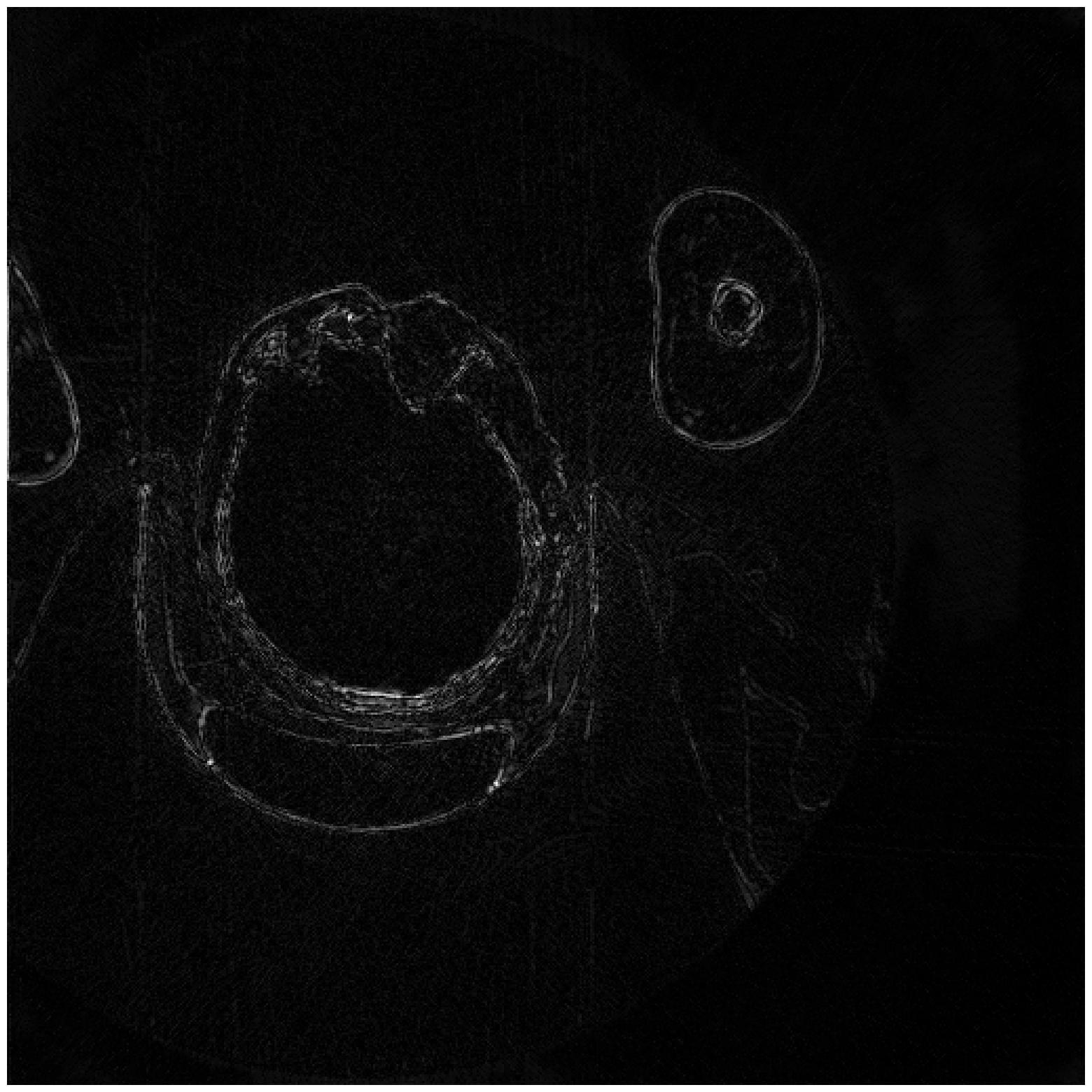}} 
	\end{varwidth}
\end{varwidth}
\caption{Generalization results of DRONE and WNet for scans that were acquired and reconstructed with different parameters compared to the ones in Fig. \ref{fig:main-results-1} (see Sec. \ref{section:generalization}). The SSIM and PSNR values with respect to the 2048-view \textbf{Cosine} filtered FBP reconstruction \enquote{Ground Truth} were added in the top-left of each image (magenta). The display window for the Abdomen results is [-350,450] HU, for the Chest results is [-1024,150] HU, and for the Head results is [-40,120] HU. The display window for the difference images is [0,400] HU.}
\label{fig:generalization}
\end{figure}

In Fig. \ref{fig:generalization}, we consider that the better generalizability performance of the DRONE model is a direct result of its iterative denoising step which can suppress overfitting artifacts generated by the trained network. As WNet lacks such an iterative step \correction{and as the training set does not cover head and abdomen data}, its results won't match the quality of those obtained with DRONE. However, the difference in computation time and network size is still the same, with DRONE taking over $70$s to finish computation while WNet under $2$s.

\subsection{Future Work} \label{section:futurework}
Embedding the reconstruction step inside the deep learning architecture and enabling the training of reconstruction parameters is a valuable tool that opens new possibilities for more complex approaches like variational layers or unrolled regularized iterative algorithms. Furthermore, the two encoder-decoder modules in WNet, can be extended with additional residual connections, as it can be seen in DD-Net, or can be replaced by competitive WGAN implementations with feature losses as a future improvement.

\section{Conclusion} \label{section:conclusion}

We propose a DL-based denoising method for Sparse-view Computed Tomography, enabling joint training of both the denoising operators and the projection filtering operation performed as part of the FBP step. In this study, we have focused on a particular SVCT setup with a relatively low number of measurements, and we have shown consistent good results compared to state-of-the-art methods. We have also shown that our method \correction{\st{is competitive}competes} with iterative algorithms with a deep learning prior like the DRONE algorithm, while featuring a significantly reduced computational cost and time.

We also show that a trained data-driven precision learning approach still has the ability to generalize quite acceptably to new data obtained with a different acquisition protocol.
\FloatBarrier

\renewcommand*{\bibfont}{\footnotesize}
\printbibliography

\end{document}